\documentclass[aps,pra,twocolumn,final,floatfix,superscriptaddress,longbibliography,10pt]{revtex4-2}

\usepackage[utf8]{inputenc}
\usepackage{physics}
\usepackage{mathtools}
\usepackage{dsfont}
\usepackage{amsmath}
\usepackage{xcolor}
\usepackage{tikz}
\usepackage{qcircuit}
\usepackage{soul}
\usepackage{bbold}
\usepackage{tabularray}
\usepackage{graphicx}
\usepackage[export]{adjustbox}
\usepackage{multirow}

\usepackage[breaklinks=true,colorlinks,citecolor=blue,linkcolor=blue,urlcolor=blue]{hyperref}

\usetikzlibrary{graphs,graphs.standard}

\begin{document}

\title{Quantum Algorithm for Green's Functions Measurements in the Fermi-Hubbard Model}

\author{Gino Bishop}
\thanks{Corresponding author: g.bishop@fz-juelich.de}
\affiliation {Institute for Quantum Computing Analytics (PGI-12), Forschungszentrum J\"ulich, 52425 J\"ulich, Germany}
\affiliation {Theoretical Physics, Saarland University, 66123 Saarbr{\"u}cken, Germany}
\affiliation{Mercedes-Benz AG, Stuttgart, Germany}

\author{Dmitry Bagrets} 
\affiliation{Institute for Quantum Computing Analytics (PGI-12), Forschungszentrum J\"ulich, 52425 J\"ulich, Germany}
\affiliation{Institute for Theoretical Physics, University of Cologne, 50937 K{\"o}ln, Germany}

\author{Frank K. Wilhelm} \affiliation{Institute for Quantum Computing Analytics (PGI-12), Forschungszentrum J\"ulich, 52425 J\"ulich, Germany} \affiliation{Theoretical Physics, Saarland University, 66123 Saarbr{\"u}cken, Germany}
\date{\today}

\begin{abstract}
    In the framework of the hybrid quantum-classical variational cluster approach (VCA) to strongly correlated electron systems one of the goals of a quantum subroutine is to find single-particle correlation functions of lattice fermions in polynomial time. Previous works suggested to use variants of the Hadamard test for this purpose, which requires an implementation of controlled single-particle fermionic operators. However, for a number 
    of locality-preserving mappings to encode fermions into qubits, a direct construction of such operators is not possible. 
    In this work, we propose a new quantum algorithm, which uses an analog of the Kubo formula adapted to a quantum circuit simulating the Hubbard model. It allows to access the Green's function of a cluster directly using only bilinears of fermionic operators and circumvents the usage of the Hadamard test. We test our new algorithm in practice by using open-access simulators of noisy IBM superconducting chips. 
\end{abstract}

\maketitle

\section{Introduction}
Strongly correlated electron materials exhibit exotic phenomena such as high-temperature superconductivity \cite{Dagotto1994} and Mott-insulating phases \cite{Moeller1995}. Investigating these effects and their origins is therefore crucial for promoting sophisticated material design \cite{Andrei2020}, building high-fidelity superconducting qubits \cite{Siddiqi2021} and advanced energy storage \cite{energy_storage}. 
Strong correlation arises due to Coulomb interaction between electrons and can be captured within the Fermi-Hubbard model~\cite{Society1963}. With adding strong disorder, this model demonstrates 
yet another intriguing phenomenon, the many-body localization (MBL), 
both in one- and two-dimensions~\cite{Schreiber:2015, Bordia:2016, Bordia:2017}.

While the Fermi-Hubbard model comprehensively incorporates electronic correlations, it is not exactly solvable beyond one dimension. Numerical methods face an inherently exponential demand in computational resources, when transitioning to larger system sizes. The landmark paper on quantum supremacy \cite{supremacy} has sparked great interest in the research field of quantum computation, which has now advanced to a playground for sophisticated hardware and algorithms. Strategies for solving the Fermi-Hubbard model on a quantum computer have been proposed recently~\cite{Cade2020, Wecker2015}. Among them the cluster perturbation methods~\cite{Senechal2002} play a prominent role. 
They are capable of mitigating computational demands by dividing a lattice system into arbitrarily small, identical and disjoint clusters. The idea is to solve one of the many clusters and extrapolate the result to the full system in a self-consistent fashion. Specifically, the variational cluster approach (VCA) \cite{Potthoff2003} can be used to relate the free energy of a microscopic cluster to the grand canonical potential of a macroscopic system. When the latter is found, it provides an access to the phase diagram of a given material. As the crucial step, the VCA scheme involves an evaluation of the Green's function which describes one-particle correlations in the interacting system. In practical terms it amounts to an evaluation of the Green's function using relatively small quantum chips, which are build up of as many qubits as the number electronic orbitals contained in a cluster, plus one ancilla qubit. 

A common strategy for evaluating the Green's function of a correlated system relies on the Hadamard test~\cite{Somma2002, Dallaire-Demers2016, Dallaire-Demers20162, Bauer:2016, Libbi:2022}. A key requirement of this approach is the implementation of controlled single-particle fermionic operators. 
To that end, a mapping from the fermion to qubit Hilbert space needs to be chosen that allows for a single fermion representation on a quantum computer; a well-known example is the Jordan-Wigner transformation.

Recently, local fermion-to-qubit mappings that employ additional ancilla qubits have gained popularity~\cite{Verstraete2005, Whitfield2016, Steudner2019, BRAVYI2002210, Zhang:2019, Kanav:2019, bosonization, Nys2023}, including the so-called 'compact' encoding~\cite{DerbyKlassen}. These methods are particularly advantageous in two or higher dimensions and can be categorized into two main types.

The first approach optimizes the Jordan-Wigner transformation by introducing additional stabilizers. In these methods~\cite{Verstraete2005, Whitfield2016, Steudner2019}, both even and odd subspaces of the fermionic Hilbert space remain accessible.
The second class of methods~\cite{BRAVYI2002210, Zhang:2019, Kanav:2019, bosonization, Nys2023} has a distinct feature: they construct products of an {\it even} number of fermionic operators, such as bilinears, while single fermionic operators (in the most cases) remain inaccessible. As a result, while these latter approaches are promising for efficiently generating the evolution operator of the Fermi-Hubbard model, they pose a fundamental challenge for implementing the Hadamard test.
	
In this work, we propose a new quantum algorithm, which is rooted in the Kubo formula of linear response theory~\cite{altland_simons_2010}, and adapted here for quantum circuits.
This algorithm, which we coin as 'direct measurement', allows to access the Green's function using only bilinears of fermionic operators. 
Its construction is based on the algebra of Majorana operators and as such 
it is applicable to any local fermion-to-qubit mapping schemes. 
Thereby, it is irrelevant whether a particular mapping encodes the even and/or odd fermionic Hilbert subspace --- our algorithm is agnostic to those choices. Furthermore, the new algorithm does not require controlled single-particle fermionic operators.

This work is structured as follows. Section \ref{sec:VCA} briefly recapitulates the main idea of variational cluster approach (VCA) which may be viewed as a motivation to the efforts of investigating Green's functions. 
In section \ref{sec:CROTHB} we review how a unitary time evolution of the Fermi-Hubbard model can be encoded into a quantum circuit, using both the Jordan-Wigner and one of possible locality-preserving mappings~\cite{Nys2023}. We further introduce a new powerful algorithm inspired by linear response theory that superimposes the common Hadamard test for measuring the Green's function and which is agnostic to the details of a fermion-to-qubit encoding scheme. We then illustrate the proposed algorithm by discussing resulting circuits in details for two chosen mappings. 

Section \ref{sec:two_site_dimer} discusses a toy model of a two-site dimer, which we use to demonstrate the advantage of new algorithm. 
In Section \ref{sec:results} we present our results of a numerical verification of the algorithm, 
obtained with the help of open-access simulators of noisy IBM  superconducting chips.
We argue, that a scheme based on a direct Green's function measurement is a viable, potentially more powerful alternative to established methods such as the Hadamard test 
as it does not require a construction of controlled single fermionic operators. Thereby it can be adapted to any fermion-to-qubit encoding beyond the common Jordan-Wigner transformation.

\section{Fermi-Hubbard model within the VCA}
\label{sec:VCA}
In this introductory section we outline the basic idea behind the variational cluster approach (VCA)~\cite{Potthoff2003} and give a context for which the Fermi-Hubbard model is utilized. 
The recapitulation mainly serves for the purpose of demonstrating the usefulness of the 
quantum algorithms aimed at finding the correlation functions of moderately large clusters and their potential speed-up over purely classical methods of computation. We also introduce the Hamiltonian of the Hubbard model and the notation to be used across the paper.

The VCA is a method that allows for solving many-body systems in a self-consistent manner. In general, we assume that a many-body system is described by a lattice Hamiltonian $H$ of macroscopic size. 
While the number of qubits needed to encode the full Hilbert space of $H$ scales linearly with the number of sites, the VCA enables one to reduce the number of required qubits by investigating only a small, representative subset, i.e. a cluster, of the full lattice. These clusters are disjoint, identical copies of each other, whose Hamiltonian is denoted as $H'$. Since the cluster acts as a proxy to the full system, meaningful investigations can be carried out with a relatively small quantum chip. 

\begin{figure}[t]
	\centering
	\includegraphics[scale=0.165]{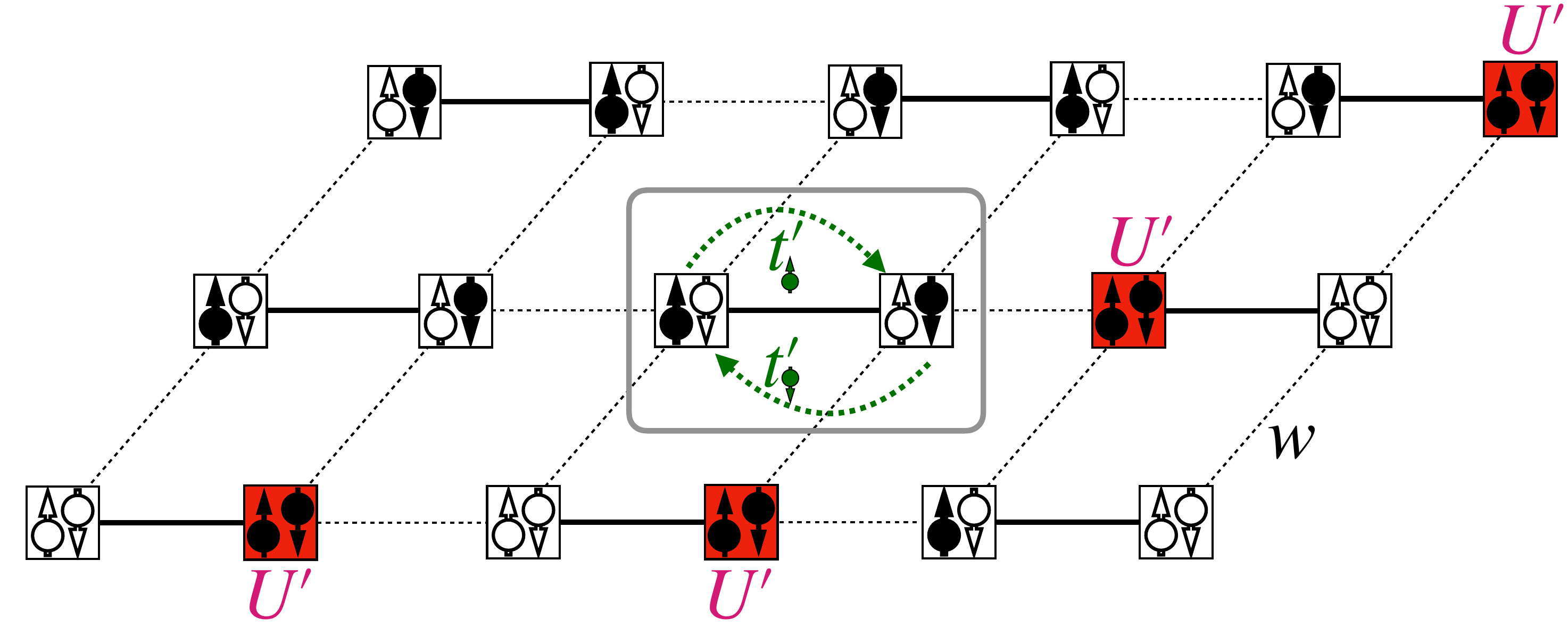}
	\bigskip
	\caption{(Color online) 2D lattice representing the Fermi-Hubbard model at half-filling. Each site can be occupied by at most two fermions. Dashed lines show the allowed hopping to neighboring sites. Two fermions occupying a single site experience a Coulomb repulsion $U$.}
	\label{fig:fh_lattice}
\end{figure} 

The Fermi-Hubbard model \cite{hubbard}, which is widely used to describe the physics of strongly correlated electrons, is a prototypical example where VCA can be successfully applied. 
In more concrete terms, the Hamiltonian of the Fermi-Hubbard model takes the form $H=H_0(\textbf{t})+H_1(\textbf{U})$. It is a sum of a kinetic short-range hopping term characterized by amplitudes $\textbf{t}$ and
a repulsive on-site interaction of strength $\textbf{U}$: 
\begin{equation}
\hat{H}=   \sum_{\langle i,j \rangle, \sigma} t_{ij} \hat{c}_{i,\sigma}^\dagger \hat{c}_{j, \sigma}+  \sum_i U_i \hat{n}_{i, \uparrow} \hat{n}_{i, \downarrow}.
\label{eq:FH_Hamiltonian}
\end{equation}
 Here, the operators ${c}_{i,\sigma}$ and ${c}^\dagger_{i,\sigma}$ destroy or create an electron with spin $\sigma$ on the $i$-th site, respectively,
 $\hat{n}_{i, \sigma} =  {c}^\dagger_{i,\sigma} {c}_{i,\sigma}$ are number operators, and the summation in the kinetic term goes over
 nearest neighbours.

A microscopically small cluster described by $H'$ cannot lead to long-range effects such as magnetism and superconductivity, which arise in the macroscopic system described by the full $H$. In order to impose these effects in the cluster, we can add symmetry breaking terms to the Hamiltonian which may promote different superconducting, ferromagnetic or charge-density orders. A connection between micro- and macroscopic systems is then established via their grand potentials. 
For mathematical details of this approach we refer the reader to Appendix~\ref{app:VCA_approach}. In the following section we proceed by constructing the unitary evolution operator $U(\tau) = e^{- i H' \tau}$ of the cluster
using a universal set of elementary gates.

\begin{figure*}[t!]
	\centering
	\includegraphics[scale=0.22]{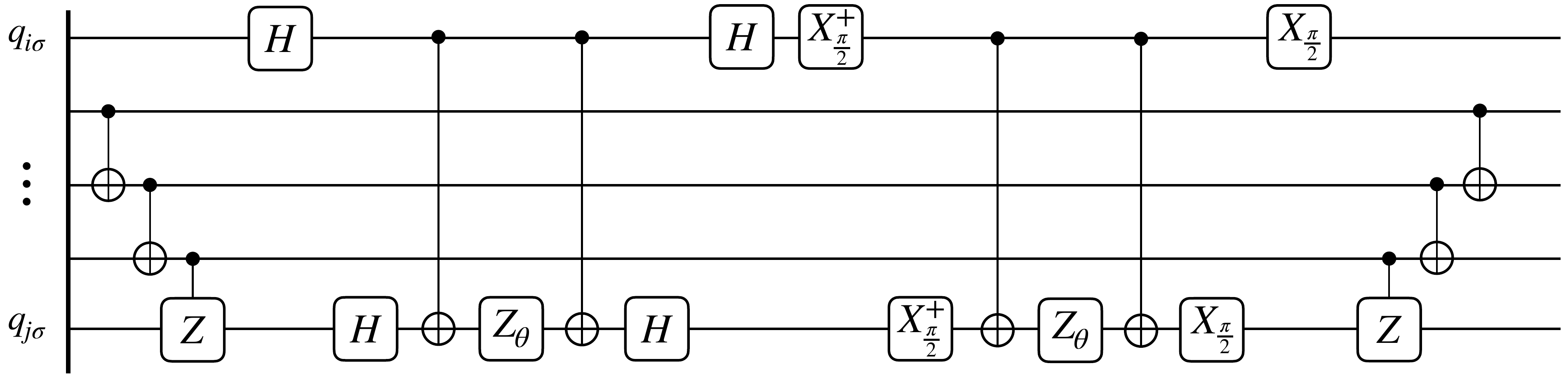}
	\caption{Hopping circuit to evaluate the term $t_{ij} (c^\dagger_{i, \sigma} c_{j, \sigma} + \text{h.c.})$  for a time step $\Delta \tau$ with
		angle $\theta = t_{ij} \Delta \tau $.}
	\label{fig:circuit_hopping}
\end{figure*}

\section{Circuit representation of the Hubbard model}
\label{sec:CROTHB}
    
We start this section by reviewing well known results in the literature~\cite{Wecker2015} on how a unitary time evolution of the Fermi-Hubbard
model can be represented by a quantum circuit and further on introduce an algorithm for the Green's function measurement based on the linear response theory.

\begin{figure}[b]
	\centering
	\includegraphics[scale=0.10]{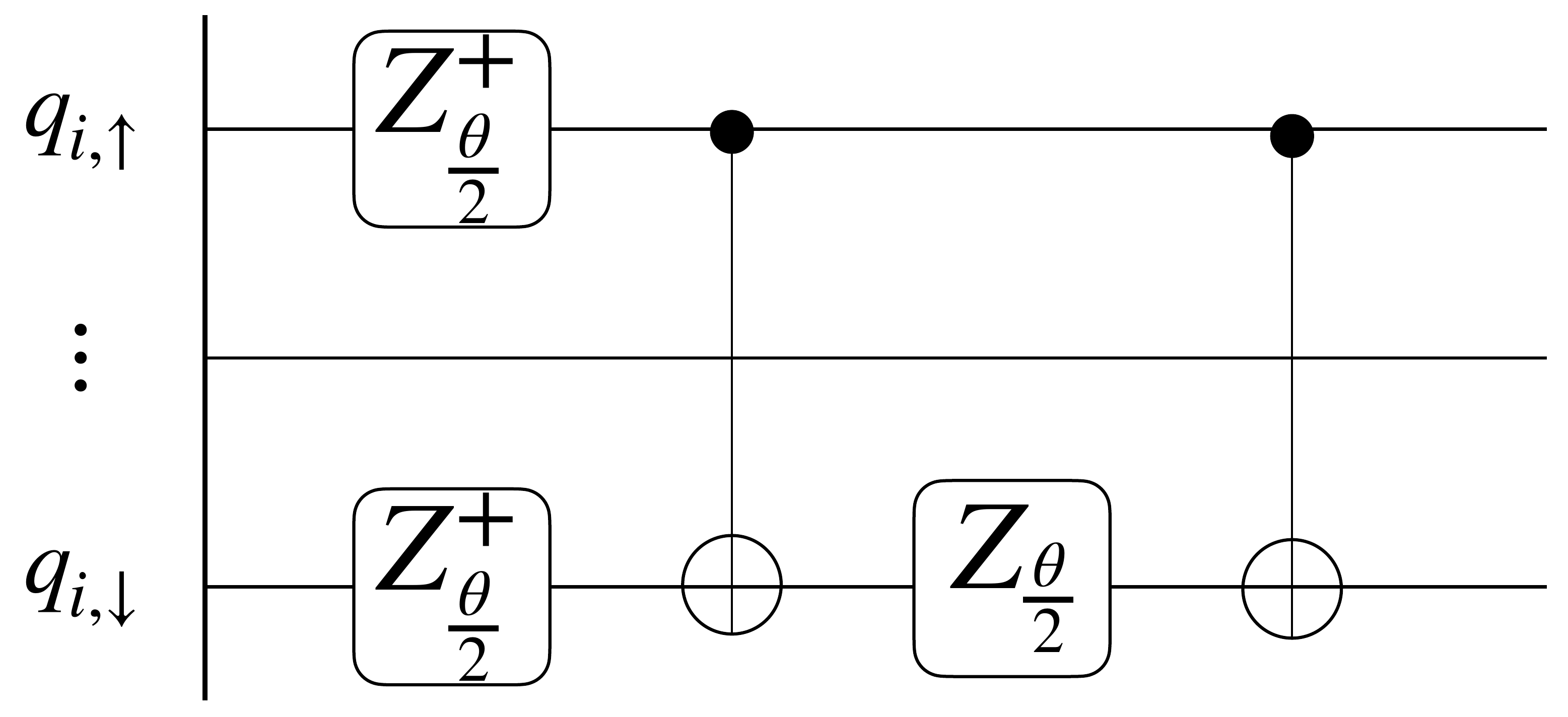}
	\caption{Repulsion circuit to evaluate the term $ U_i n_{i, \uparrow} n_{i, \downarrow}$ for a time step $\Delta \tau$ with angle
    $\theta = U_i \Delta \tau$.}
	\label{fig:circuit_repulsion}
\end{figure}

\subsection{Quantum circuits for hopping and repulsion terms}
The circuit representation of any fermionic Hamiltonian begins by choosing a fermion-to-qubit mapping. Among these, the Jordan-Wigner scheme is one of the most prominent examples, known for its simplicity and straightforward implementation --- and for the presence of Jordan-Wigner strings. In low-dimensional, small systems, it is a perfectly valid method. Yet, unfortunate cases of nearest-neighbor interactions may trigger local interactions that propagate through the full system in terms of $Z$-strings.
As these additional terms contribute to errors in quantum simulations, it is often advisable to choose a mapping that optimally balances qubit overhead, the ability to preserve locality and overall suitability for the specific problem at hand.

In this regard, locality-preserving mappings, aiming to
avoid the constructing of long Jordan-Wigner strings, provide a powerful alternative.
Famously, the Bravyi-Kitaev mapping~\cite{BRAVYI2002210} is less prone to non-locality. In this early landmark paper, it was discussed that only products (or sums thereof) of an even number of fermionic operators span the algebra of physical operators. It follows that for a number of more advanced 
mappings~\cite{Zhang:2019, Kanav:2019, bosonization, Nys2023}, single fermion operators 
can not be straightforwardly constructed either. As the latter 
are required by the Hadamard test to evaluate the Green's function, 
the VCA approach becomes harder to implement.
It is this fact that renders our new method of a direct measurement of a response function more physical and versatile as it can be applied to mappings beyond Jordan-Wigner. In the following we show the quantum circuits for our new algorithm resulting from the conventional (Jordan-Wigner) mapping, as well as from 
the novel locality-preserving mapping~\cite{Nys2023}. 

\subsubsection{Jordan-Wigner mapping}
\label{sec:QC_hoppings_and_U_terms}
To construct an evolution operator of the cluster Hubbard Hamiltonian related to a single Trotter step,  
one needs to map fermionic operators to the qubit ones. This can be achieved in two stages. First, we introduce \textit{Majorana fermions},
$x_{i\sigma} = c_{i\sigma} + c_{i\sigma}^\dagger$ and $y_{i\sigma} = i(c_{i\sigma} - c_{i\sigma}^\dagger)$, which are Hermitian operators.
They obey to the anti-commutation relations
\begin{eqnarray}
\{x_{i\sigma}, x_{j\sigma'}\} &=& \{y_{i\sigma}, y_{j\sigma'}\} = 2\delta_{ij} \delta_{\sigma \sigma'}, \nonumber \\
\label{eq:Majorana_brackets}
\{x_{i\sigma}, y_{j\sigma'}\} &=& 0.
\end{eqnarray}
At the second stage, the Jordan-Wigner transformation is used
to represent $x_{i\sigma}$ and $y_{i\sigma}$ via the following sequences of $X$-, $Y$ and $Z$-gates:
\begin{equation}
\begin{aligned}
\label{eq:JW_transform}
x_{i \uparrow} &= \mathbb{1}^{\otimes 2(N_c-i)+1}\otimes X \otimes Z^{\otimes2(i-1)}, \\
x_{i \downarrow} &= \mathbb{1}^{\otimes 2(N_c-i)}\otimes X \otimes Z^{\otimes2i-1},\\
y_{i\uparrow} &= -\mathbb{1}^{\otimes 2(N_c-i)+1}\otimes Y \otimes Z^{\otimes2(i-1)},\\
y_{i\downarrow} &= -\mathbb{1}^{\otimes 2(N_c-i)}\otimes Y \otimes Z^{\otimes2i-1},
\end{aligned}
\end{equation}
which guarantees satisfiability of the anti-commutation relations~\eqref{eq:Majorana_brackets}. 
Consequently, we define the correlation function for original fermions,
\begin{equation}
\label{eq:superposition_pauli}
    i G_{ij}^{\sigma \sigma'}(\tau) = \begin{pmatrix}
        \langle c_{i\sigma}(\tau) c_{j\sigma'}^\dagger(0)\rangle &  \langle c_{i\sigma}(\tau) c_{j\sigma'}(0)\rangle \\
         \langle c_{i\sigma}^\dagger(\tau) c_{j\sigma'}^\dagger(0)\rangle  &  \langle c_{i\sigma}^\dagger(\tau) c_{j\sigma'}(0)\rangle
    \end{pmatrix},
\end{equation}
and for Majorana ones
\begin{equation}
\label{eq:g_Majorana}
   i g_{ij}^{\sigma \sigma'}(\tau) =
    \begin{pmatrix}
        \langle x_{i\sigma}(\tau) x_{j\sigma'}(0)\rangle & \langle x_{i\sigma}(\tau) y_{j\sigma'}(0) \rangle \\
        \langle y_{i\sigma}(\tau) x_{j\sigma'}(0) \rangle  & \langle y_{i\sigma} (\tau) y_{j\sigma'}(0) \rangle
    \end{pmatrix},
\end{equation}
then the two are related by a unitary transformation
\begin{equation}
    G_{ij}^{\sigma \sigma'}(\tau) = \frac 12 M^\dagger g_{ij}^{\sigma \sigma'}(\tau) M, \quad 
    M = \frac{1}{\sqrt{2}} \begin{pmatrix}
        1 & 1 \\ i & -i
    \end{pmatrix}.
\end{equation}
For the sake of generality, we do not imply any time ordering in the definition of the Green's functions.

Following \cite{Wecker2015}, we then present quantum circuits for a time step $\Delta \tau$ for both hopping and repulsion as per the Fermi-Hubbard model, Eq. (\ref{eq:FH_Hamiltonian}). A circuit for the hopping term is shown in Fig. \ref{fig:circuit_hopping} and a circuit for the repulsion can be seen in Fig. \ref{fig:circuit_repulsion}. Here, the gate $X_{\pi/2}$ here refers to the Y-basis change gate. A complete list of applied gates in matrix representation is presented in Table \ref{tab:applied_gates} around Appendix~\ref{app:Simulation_Details}.

For our subsequent discussion of the Green's function measurement scheme (see Sec.~\ref{direct_measurement}) it is instructive to rationalize the circuit behind the hopping term, shown in Fig.~\ref{fig:circuit_hopping}. To this end we note that its representation in terms of Majorana fermions
reads
\begin{equation}
\label{eq:h_ij_f}
h_{ij}^{\sigma \sigma} = c_{i\sigma}^\dagger c_{j\sigma} + {\rm h.c.} = \frac i2 \left( y_{i\sigma} x_{j\sigma} - x_{i\sigma} y_{j\sigma} \right).
\end{equation}
The Jordan-Wigner transformation~(\ref{eq:JW_transform}) reduces the above operator to
\begin{equation}
\label{eq:h_ij}
  h_{ij}^{\sigma \sigma} = \frac 12 \left( X_m X_n +  Y_m Y_n\right) Z_{\rm JW}(m,n),
\end{equation}
where $m= 2i + (1-\sigma)/2$, $ n = 2 j + (1-\sigma)/2$ and $Z_{\rm JW}(m,n)$ stands for the Jordan-Wigner string,
\begin{equation}
   Z_{\rm JW}(m,n) = \bigotimes_{k=m+1}^{n-1} Z_k,
\end{equation}
with Pauli operators $X_k$, $Y_k$ and $Z_k$ acting on $k$-th qubit.
We can then introduce unitary Clifford gates $S_{mn}$ acting on all qubits $k$ with $m < k < n$ (its equivalent circuit is shown in Fig.~\ref{fig:definition_S_mn}), for which the role is 
to eliminate the Jordan-Wigner string and simplify~(\ref{eq:h_ij}) to
\begin{align}
    & h_{ij}^{\sigma \sigma} = \frac 1 2 S_{mn}^\dagger  \left( X_m X_n +  Y_m Y_n \right)   S_{mn}.
    \label{eq:S_mn}
\end{align}
The $XX$- and $YY$-terms above commute, so that an evolution operator generated by $h_{ij}^{\sigma \sigma}$ naturally splits into the product of two.
Subsequent unitary transformations using single qubit gates $H$ and $X_{\pi/2}$ transform each term of the sum in Eq.~(\ref{eq:S_mn}) to the product
$Z_m \otimes Z_n$. After that, the unitary evolution corresponding to a single Trotter step of a hopping operator with an angle 
$\theta = t_{ij} \Delta \tau$ is realized with the help of \text{Z}-rotations and additional similarity transformations with \text{CNOT} gates, as shown in Fig.~\ref{fig:circuit_hopping}.

\begin{figure}[t]
	\centering
	\includegraphics[scale=0.27]{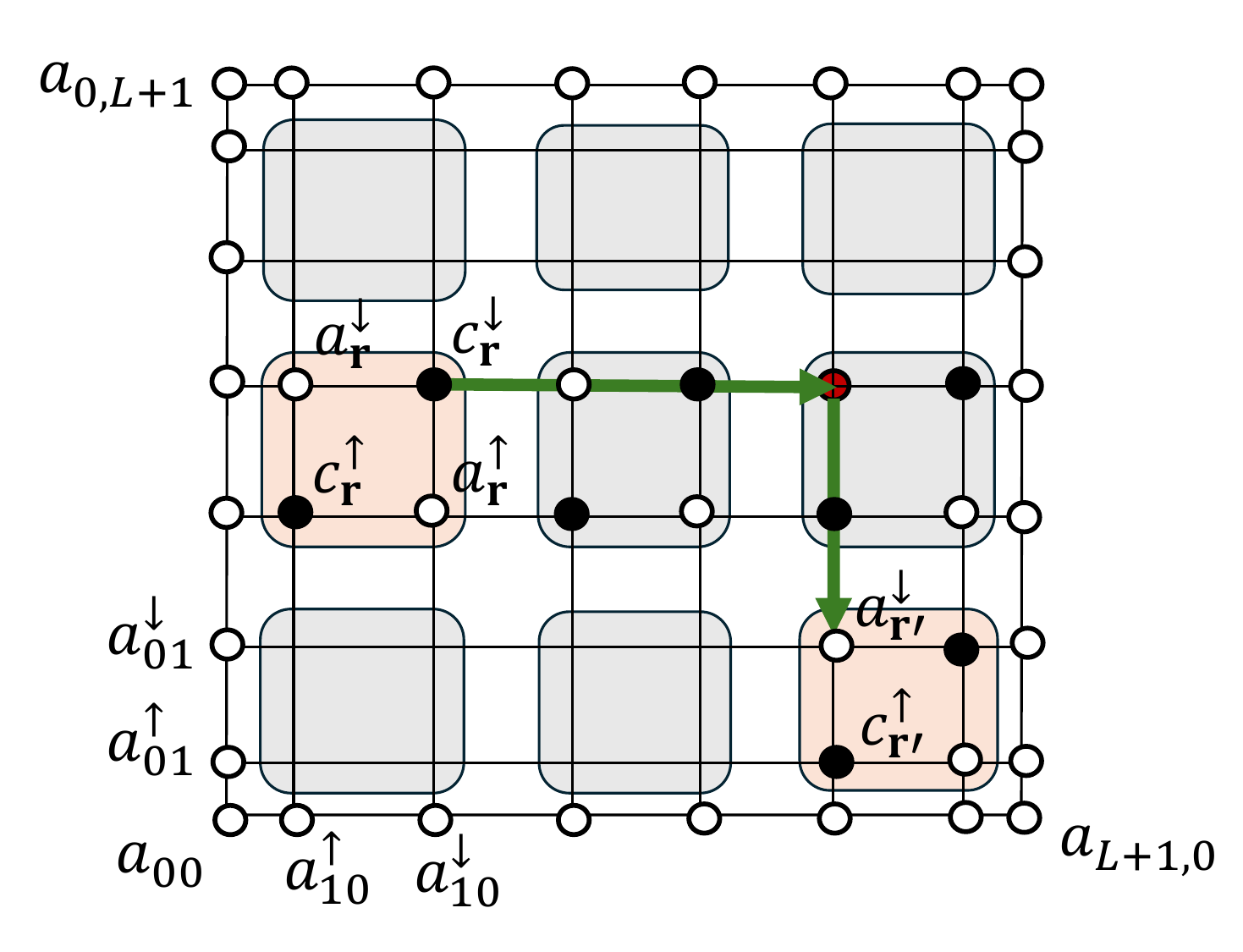}
	\caption{(Color online) Cluster of physical (black) and auxiliary (white) fermions used to implement quantum-classical variational cluster approach (VCA) for the 2D Hubbard model. }
	\label{fig:2DCluster}
\end{figure}

\begin{figure}[t]
	\centering
	\includegraphics[scale=0.32]{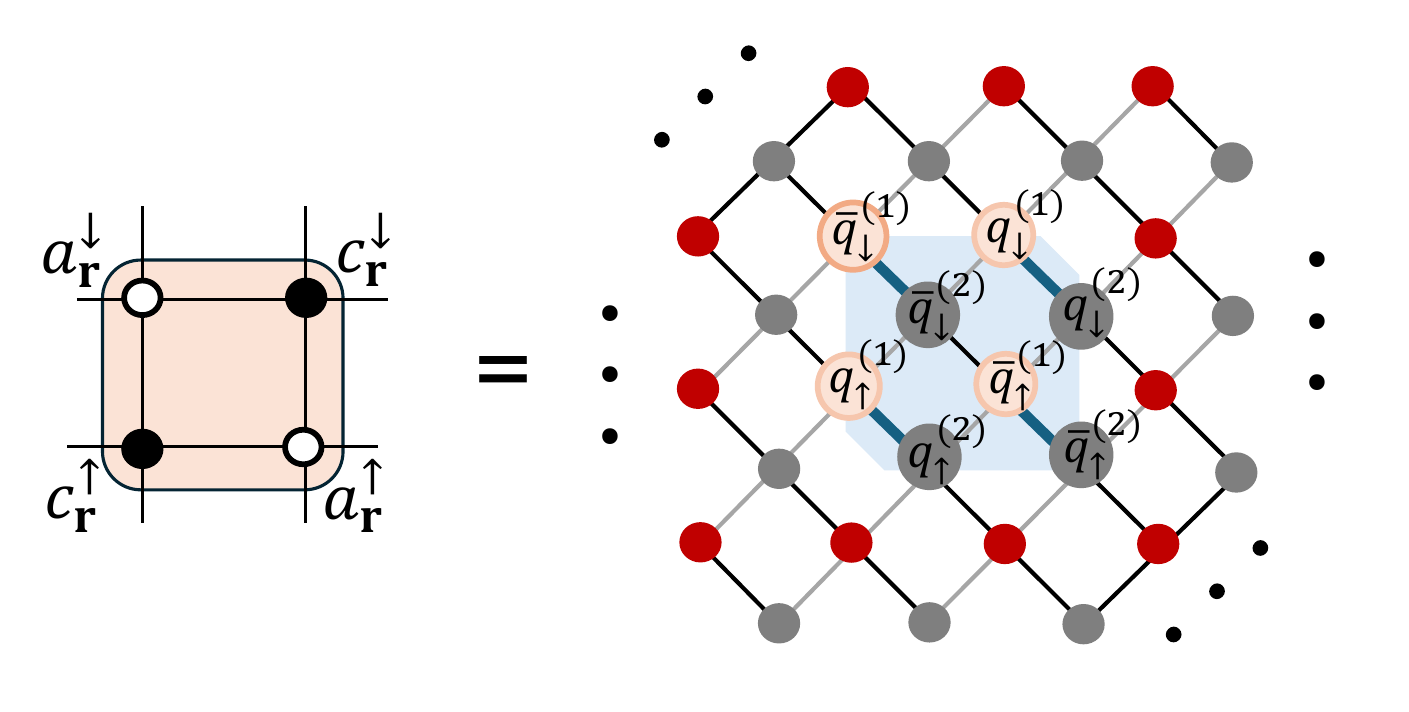}
	\caption{(Color online) Encoding four fermions comprising an elementary unit cell of the VCA cluster in terms of eight qubits. We denote by $q_{{\bf r}\sigma}$ the qubits associated with physical fermions $c_{\bf r}^\sigma$, and by $\overline{q}_{{\bf r}\sigma}$ the ones related to auxiliary fermions $a_{\bf r}^\sigma$ used for the measurement protocol. The pair of qubits $(q^{(1)}_\uparrow, q^{(2)}_\uparrow)$ is used to encode the Hilbert space of the physical fermion $c_{\bf r}^\uparrow$, where a spacial index ${\bf r}$ is omitted for brevity.}
	\label{fig:Unit_Cell}
\end{figure}

\subsubsection{Locality-preserving mappings}
\label{sec:QC_hoppings_and_U_terms_locality_preserving}
In this subsection we follow the approach of Refs.~\cite{Nys2022variational, Nys2023} to 
design the quantum algorithm for the Fermi-Hubbard model, which
is based on the local fermion-to-qubit mapping by Li and Po~\cite{bosonization}
for spinful fermions in two dimensions. 
In the framework of their scheme, we encode an $L \times L$ physical cluster as the
extended cluster of size $(2L+2) \times (2L+2)$ incorporating both physical ($c_{\bf r}^\sigma$) and auxiliary ($a_{\bf r}^\sigma$) fermions as shown in Fig.~\ref{fig:2DCluster}.

Each unit cell, depicted by either grey or orange squares, contains two physical fermions (located at black sites) and two auxiliary fermions (located at white sites). It turns out that such choice simplifies the implementation of the quantum algorithm for the Green's function measurement.  Additionally, it produces a square shape of a cluster with lengths $L_x=L_y = 2(L+1)$, enabling an efficient construction of the physical vacuum state within the local fermion-to-qubit mapping. By construction, 
the 'bosonization' scheme of Ref.~\cite{bosonization} assumes a torus geometry with
periodic boundary conditions. The latter is needed to formulate the non-contractible
Wilson loop constrains. To comply this periodicity restrictions with the VCA approach, 
for which open boundary conditions are assumed, we
add additional auxiliary fermions to all boundaries of the cluster.
By design, these auxiliary fermions are not connected by hopping terms --- neither
between each other nor to physical fermions. In this regard they are virtual and only become important
for the vacuum-state preparation on a 2D torus within the scheme of Li and Po, 
see more details on that in Ref.~\cite{Nys2023}.
With this trick, an implementation of any physical variational ground state of a cluster, free from periodicity constraints, becomes possible.

In the local scheme of fermion-to-qubit mapping, one doubles the number of qubits to encode $2^{2 L_x L_y}$ physical states, cf. Fig.~\ref{fig:Unit_Cell}. In this way each physical fermion mode $c_{\bf r}^\sigma$ is associated with a pair of qubits $(q^{(1)}_{{\bf r}\sigma}, q^{(2)}_{{\bf r}\sigma})$; and the auxiliary fermion mode $a_{\bf r}^\sigma$ is mapped onto the pair $(\bar q^{(1)}_{{\bf r}\sigma}, \bar q^{(2)}_{{\bf r}\sigma})$. Correspondingly, we denote Pauli operators acting on qubits $q^{(s)}_{{\bf r}\sigma}$ by $X^{(s)}_{{\bf r}\sigma}$, $Y^{(s)}_{{\bf r}\sigma}$ and $Z^{(s)}_{{\bf r}\sigma}$, with $s=1,2$. Similarly, Pauli matrices $\bar X^{(s)}_{{\bf r}\sigma}$, $\bar Y^{(s)}_{{\bf r}\sigma}$ and $\bar Z^{(s)}_{{\bf r}\sigma}$ act on auxiliary qubits $\bar q^{(s)}_{{\bf r}\sigma}$. 

\begin{figure}[t]
	\centering
	\includegraphics[scale=0.17]{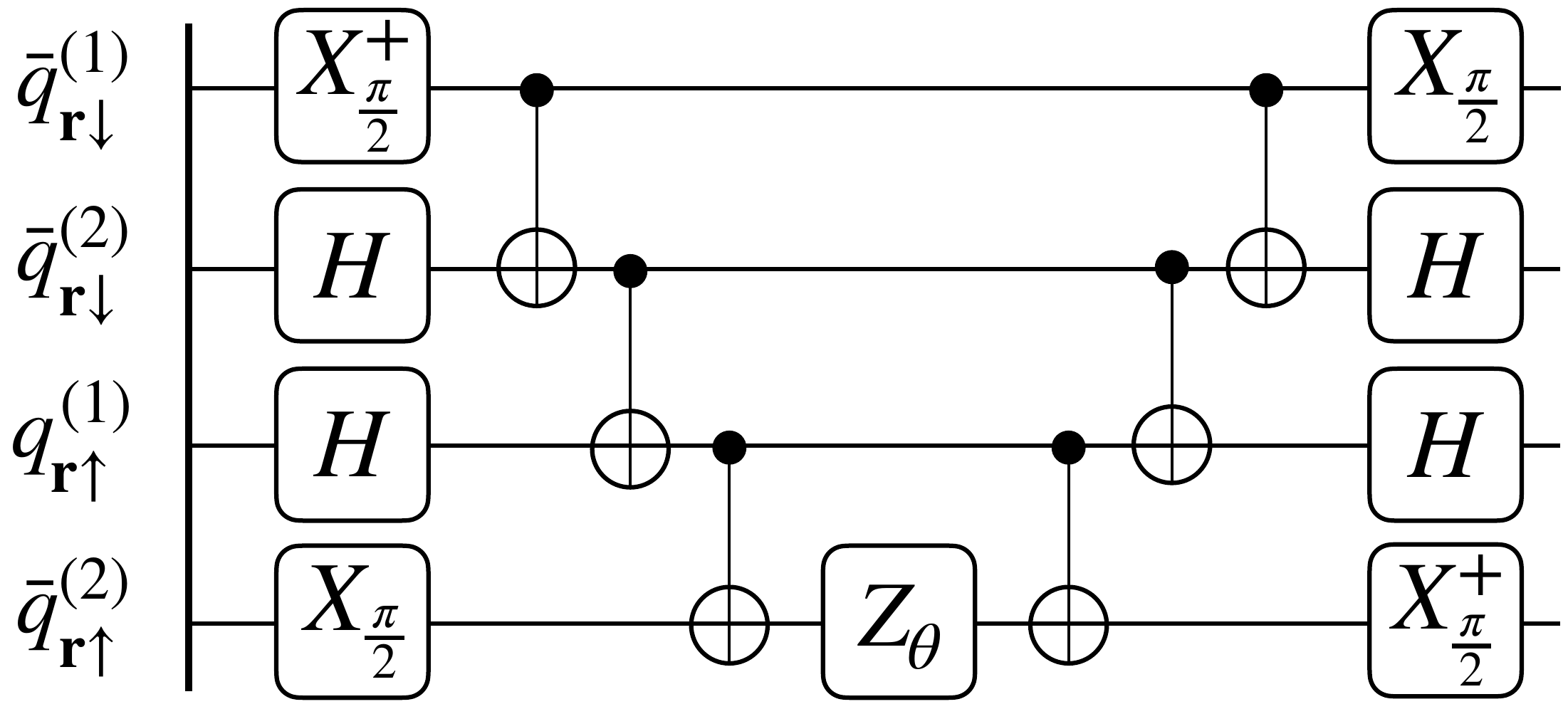}
	\caption{Quantum circuit realizing the unitary operator $e^{- i \theta A_{\bf r}^\uparrow}$. }
	\label{fig:Perturbation_Circuit}
\end{figure}

\begin{figure*}[t]
	\centering
	\includegraphics[scale=0.22]{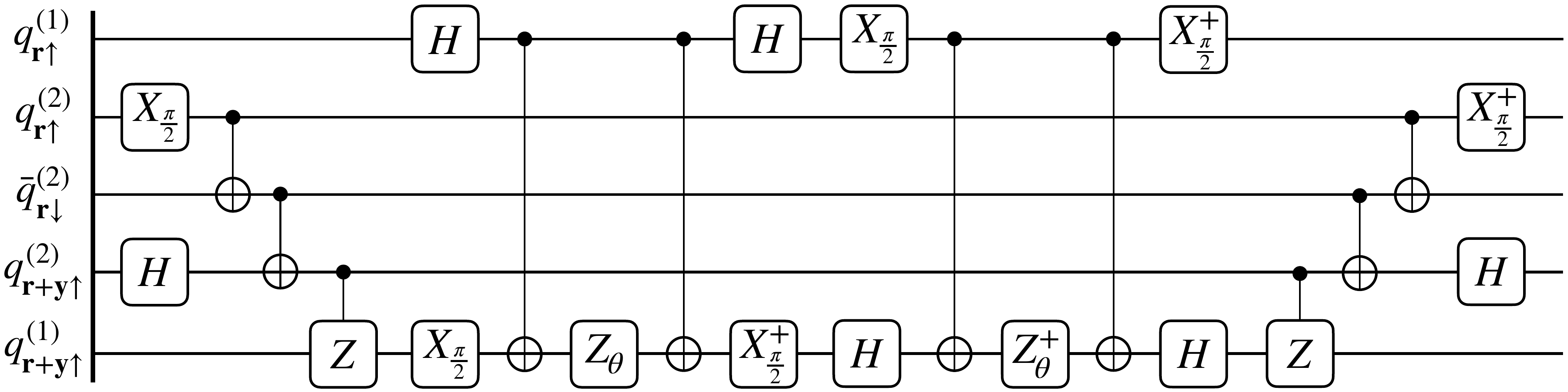}
	\caption{Quantum circuit realizing one Trotter step associated with the hopping operator $T_{{\bf r}\uparrow}^y$. }
	\label{fig:Hopping_Ty}
\end{figure*}

Following the bosonization approach of ref. \cite{bosonization}, we discuss elementary quantum circuits that correspond to different hopping terms. Similar to our consideration in previous sections we are going to use Majorana decomposition of fermionic creation and destruction operators. Namely, we write $c_{\bf r}^\sigma = \frac 12 ( x_{\bf r}^\sigma  - i y_{\bf r}^\sigma )$ for physical fermions and $a_{\bf r}^\sigma = \frac 12 ( \bar x_{\bf r}^\sigma  - i \bar  y_{\bf r}^\sigma )$ for auxiliary ones. The simplest hopping operator to consider is
\begin{equation}
A_{\bf r}^\uparrow = \frac{i}{2} y_{\bf r}^\uparrow \bar x_{\bf r}^\downarrow,
\end{equation}
When it is placed at a unit cell ${\bf r'}$, it can be used for Green's function measurements 
(discussed in more detail later in Sec.~\ref{direct_measurement}), where it acts as a source term (see Fig.~\ref{fig:2DCluster}).
The qubit representation of such hopping operator reads, cf. \cite{Nys2023}, 
\begin{equation}
\label{eq:A_source}
     y_{\bf r}^\uparrow \bar x_{\bf r}^\downarrow   \to - X_{{\bf r} \uparrow }^{(1)} \bar Y_{{\bf r} \downarrow }^{(1)} \times 
     Y_{{\bf r} \uparrow }^{(2)} \bar X_{{\bf r} \downarrow }^{(2)}. 
\end{equation}

The quantum circuit realizing a unitary evolution with the generator $A_{\bf r}^\uparrow$ is presented in Fig.~\ref{fig:Perturbation_Circuit}. It can be motivated in the following way: First, the unitary transformation with $H$ and $X_{\pi/2}$ gates brings $A_{\bf r}^\uparrow$ to the form $\bar Z_{{\bf r} \downarrow }^{(1)} \bar Z_{{\bf r}\downarrow }^{(2)} Z_{{\bf r} \uparrow }^{(1)} Z_{{\bf r} \uparrow }^{(2)} $. Further simplification is based on a sequential utilization of the similarity transformation with {\rm CNOT} gates: 
\begin{equation}
 {\rm CNOT}^{(ij)}\, (Z^{i}\otimes Z^{j})\, {\rm CNOT}^{(ij)} = Z^{i},  \qquad i \neq j.  
\end{equation}

\vspace{1em}
\paragraph{Hopping in $x$-direction:} 
Next, we construct quantum circuits for hopping evolution operators of physical fermions. We start by discussing hoppings in $x$-direction. 
The operator of interest is
\begin{equation}
    T_{{\bf r}\sigma}^x = c_{\bf r}^{\sigma\dagger} c_{\bf r+x}^\sigma + c_{\bf r +x}^{\sigma\dagger} c_{\bf r}^\sigma= 
    \frac{i}{2}\left( y_{\bf r}^{\sigma} x_{\bf r + x}^{\sigma}  -  x_{\bf r}^{\sigma} y_{\bf r + x}^{\sigma}\right).
\end{equation}
Specifically, we concentrate on the spin up hopping operator $T_{{\bf r}\uparrow}^x$. On a cluster with physical and auxiliary
fermions, such hopping operator corresponds to the next-to-nearest hopping. In order to express it in terms of qubit gates 
using the bosonization rules for elementary hopping operators (see Appendix~\ref{app:bosonization_rules}), we equivalently rewrite it as 
\begin{equation}
\label{eq:Tx_via_a}
    T_{{\bf r}\sigma}^x =
    \frac{i}{2}\left( y_{\bf r}^{\uparrow} {\bar x}_{\bf r}^{\uparrow} \cdot \bar x_{\bf r}^{\uparrow} x_{\bf r + x}^{\uparrow}  
    -  x_{\bf r}^{\uparrow} {\bar x}_{\bf r}^{\uparrow} \cdot \bar x_{\bf r}^{\uparrow} y_{\bf r + x}^{\uparrow}  \right),
\end{equation}
where we have used that ${\bar x}_{\bf r}^{\uparrow\,2} = 1$. Using relations~(\ref{eq:x_axis_maps}), one then
arrives at the following representation of $T_{{\bf r}\uparrow}^x$ in terms of Pauli matrices,
\begin{equation}
    T_{{\bf r}\uparrow}^x \mapsto \frac 12 \left(  X_{{\bf r}\uparrow}^{(1)}  \bar Z_{{\bf r}\uparrow}^{(1)}  X_{{\bf r+x}\uparrow}^{(1)} + 
       Y_{{\bf r}\uparrow}^{(1)}  \bar Z_{{\bf r}\uparrow}^{(1)} Y_{{\bf r+x}\uparrow}^{(1)}  \right) \otimes 
       \bar Z_{{\bf r}\uparrow}^{(2)}  Z_{{\bf r + x}\uparrow}^{(2)}.  
\end{equation}
The construction of the $T_{{\bf r}\downarrow}^x$ operator can be done along the same lines with minor changes. Importantly, the expression for $ T_{{\bf r}\uparrow}^x$ formally coincides with Eq.~\ref{eq:h_ij} with qubits $q_m=q_{{\bf r}\uparrow}^{(1)}$, $q_n=q_{{\bf r+x}\uparrow}^{(1)}$ and the Jordan-Wigner string of length three given by $Z_{\rm JW}=\bar Z_{{\bf r}\uparrow}^{(1)} \bar Z_{{\bf r}\uparrow}^{(2)}  Z_{{\bf r + x}\uparrow}^{(2)}$.
Therefore, the quantum circuit to realize an elementary Trotter step associated with the operator $T_{{\bf r}\uparrow}^x$ is identical to the one shown in Fig.~\ref{fig:circuit_hopping}.
At variance with the fermion-to-qubit mapping based on Jordan-Wigner transformation, the hopping circuit involves only five neighboring qubits
with the fixed lengths of JW strings, which is the strong advantage of this local scheme in application to two-dimensional clusters.

\vspace{1em}
\paragraph{Hopping in $y$-direction:}
Hereinafter we discuss the hopping operators in $y$-direction defined as 

\begin{equation}
        T_{{\bf r}\sigma}^y = c_{\bf r}^{\sigma\dagger} c_{\bf r+y}^\sigma + c_{\bf r +y}^{\sigma\dagger} c_{\bf r}^\sigma= 
    \frac{i}{2}\left( y_{\bf r}^{\sigma} x_{\bf r + y}^{\sigma}  -  x_{\bf r}^{\sigma} y_{\bf r + y}^{\sigma}\right).
\end{equation}
To obtain its form in terms of elementary qubit operations, we follow the same idea as above by representing $T_{{\bf r}\sigma}^y$ as the hopping via an intermediate auxiliary fermion $a_{\bf r}^\sigma$, see Eq.~(\ref{eq:Tx_via_a}). Equipped with a set of rules for elementary hoppings, cf. Appendix I, 
one then obtains the following identification.
\begin{equation}
\label{eq:Ty_up}
       T_{{\bf r}\uparrow}^y \mapsto \frac 12 \left(  X_{{\bf r}\uparrow}^{(1)} Y_{{\bf r+y}\uparrow}^{(1)} 
       - Y_{{\bf r}\uparrow}^{(1)}  X_{{\bf r+y}\uparrow}^{(1)}  \right) \otimes 
        Y_{{\bf r}\uparrow}^{(2)}  \bar Z_{{\bf r}\downarrow}^{(2)} X_{{\bf r + y}\uparrow}^{(2)}.  
 \end{equation}
Once again, minor changes are required to derive spin down hopping $T_{{\bf r}\downarrow}^y$. Namely, all spins in Eq.~(\ref{eq:Ty_up})
need to be reverted, except of the operator $\bar Z_{{\bf r}\downarrow}^{(2)}$, which has to be substituted by $\bar Z_{{\bf r+y}\uparrow}^{(2)}$. 

The unitary circuit realizing an elementary Trotter step generated by $\theta \cdot T_{{\bf r}\uparrow}^y$ is presented in Fig.~\ref{fig:Hopping_Ty}.
It can be rationalized as follows. First, a similarity transformation with the unitary $X_{\pi/2}^\dagger \otimes H$, acting, respectively, on qubits
$q_{{\bf r}\uparrow}^{(2)}$ and $q_{{\bf r + y}\uparrow}^{(2)}$, maps a triplet of operators applied to auxiliary qubits in (\ref{eq:Ty_up}) onto the
Jordan-Wigner string $Z_{\rm JW}=Z_{{\bf r}\uparrow}^{(2)}  \bar Z_{{\bf r}\downarrow}^{(2)} Z_{{\bf r + y}\uparrow}^{(2)}$.
As the result, the transformed operator $T_{{\bf r}\sigma}^y$ shares the same structure with  
the hopping operator $h_{ij}^{\sigma\sigma}$ --- now built on a pair of qubits $q_m=q_{{\bf r}\uparrow}^{(1)}$, $q_n=q_{{\bf r+y}\uparrow}^{(1)}$
and joined by the string $Z_{\rm JW}$. Thus the quantum circuit in Fig.~\ref{fig:Hopping_Ty} mirrors 
the most general form shown in Fig.~\ref{fig:circuit_hopping}.

\vspace{1em}
\paragraph{Repulsion:}
The elementary circuit realizing the repulsion term $U_i n_{i\uparrow} n_{i\downarrow}$ remains the same as shown in Fig.~\ref{fig:circuit_repulsion}, provided qubits $q_{i\sigma}$ are understood as $q^{(1)}_{{\bf r}\sigma}$.

\subsection{Variational Hamiltonian ansatz (VHA) for the ground state}
The correlation functions defined by Eq. (\ref{eq:superposition_pauli}) presume the average over the equilibrium density matrix. 
At zero temperature one is required to start from the ground state of the cluster at time $\tau = 0$. 
For this reason we will briefly review the variational Hamiltonian ansatz (VHA), \cite{Reiner2019}, which is used to construct the ground state. 

The defining idea of VHA is to find a unitary operator $U(\theta)$, such that under variations of the parameters $\theta_i$ from the set $\theta$ one minimizes the energy expectation value $\langle \Psi_0 \lvert U^\dagger(\theta) H U(\theta) \lvert \Psi_0 \rangle$, where $\lvert \Psi_0 \rangle$ is a guess state which can be prepared easily. 
Identifying an underlying operator $H$ as the sum of $p$ independent terms $H = \sum_{j=1}^pH_j$, the operator $U(\theta)$ is defined over $n$ steps as
\begin{equation}
    U(\theta) = \prod_{k=1}^n \prod_{j=1}^p e^{-i\theta_{j,k}H_j}.
    \label{eq:VHA_theory}
\end{equation}
In each step $k$, the $p$ parameters are updated until energy measurements on $\lvert \Psi \rangle = U(\theta)\lvert \Psi_0 \rangle$ yield minimum values. An example for a rule set governing the update of $\theta_{j,k}$ can be found in \cite{Piskor2021}.

In the following, we discuss the details of implementing the VHA, assuming that the quantum circuits in question are constructed using the Jordan-Wigner transformation. We will rely on these results later when performing noisy simulations of Green's function measurements. On the other hand, the practical implementation of local fermion-to-qubit mapping schemes, even for moderately sized clusters, remains challenging. For an analysis of the VHA circuit depth within the locality-preserving mapping scheme discussed in Sec.~\ref{sec:QC_hoppings_and_U_terms_locality_preserving}, we refer readers to the original Ref.~\cite{Nys2023}.

\subsection{Measuring ground state energy}
\label{ch:measurement_GS}

\begin{figure}[t]
	\centering
	\includegraphics[scale=0.08]{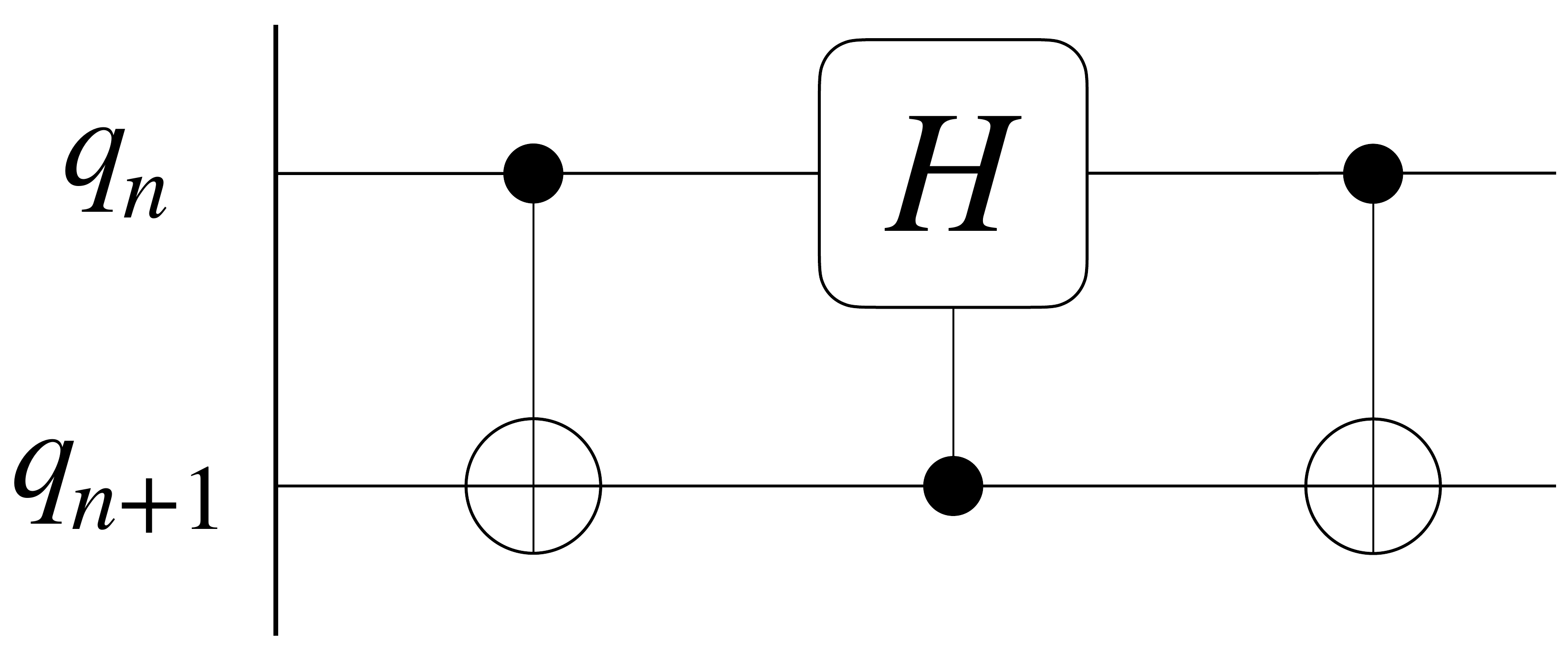}
	\caption{Circuit for diagonalizing $\frac{1}{2} (X_n X_{n+1} + Y_n Y_{n+1})$ into $\lvert 01\rangle \langle 01\lvert - \lvert 10 \rangle \langle 10 \lvert$.}
	\label{fig:stasja_measurement}
\end{figure}

Finding the minimum expectation value $\langle\Psi_0\lvert U^\dagger(\theta)HU(\theta)\lvert\Psi_0\rangle$ as pointed out in the previous section requires energy measurements. From Eq.~(\ref{eq:FH_Hamiltonian}) we find terms for hopping and repulsion, which in the following we refer to as $H_0$ and $H_U$, respectively. Hence, schedules for measurement of hopping and repulsion energy, cf.\ \cite{Cade2020}, are reviewed. 

Measurements of repulsion energy are done by measuring each qubit in the computational basis. Since we operate within the Jordan-Wigner framework, repulsion terms are mapped to the matrix $\lvert11\rangle\langle11\lvert_{mn}$, where $m,n$ correspond the inspected orbitals belonging to one parent site.
Hence, the energy equals the probability to find both qubits to be in state $\lvert1\rangle$. 

Measurements of kinetic energies depend on the hopping direction, which is rooted in the way how the Jordan-Wigner strings are chosen in the
mapping~(\ref{eq:JW_transform}). For one of the possible choices, horizontal hoppings are considered less costly in terms of gate depth since Jordan-Wigner strings can be neglected, whereas vertical hoppings may lead to long Jordan-Wigner strings. 

Following \cite{Cade2020}, horizontal hopping map to the matrix $\frac{1}{2} (X_n X_{n+1} + Y_n Y_{n+1})$. In order to perform computational basis measurements, the unitary that diagonalizes $\frac{1}{2} (X_n X_{n+1} + Y_n Y_{n+1})$ into $\lvert 01\rangle \langle 01\lvert - \lvert 10 \rangle \langle 10 \lvert$ is shown in Fig. \ref{fig:stasja_measurement}. Desired energy expectation is thus the probability of measuring $\lvert 01\rangle $ minus the probability of measuring $\lvert 10\rangle$. 

\begin{figure}[t]
	\centering
	\includegraphics[scale=0.15]{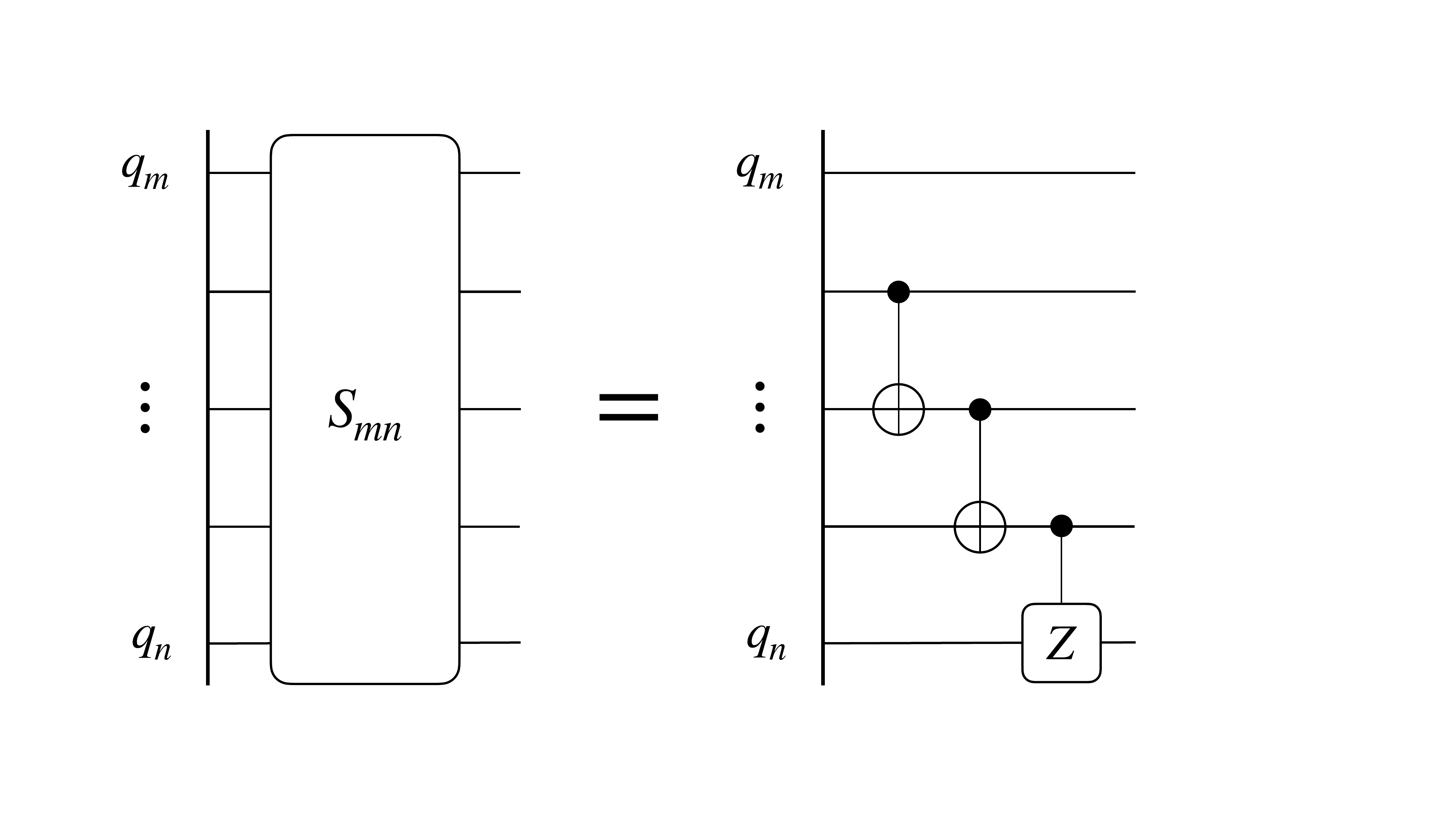}
	\caption{Definition of the operator $S_{mn}$, with the purpose to remove Jordan-Wigner strings. It is non-trivial for $m-n\geq 2$ and we set $S_{n+1,n} = \mathds{1}$.}
	\label{fig:definition_S_mn}
\end{figure}

\begin{figure}[t]
	\centering
	\includegraphics[scale=0.16]{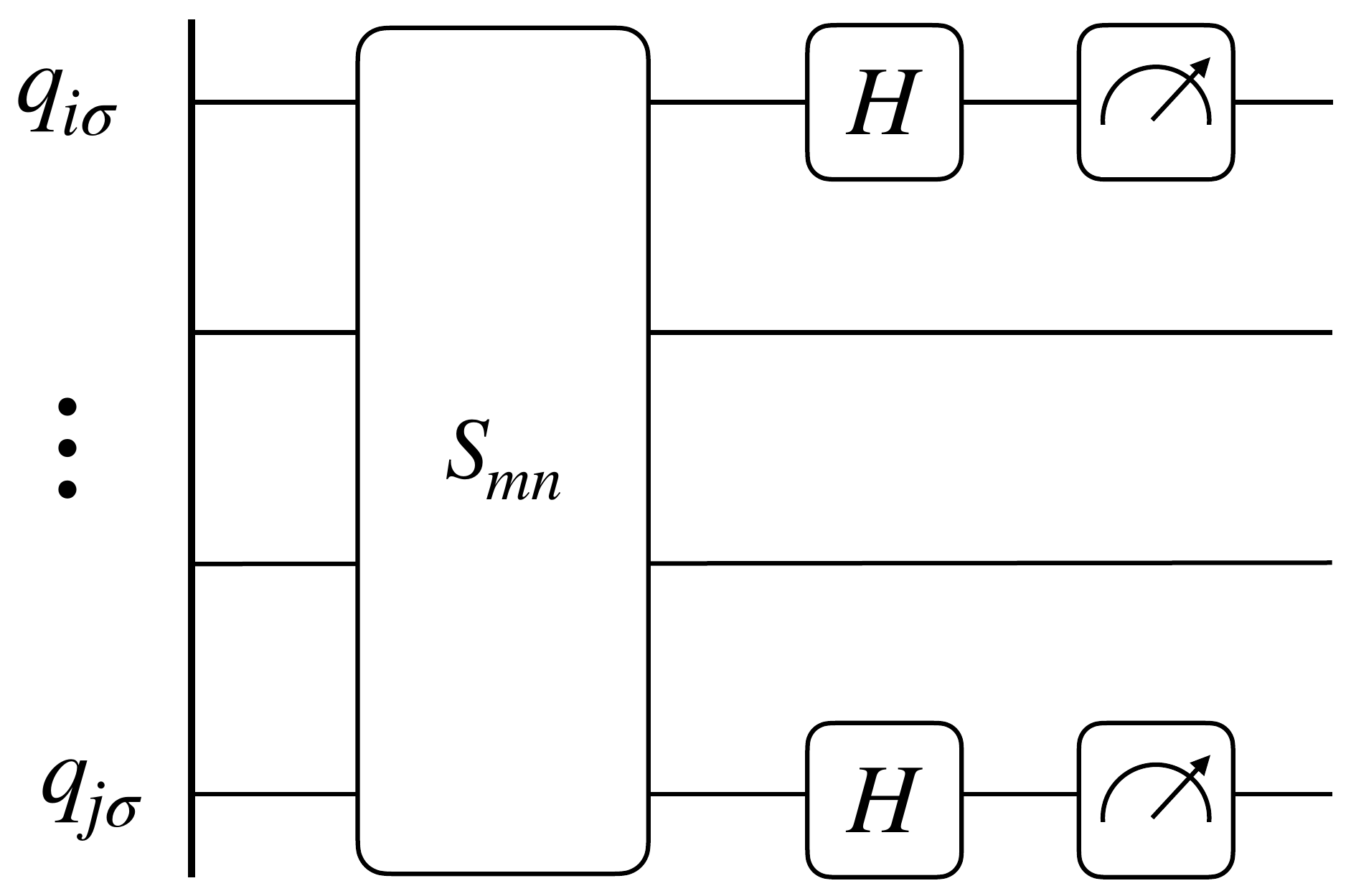}
	\caption{Measurement of $ i y_{i\sigma}x_{j\sigma}$ term which is reduced to 
    $ \langle {Z}_{i\sigma} {Z}_{j\sigma'} \rangle$ after a similarity transformation.}
	\label{fig:S_mn_with_measurement}
\end{figure}

On the other hand, a kinetic energy term $h_{ij}^{\sigma\sigma'}$ describing vertical hopping is mapped to the operator~(\ref{eq:h_ij}), which contains an additional Jordan-Wigner string. As discussed in subsection~\ref{sec:QC_hoppings_and_U_terms}, the latter can be eliminated by similarity 
transformation via the unitary $S_{mn}$ shown in Fig.~\ref{fig:definition_S_mn}. Afterwards, one can measure two terms in $h_{ij}^{\sigma\sigma'}$ separately.
The quantum circuit to implement the measurement of the first term, $ i y_{i\sigma}x_{j\sigma'}$, 
is given in Fig.~\ref{fig:S_mn_with_measurement}. It is based on the following similarity transformation of this operator, i. e. 
\begin{eqnarray}
     i y_{i\sigma}x_{j\sigma'} &=& {\textbf S}_{mn}^\dagger\ ({Z}_m {Z}_n) {\textbf S}_{mn}, \nonumber \\
     {\textbf S}_{mn} &=& (\text{H}_m \text{H}_n) \, S_{mn}.
\end{eqnarray}
The expectation value of this operator is then reduced to the average parity of qubits $m$ and $n$.
A measurement of the second term in the hopping term, $ -i x_{i\sigma}y_{j\sigma'}$, is implemented along the same lines with the only difference
that the Hadamard gate is replaced by ${X}_{\pi/2}$.

\subsection{Green’s function measurement}
We present two routes to measure the Green's function. The orthodox way goes back to the Hadamard test 
circuit originally proposed in Ref.~\cite{Somma2002}, see Fig.~\ref{fig:correlation_circuit}. Another one makes use of linear response theory, particularly the Kubo formula, which we present afterwards.

\subsubsection{The Hadamard test}

\begin{figure}[t]
	\centering
	\includegraphics[scale=0.135]{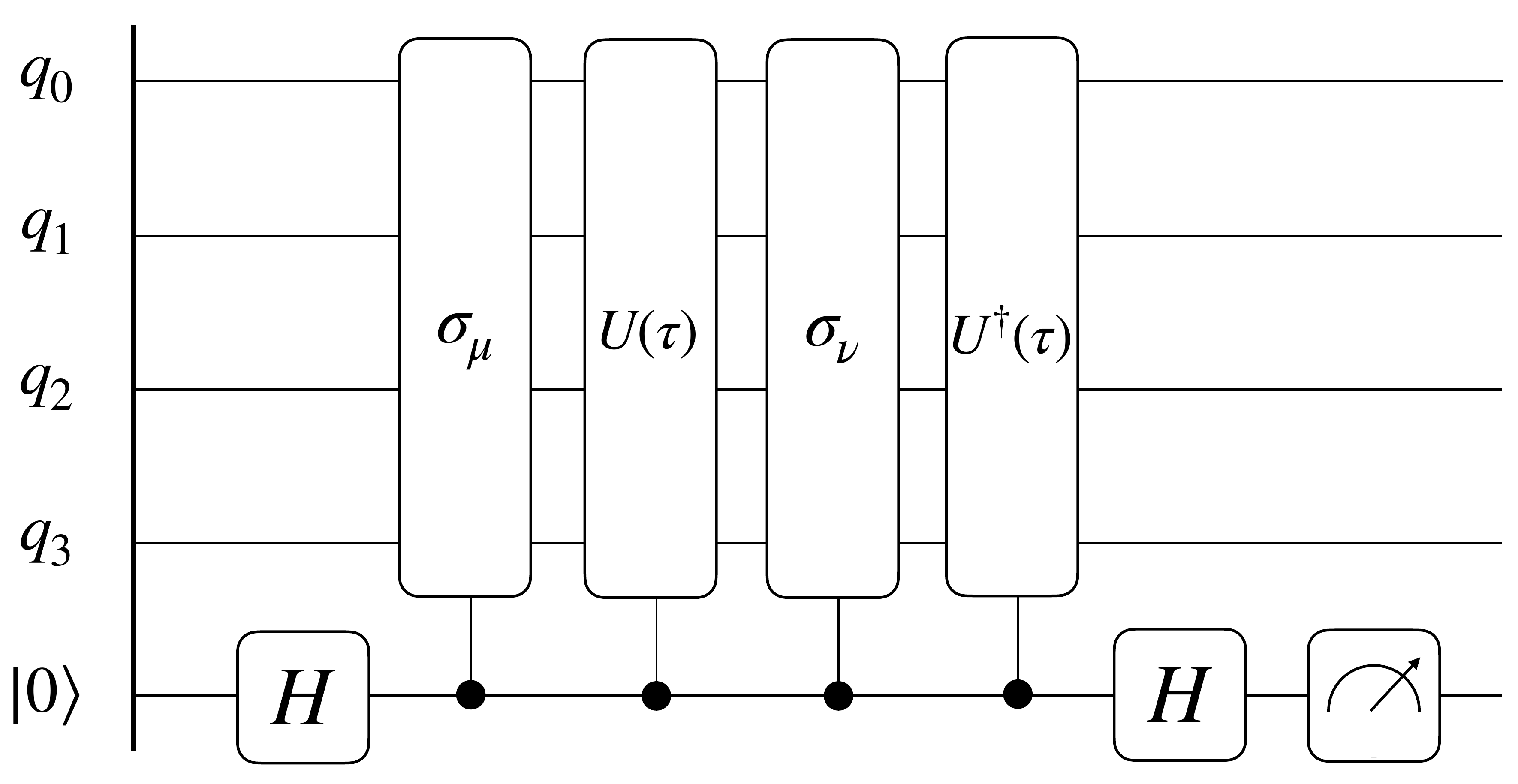}
	\caption{Quantum circuit for measuring correlation functions $C_{\mu \nu}$. The first four qubits $q_0, \cdots, q_3$ represent the physical system, whereas the last qubit in state $\lvert 0 \rangle$ represents the control qubit.}  
	\label{fig:correlation_circuit}
\end{figure}

The most standard Hadamard test uses a quantum circuit with a single ancilla which generates a random binary variable ($\pm 1$) whose average value gives ${\rm Re}\langle \Psi_* | {\cal U} |\Psi_*\rangle$. Here $|\Psi_*\rangle$
denotes an initial quantum state and ${\cal U}$ is an arbitrary unitary acting on $|\Psi_*\rangle$. In application to the Green's function measurement within the VCA framework, the wave function $|\Psi_*\rangle$ represents
an (approximate) ground state of a Hubbard cluster and one sets ${\cal U} = U^\dagger (\tau) \sigma_\nu U(\tau) \sigma_\mu \equiv \sigma_\nu(\tau) \sigma_\mu$, where $\sigma_\mu$ may refer
to any of the Hermitian Majorana operators $x_\mu$, $y_\mu$. The averaged real part of~${\,\cal U}$ then coincides with the retarded correlator of two Majoranas, that is 
\begin{equation}
{\rm Re}[i g_{\nu\mu}(\tau)] = \frac 12 \langle \Psi_* |  \{\sigma_\nu(\tau) , \sigma_\mu\}  |\Psi_*\rangle.
\end{equation}
The corresponding circuit to evaluate the Green's function within the above scheme is given in Fig. \ref{fig:correlation_circuit}. Here $q_0, \cdots, q_3$ are qubits representing a small physical system (e. g. the four qubit cluster),
while the last qubit is an ancilla whose initial state is $\lvert 0 \rangle$. Within the logic of Hadamard test the correlation function $g_{\mu\nu}(\tau)$ can be then estimated as
\begin{equation}
\text{Re}[ i g_{\nu\mu}(\tau)] = P_{\nu\mu}(\mathcal{M}=0, \tau) - P_{\nu \mu}(\mathcal{M}=1, \tau),
\label{eq:probability}
\end{equation}
where $P_{\nu\mu}(\mathcal{M}, \tau)$ denotes the empirical probability of measuring the ancilla in the state $\mathcal{M}$ at time $\tau$.  

The Hadamard test circuit requires four blocks of controlled evolution, which include computationally costly two-qubit gates~\footnote{
	One can also argue that it suffices to control only measured unitary operators $\sigma$, while forward and backward evolution operators can be left
	uncontrolled~\cite{Wecker2015}. Such simplification however is based on the identity $U(\tau) U^\dagger(\tau) = \mathds{1}$, which 
	tends to be violated in the case of NISQ devices operating for a sufficiently long time.
}. 

\begin{figure}[b]
	\centering
	\includegraphics[scale=0.19]{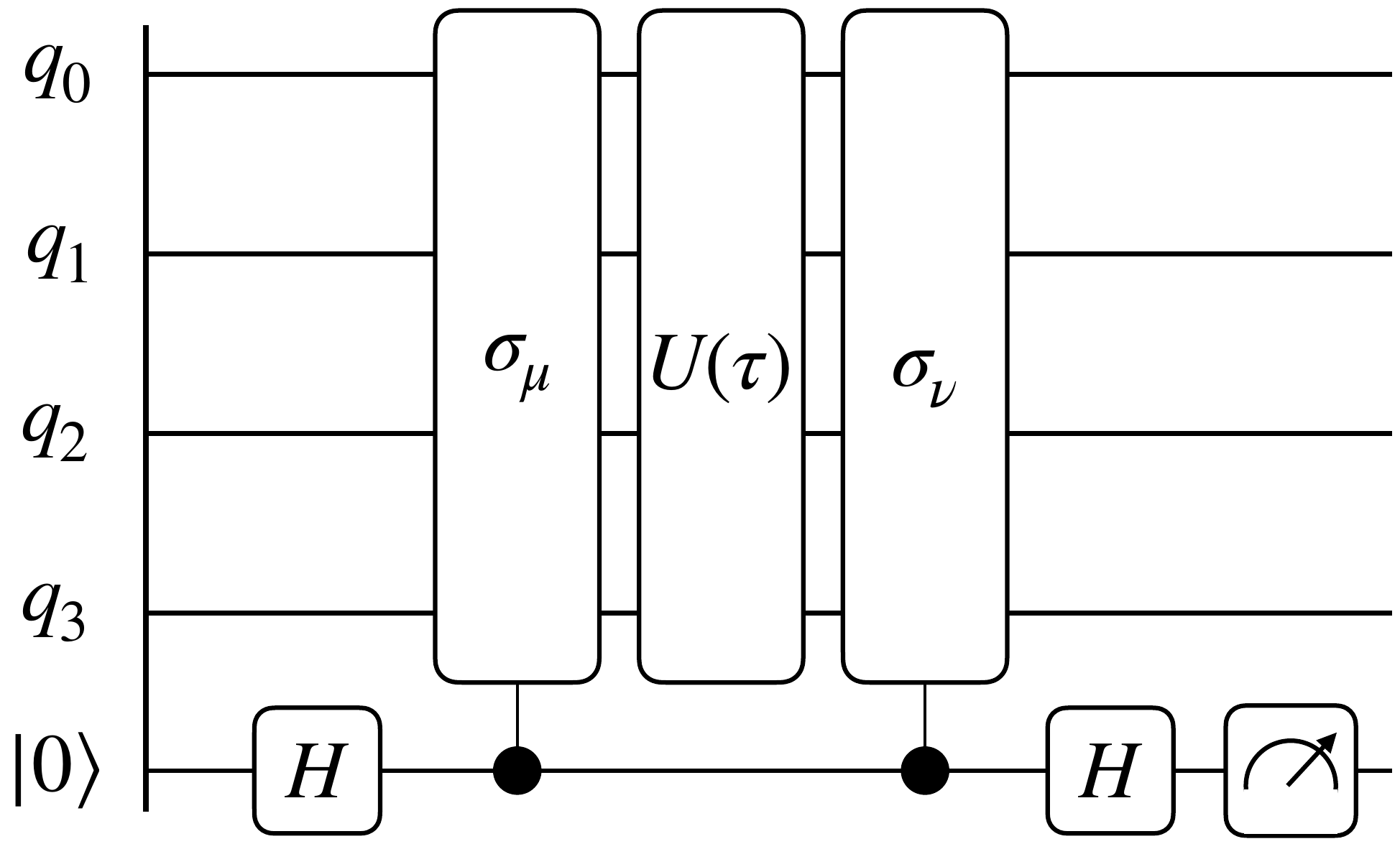}
	\caption{The advanced Hadamard test only requires one evolution $U(\tau)$ without any control ancillas.}
	\label{fig:advanced_hadamard_test}
\end{figure}

Recently, Endo et al., \cite{Endo2020}, presented a more elegant version of the Hadamard test (the latter finds its roots in the pioneering work~\cite{Somma2002}), whose circuit is presented in Fig.~\ref{fig:advanced_hadamard_test}. It does not require a controlled evolution and neither does it require an evolution of the form $U^\dagger(\tau)$ (cf. Ref.~\cite{Libbi:2022} for further generalizations). 
On the other hand, the possibility of constructing controlled single-particle fermion operators is a defining feature of their algorithm. 

In the following subsection we discuss the new direct measurement scheme which is 
more physical in a sense that the latter can be applied to more sophisticated, particularly locality-preserving 
mappings (e.g. the one discussed in Sec.~\ref{sec:QC_hoppings_and_U_terms_locality_preserving}), 
that do not allow for single Majorana fermion operators.

\subsubsection{Direct measurement}
\label{direct_measurement}
In this subsection we propose an alternative approach to evaluate the Green's function on a quantum computer, which we refer to in what follows as the direct measurement. We will see, that this approach can reduce the depth of the circuit by at least a factor of two.
Importantly, it requires no assumption on the initial density matrix $\rho_0$ of a simulated many-body system and, 
in contrast with the Hadamard test, relies merely on the forward uncontrolled evolution operator $U(\tau)$.
Furthermore, we use the two-site dimer toy model to demonstrate the direct measurement and probe its potential for being a superior method for accessing the system's Green's function.

Exploiting Green's functions for calculating observables and order parameters is motivated by linear response theory (cf. Appendix~\ref{appendix_kubo}). Let $H'(t)=H + V(t)$ be our system of interest, with $H$ being the stationary, time-independent part and $V(t)=\sum_i\Phi_i(t){A}_i$ 
being the time-dependent perturbation, whose exact form we specify momentarily.  Linear response theory describes how the system reacts to a given perturbation $V(t)$, where $\Phi_i(t)$ is the interaction strength of operator ${A}_i$. 
If we assume a sufficiently weak perturbation, the change in an expectation value of any Heisenberg operator $\tilde{{A}}_i(t)$ defined
relative to the full Hamiltonian $H'(t)$ is linear in the perturbing source $\Phi(t)$. This is formulated as
\begin{equation}
\delta \langle \tilde{{A}}_i(t) \rangle = \sum_j \int \text{d}t' \chi_{ij}(t;t') \Phi_j(t'),
\end{equation}
where
$\chi_{ij}(t;t')$ is the response function given by
\begin{equation}
\label{eq:response_function}
    \chi_{ij}(t, t') = -i\theta(t-t') \langle [A_i(t), A_j(t')] \rangle.
\end{equation}
Here, operators $A_j(t)$ evolve under the action of
the non-perturbed Hamiltonian $H$ and, as before, is averaged over an initial density matrix $\rho_0$.
In particular, for the perturbation localized in time at time $t'$, one writes $\Phi_j(t) = \Phi_j\, \delta(t-t')$, and arrives at
the relation
\begin{equation}
\label{eq:dA_i}
\delta \langle  \widetilde {A}_i(t)\rangle = \sum_j\chi_{ij}(t;t') \Phi_j.
\end{equation}
which can be used to extract the response function in the demonstration. One assumes here that $\Phi_j$ is relatively small 
so that non-linear effects can be disregarded.

\begin{figure*}[t]
	\centering
	\includegraphics[scale=0.23]{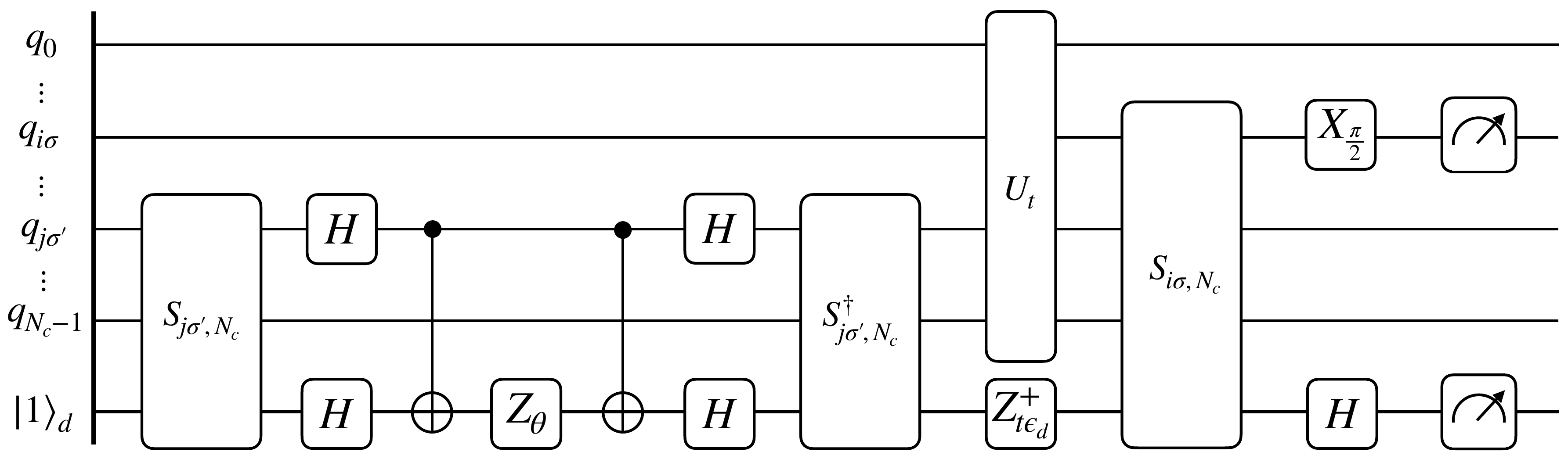}
	\caption{Measurement of the correlator $\langle \{ x_{i\sigma}(\tau), y_{j\sigma'} (0) \} \rangle$. The unitary $U_t$ refers to the Trotterized evolution under the Hubbard cluster Hamiltonian~(\ref{eq:FH_Hamiltonian}). While the $Z$-rotation with the angle $\theta=\Phi_j^{\sigma'}$ is due to perturbation, the $Z^\dagger$-rotation with the angle $t\epsilon_d$ accounts for a free evolution of the $d$-fermion.}
	\label{fig:measurement_full_correlator}
\end{figure*}

To adapt this general idea to the Green's function measurement of $c$-fermions in our system, we couple them to an auxiliary spinless $d$-fermion 
with the Hamiltonian $H_d = \epsilon_d d^\dagger d$ and introduce the hopping operator 
\begin{equation}
    A_j^{\sigma} = d^\dagger c_{j\sigma} + c_{j\sigma}^\dagger d
    \label{eq:hopping_direct_meas}
\end{equation}
acting on the $j$-th fermion, while the ancilla qubit stores the state of the $d$-fermion. The explicit form of the perturbation $V(t)$ then reads
\begin{equation}
V(t) = \sum_j\Phi_j^{\sigma}(t)A_j^{\sigma} = \sum_j\Phi_j^{\sigma}(t)[d^\dagger c_{j\sigma} + c_{j\sigma}^\dagger d],
\label{eq:perturbation_direct_measurement}
\end{equation}
where again $\Phi_j^\sigma(t)$ is the external field acting on the $j$-th fermion with spin $\sigma$. 
Furthermore, considering the response function (\ref{eq:response_function}) we may assume $t'=0$ and $t>t'$ such that we can neglect the $t'$-dependence. 
The commutator then becomes
\begin{equation}
\label{eq:chi_ij_t}
    \chi_{ij}^{\sigma\sigma'}(t) = -i \Big(\langle A_i^{\sigma}(t)A_j^{\sigma'} \rangle - \langle A_j^{\sigma'} A_i^{\sigma}(t) \rangle \Big).
\end{equation}
Using Eq.~(\ref{eq:hopping_direct_meas}), this leads to 
\begin{equation}
\begin{aligned}
    \chi_{ij}^{\sigma\sigma'}(t) = &-  i\big\langle \big(d^\dagger(t)c_{i\sigma}(t) + c_{i\sigma}^\dagger(t)d(t)\big)
    \cdot \big(d^\dagger c_{j\sigma'} + c_{j \sigma'}^\dagger d \big) \big\rangle  
    \\
    & + i\big\langle  \big(d^\dagger c_{j\sigma'} + c_{j\sigma'}^\dagger d \big) \cdot 
    \big(d^\dagger(t) c_{i\sigma}(t) + c_{i\sigma}^\dagger(t)d(t)\big) \big\rangle.
\end{aligned} \nonumber
\end{equation}
If we assume the $d$-fermion orbital to be occupied, it follows that $\langle d^\dagger (t) d \rangle \neq 0$ and $\langle d(t)d^\dagger \rangle = 0$. Hence, the above equation simplifies to 
\begin{equation}
    \chi_{ij}^{\sigma \sigma'}(t) = -i \big\langle d^\dagger(t) d \, c_{i\sigma}(t) c_{j\sigma'}^\dagger \big\rangle + 
    i \big\langle d^\dagger d(t) c_{j\sigma'} c_{i\sigma}^\dagger(t) \big\rangle.
\label{eq:chi_simplified}
\end{equation}
Wick's theorem can be used to write the four-point correlators in terms of a combination of two-point correlators. The only non-zero two-point correlators are $\langle d^\dagger(t)d \rangle$, $\langle c_i(t)c_j^\dagger \rangle$, $\langle d^\dagger d(t) \rangle$ and $\langle c_j c_i^\dagger(t) \rangle$. It then follows that
\begin{eqnarray}
    \chi_{ij}^{\sigma \sigma'}(t) = -i \big\langle d^\dagger(t) d \big\rangle \big\langle c_{i\sigma}(t) c_{j\sigma'}^\dagger \big\rangle  +i  \big\langle d^\dagger d(t)\big\rangle \big\langle c_{j\sigma'}c_{i\sigma}^\dagger(t) \big\rangle, \nonumber \\
\label{eq:chi_wicked}
\end{eqnarray}
where $\big\langle d^\dagger(t) d \big\rangle$ and $\big\langle d^\dagger d(t)\big\rangle$ equal $e^{i\epsilon_dt}$ and $e^{-i\epsilon_dt}$, respectively. Finally, we arrive at 
\begin{equation}
    \chi_{ij}^{\sigma \sigma'}(t) = -i e^{i\epsilon_dt} \big\langle c_{i\sigma}(t) c_{j\sigma'}^\dagger \big\rangle + 
     i e^{-i\epsilon_dt} \big\langle c_{j\sigma'} c_{i\sigma}^\dagger(t) \big\rangle.
    \label{eq:fermionic_commutator_response}
\end{equation}
The above relation can be represented in the equivalent form:
\begin{eqnarray}
\label{eq:chi_RK}
\chi_{ij}^{\sigma \sigma'}(t) =\sin \lambda \langle \{ c_{i\sigma}(t), c_{j\sigma'}^\dagger \} \rangle 
- i \cos \lambda  \langle [ c_{i\sigma}(t), c_{j\sigma'}^\dagger] \rangle, \nonumber \\	 
\end{eqnarray} 
where $\lambda = \epsilon_d t$. Since the energy of the $d$-fermion $\epsilon_d$ is quasi-arbitrary, 
one may vary the phase $\lambda$ to recover two independent Green's functions.
By setting $\lambda = \pi/2$ one obtains the retarded Green's function given by the anti-commutator, while
the choice $\lambda = 0$ leads to the so-called Keldysh correlator expressed via the commutator of two fermion operators.

When it comes to the actual measurement protocol using the outlined linear response scheme, it is advantageous to
perform the measurements in the Majorana basis and use the relation~(\ref{eq:superposition_pauli}) to reconstruct the Green's function of
complex fermions afterwards. As an example, let us consider the measurement of the correlator
of two Hermitian operators $x_{i\sigma}(t)$ and $y_{j\sigma'}$.
To this end we introduce two Majorana fermions, 
$x_d = d +d^\dagger$ and $y_d = i (d -d^\dagger)$, associated with the
auxiliary $d$-fermion and define hopping operators as follows, cf. Eq.~(\ref{eq:h_ij_f})
\begin{equation}
 A_i^\sigma = \tfrac i 2 x_{i\sigma}\, x_d,  \quad A_j^{\sigma'} = \tfrac i 2 y_{j \sigma'}\, x_d. 
\end{equation}
Repeating the steps leading to the intermediate result~(\ref{eq:chi_wicked}), one finds that for such choice of hopping operators $\chi_{ij}^{\sigma \sigma'}(t)$ changes to
\begin{eqnarray}
\label{eq:chi_wicked_2}
    \chi_{ij}^{\sigma \sigma'}(t) &=& - \tfrac i 4 \big\langle x_d(t) x_d \big\rangle \big\langle x_{i\sigma}(t) y_{j\sigma'}\big\rangle \nonumber \\
                 &&  + \tfrac i 4  \big\langle x_d x_d(t)\big\rangle \big\langle y_{j\sigma'} x_{i\sigma}(t) \big\rangle.
\end{eqnarray}
Additionally, the correlator of an auxiliary Majorana fermion becomes $\langle x_d(t) x_d \rangle = e^{i\epsilon_d t}$.
This means that for the Green's functions of Majorana operators we can use exactly the same final relation (\ref{eq:chi_RK}). In particular, the retarded correlator reads
\begin{equation}
\label{eq:GR_ij_t}
\frac 14 \langle \{x_{i\sigma}(t), y_{j \sigma'}\} \rangle = \chi_{ij}^{\sigma \sigma'}(t) \Bigl|_{\epsilon_d t = \pi/2}.
\end{equation} 
For the later purpose it is advantageous to rewrite the above relation as
\begin{align}
    \frac 1 2 \langle \{x_{i\sigma}(t), y_{j \sigma'}\}  \rangle =  
    \frac{\langle i x_{i\sigma} x_d \rangle_\Phi (t,\epsilon_d)}{ \sin \Phi_j^{\sigma'}}\Biggl|_{\epsilon_d t = \pi/2},
    \label{eq:measurement}
\end{align}

where $\langle \dots \rangle_\Phi$ refers to an average in the presence of a perturbation. Note, that within the linear response theory framework 
the denominator needs to be substituted by just $\Phi_j^{\sigma'}$. In this case the relation~(\ref{eq:measurement}) follows 
from Eqs.~(\ref{eq:dA_i}) and (\ref{eq:GR_ij_t}), where $\langle  A_i^\sigma \rangle  = \frac i 2 \langle i x_{i\sigma} x_d \rangle_\Phi$ and we took into account that $\langle  A_i^\sigma \rangle$ vanishes in the absence of perturbation. 
In Appendix \ref{appendix_nonlin_kubo} we evaluate the response $\langle  A_i^\sigma \rangle$ to the source field in all orders and prove the 
validity of Eq.~(\ref{eq:measurement}) at arbitrary $\Phi_j^{\sigma'}$.

Our primary focus in this paper is on the retarded Green's function given by the anti-commutator~(\ref{eq:measurement}), since
the latter eventually enters into the VCA scheme outlined in Sec.~\ref{sec:VCA}. However, the full set of possible Green's functions can be evaluated using the proposed algorithm. For instance, for the Keldysh correlator of Majoranas one can write
\begin{align}
	- \frac i2 \langle [x_{i\sigma}(t), y_{j \sigma'}]  \rangle =  
	\frac{\langle i x_{i\sigma} x_d \rangle_\Phi (t,\epsilon_d)}{ \sin \Phi_j^{\sigma'}}\Biggl|_{\epsilon_d  = 0},
	\label{eq:measurement_K}
\end{align}

Then other correlators, such as $g_{ij}^{\sigma \sigma'}(t)$ defined in Eq.~(\ref{eq:g_Majorana}), can be reconstructed from the retarded and Keldysh Green's functions. Additionally, the best choice for the strength of the perturbation is $\Phi_j^{\sigma'} = \pi/2$, which leads to the strongest response and is used for quantum computations in what follows.

\vspace{1em}

Equations~(\ref{eq:measurement}) and (\ref{eq:measurement_K}) summarize the main result of this subsection. In what follows we present our new quantum algorithm which enables 
the measurement of these correlators using both the Jordan-Wigner and one of local fermion-to-qubit mappings, discussed previously in Sec.~\ref{sec:QC_hoppings_and_U_terms_locality_preserving}. 

\paragraph{Jordan-Wigner framework:}

The quantum circuit (see Fig.~\ref{fig:measurement_full_correlator}), which accomplishes the measurement of correlators within the Jordan-Wigner mapping, can be rationalized as follows.
Right after the initialization of a quantum computer to the ground state with the help of VHA (not shown) one applies the perturbation with the potential $V(t)$. It can be achieved in one Trotter step yielding the unitary
\begin{align}
    \exp&(\tfrac 12\Phi_j^{\sigma'} y_{j \sigma'}\, x_d )  = \nonumber \\
    & S_{j\sigma', N_c}^\dagger\, \exp\left( - \tfrac i 2 \Phi_j^{\sigma'} X_{j\sigma'} X_{N_c} \right) \,S_{j\sigma', N_c}.
\end{align}
Its circuit representation is analogous to the one describing the evolution under the hopping term, see Fig.~\ref{fig:circuit_hopping}.
In this context, it entangles the ancilla qubit used to represent the $d$-fermion with the qubits' states representing the cluster. 
Subsequent independent evolutions of the cluster and the $d$-fermion are then followed by the measurement of an operator 
$A_i^\sigma = \frac i2  x_{i\sigma} x_d$.
The way to average such operator has already been described in subsection~\ref{ch:measurement_GS}: 
in a few unitary transformations $\langle A_i^\sigma \rangle$ can be related to the average parity of two qubits, $\langle Z_{i\sigma} Z_{N_c} \rangle$, 
see Fig.~\ref{fig:S_mn_with_measurement}. 

\vspace{1em}
\paragraph{Bosonization framework:}
To construct the measurement circuit for the correlates of two Majoranas, e.g. $\langle \{ x_{{\bf r}\downarrow}(\tau), y_{{\bf r'} \uparrow}(0)\} \rangle$, which generalizes the circuit shown in Fig.~\ref{fig:measurement_full_correlator}, one needs to know a representation of operators 
\begin{equation}
	\label{eq:A_operators}
   A_{\bf r}^\downarrow = \frac{i}{2} x_{\bf r}^\downarrow \bar x_{\bf r'}^\downarrow,  \qquad
   A_{\bf r'}^\uparrow = \frac{i}{2} y_{\bf r'}^\uparrow \bar x_{\bf r'}^\downarrow
\end{equation}
in terms of Pauli matrices. Note, that we have chosen the fermion $a_{\bf r'}^\downarrow$ as the auxiliary one. It is located 
in the direct neighborhood of the physical fermion $c_{\bf r'}^\downarrow$, see Fig.~\ref{fig:2DCluster};
therefore the operator $A_{\bf r'}^\uparrow$ becomes local. The latter has been already analyzed, see Eq.~(\ref{eq:A_source}), and
the corresponding unitary circuit is presented in Fig.~\ref{fig:Perturbation_Circuit}. In what follows,
we discuss the non-local bilinear $A_{\bf r}^\downarrow$. Its representation involves the Jordan-Wigner string connecting 
sites associated with operators $c_{\bf r}^{\downarrow}$ and $a_{\bf r'}^{\downarrow}$, visualized by a green arrow in Fig.~\ref{fig:2DCluster}.

\begin{figure*}[t]
	\centering
    \includegraphics[scale=0.22]{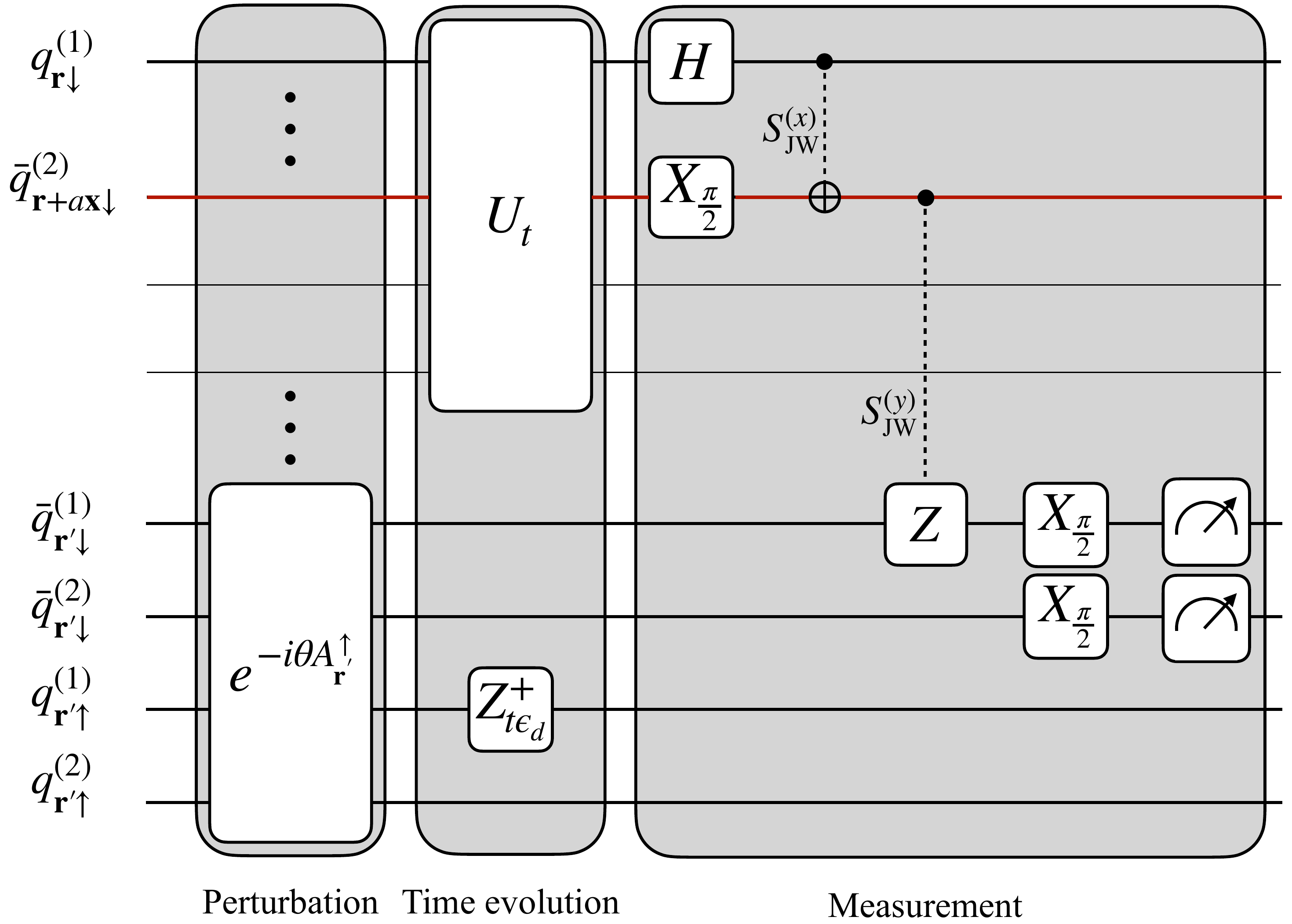}
	\caption{(Color online) Quantum circuit realizing the measurement of the correlator $\langle \{ x_{{\bf r}\downarrow}(\tau), y_{{\bf r'} \uparrow}(0)\} \rangle$, partitioned in a perturbation stage, time evolution and finally a measurement stage. 
	This circuit superimposes the one shown in Fig.~\ref{fig:measurement_full_correlator}
	in the case when a locality-preserving mapping to encode fermion operators in terms of qubits is used. 
	The final measurement block is used to estimate the average $\langle A_{\bf r}^\downarrow \rangle$ defined in Eq.~(\ref{eq:A_operators}). 
	Here two unitary operators, $S^{(x)}_{\rm JW}$ and $S^{(y)}_{\rm JW}$,
	which are used to remove the JW strings, jointly operate on the qubit $\bar q^{(2)}_{{\bf r} + a {\bf x} \downarrow}$ (red line). The latter is used to encode the auxiliary Majorana $a_{{\bf r} + a {\bf x}}^{\downarrow}$ at the 'corner' site where vertical and horizontal JW strings meet, see Fig.~\ref{fig:2DCluster}.
}
	\label{fig:Measurement_XY}
\end{figure*}

\begin{figure}
	\centering
    \includegraphics[scale=0.126]{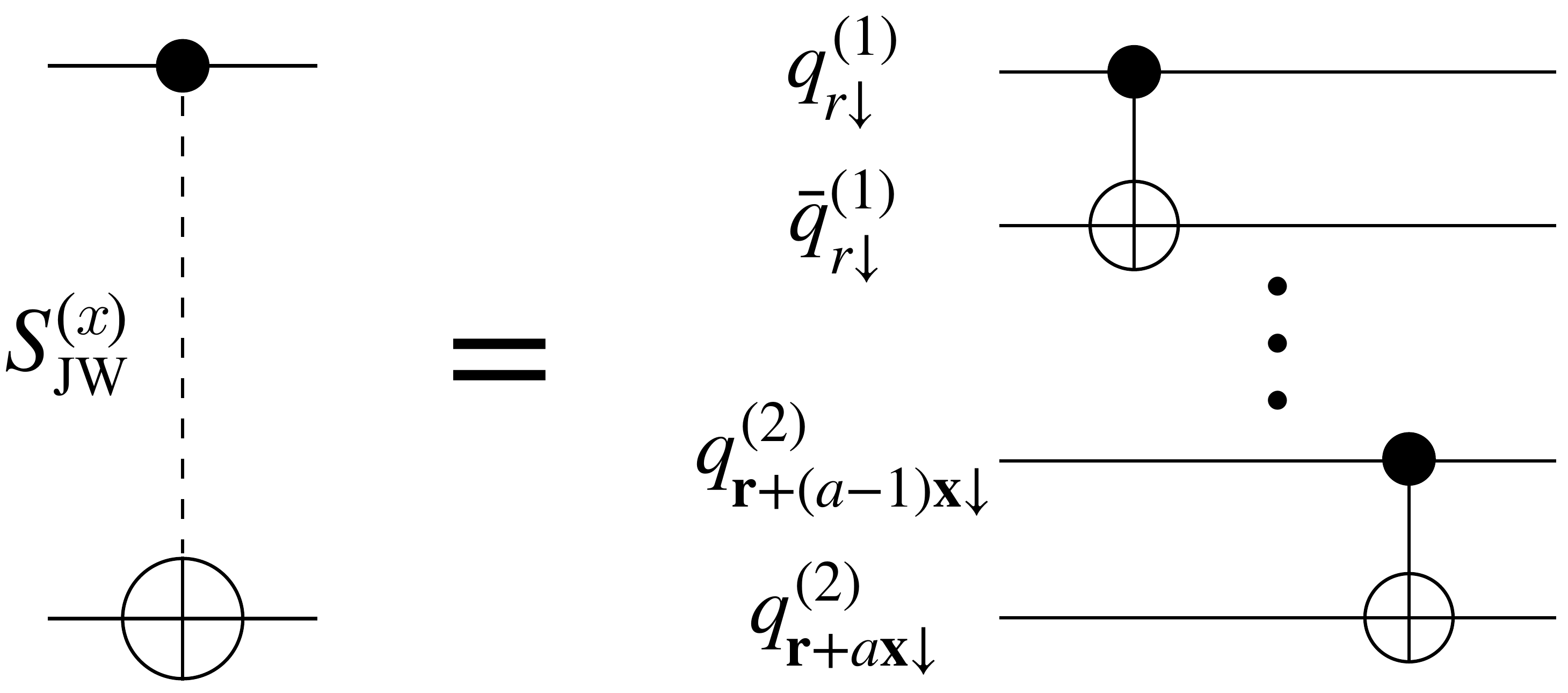}
	\caption{Expansion of the $S_{\text{JW}}^{(x)}$ circuit element, for which the role is to remove JW string in $x$-direction.}
	\label{fig:Measurement_X}
\end{figure}

\begin{figure}
	\centering
    \includegraphics[scale=0.12]{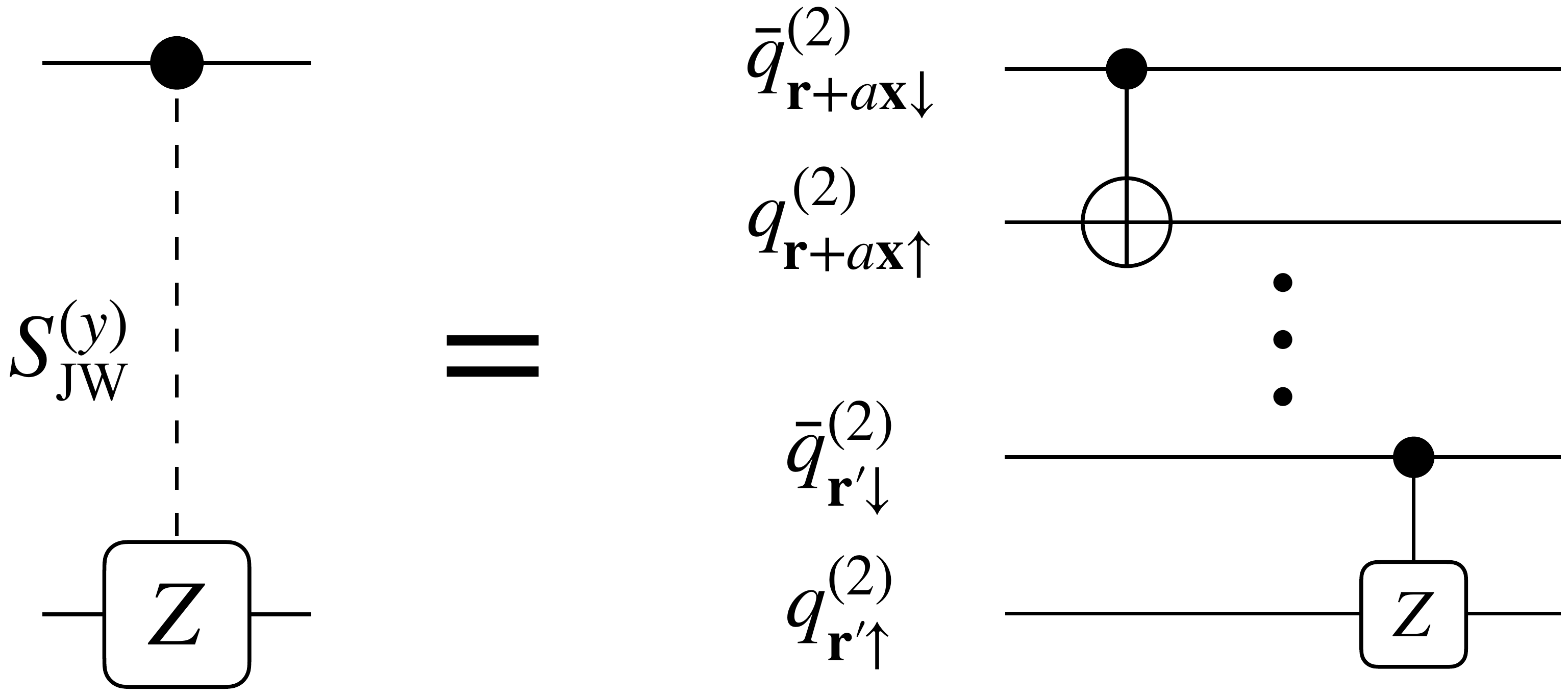}
	\caption{Expansion of the $S_{\text{JW}}^{(y)}$ circuit element, for which the role is to remove JW string in $y$-direction.}
	\label{fig:Measurement_Y}
\end{figure}

Let ${\bf r'} = {\bf r}  + a {\bf x} - b {\bf y}$, where $a$ and $b$ be positive integers. One can use a corner site located
at ${\bf r}  + a {\bf x}$ (shown in red in Fig:~\ref{fig:2DCluster}) to split $A_{\bf r}^\downarrow$ into a
product of two bilinears, 
\begin{equation}
   A_{\bf r}^\downarrow = -\frac i 2 (x_{\bf r}^\downarrow \bar x_{{\bf r} + a {\bf x}}^\downarrow) \times (\bar x_{\bf r'}^\downarrow \bar x_{{\bf r} + a {\bf x}}^\downarrow ).
\end{equation}
Each pair of Majorana operators above can in turn be bosonized following the same route as how hopping operators $T_{{\bf r}\sigma}^{x(y)}$ were constructed in \ref{sec:QC_hoppings_and_U_terms_locality_preserving}. Specifically, for the horizontal hopping over a distance $a$ one obtains
\begin{eqnarray}
\label{eq:X_hop}
    i\, (x_{\bf r}^\downarrow \bar x_{{\bf r} + a {\bf x}}^\downarrow) = - X_{{\bf r} \downarrow}^{(1)} \times Z_{\rm JW}^{(x)} \times
    \bar Y_{{\bf r} + a {\bf x} \downarrow}^{(1)} \, \bar Z_{{\bf r} + a {\bf x} \downarrow}^{(2)},
\end{eqnarray}
where the horizontal Jordan-Wigner string is defined by 
\begin{equation}
    Z_{\rm JW}^{(x)} = \prod_{n=1}^{a-1}\left( \bar Z_{{\bf r} + n {\bf x} \downarrow}^{(1)}  \bar Z_{{\bf r} + n {\bf x} \downarrow}^{(2)} 
    Z_{{\bf r} + n {\bf x} \downarrow}^{(1)}  Z_{{\bf r} + n {\bf x} \downarrow}^{(2)}\right),
\end{equation}
and spans both sets of qubits. Analogous considerations in the case of vertical hopping operator over a distance $b$ yield 
\begin{equation}
\label{eq:Y_hop}
    i\,(\bar x_{\bf r'}^\downarrow \bar x_{{\bf r} + a {\bf x}}^\downarrow ) = (-1)^{b-1}  \bar Y_{{\bf r'}\downarrow}^{(1)}    
    \bar Y_{{\bf r} +  a {\bf x} \downarrow}^{(1)} \times \bar Y_{{\bf r'}\downarrow}^{(2)} \, Z_{\rm JW}^{(y)} \,
    \bar X_{{\bf r} +  a {\bf x} \uparrow}^{(2)}. 
\end{equation}
The vertical Jordan-Wigner string spans the second set of qubits only,
\begin{equation}
    Z_{\rm JW}^{(y)} =   Z_{{\bf r} + a {\bf x} \uparrow}^{(2)} 
    \prod_{n=1}^{b-1}\left(\bar Z_{{\bf r} + a {\bf x}  -  n  {\bf y} \downarrow}^{(2)} 
    Z_{{\bf r} + a {\bf x} - n  {\bf y} \uparrow}^{(2)}\right).
\end{equation}
Combining intermediate results (\ref{eq:X_hop}) and (\ref{eq:Y_hop}) leads to a Pauli representation of the desired operator we wish to measure:
\begin{equation}
    A_{\bf r}^\downarrow = -\frac {(-1)^{b} }{2} X_{{\bf r} \downarrow}^{(1)} \times ( Z_{\rm JW}^{(x)} \, \bar Y_{{\bf r} + a {\bf x} \downarrow}^{(2)}
    \,  Z_{\rm JW}^{(y)} ) \times \bar Y_{{\bf r'}\downarrow}^{(1)} \bar Y_{{\bf r'}\downarrow}^{(2)}.
\end{equation}
For the sake of clarity operators are now ordered along the path going from ${\bf r}$ to  ${\bf r'}$ as shown in Fig.~\ref{fig:2DCluster}. 

In order to efficiently measure an average $\langle A_{\bf r}^\downarrow\rangle$, one can first apply simple unitary rotations $H$ and $X_{\pi/2}$
acting on qubits $q_{{\bf r} \downarrow}^{(1)}$ and $\bar q_{{\bf r} + a {\bf x} \downarrow}^{(2)}$, respectively. They will transform
the operator $A_{\bf r}^\downarrow$ into the type of expression \ref{eq:h_ij}, namely
\begin{equation}
    \tilde A_{\bf r}^\downarrow =  -\frac {(-1)^{b} }{2} ( \tilde Z_{\rm JW}^{(x)} \,  \tilde Z_{\rm JW}^{(y)} ) \times \bar Y_{{\bf r'}\downarrow}^{(1)} \bar Y_{{\bf r'}\downarrow}^{(2)}.
\end{equation}
The updated Jordan-Wigner strings read 
\begin{align}
    \tilde Z_{\rm JW}^{(x)} &=   Z_{{\bf r} \downarrow}^{(1)}  Z_{\rm JW}^{(x)}, \\ \qquad 
    \tilde Z_{\rm JW}^{(y)} &=   \bar Z_{{\bf r} + a {\bf x} \uparrow}^{(2)}  Z_{\rm JW}^{(y)} =
    \prod_{n=0}^{b-1}\left(\bar Z_{{\bf r} + a {\bf x}  -  n  {\bf y} \downarrow}^{(2)}   Z_{{\bf r} + a {\bf x} - n  {\bf y} \uparrow}^{(2)}\right).
\end{align}
After the above transformation one may use the same scheme as it was previously discussed in \ref{ch:measurement_GS}. Similarity transformations with operators $S_{\rm JW}^{x,y}$ alongside two subsequent single qubit rotations with $X_{\pi/2}$ reduce $\tilde A_{\bf r}^\downarrow $ to the product of just two Pauli $Z$ operators,
\begin{equation}
    \tilde A_{\bf r}^\downarrow = {\cal S}^\dagger ( \bar Z_{{\bf r'}\downarrow}^{(1)} \bar Z_{{\bf r'}\downarrow}^{(2)}) {\cal S},  \qquad {\cal S} =   (\bar X_{\tfrac \pi 2})_{{\bf r'}\downarrow}^{(1)} (\bar X_{\tfrac \pi 2})_{{\bf r'}\downarrow}^{(2)} S_{\rm JW}^{(y)} S_{\rm JW}^{(x)}.
\end{equation}
The resulting quantum circuit realizing such measurement is shown in Figs.~\ref{fig:Measurement_XY}, \ref{fig:Measurement_X}, \ref{fig:Measurement_Y}, where we have defined operators $S_{\rm JW}^{x,y}$ --- their role is to remove Jordan-Wigner strings acting along horizontal and vertical directions. 
Due to the local type of the fermion-to-qubit mapping, the number of {\rm CNOT} gates used in the ${\cal S}$ operator scales as $4a+2b$.

\section{Two-site dimer model }
\label{sec:two_site_dimer}
In this section we introduce a two-site dimer \cite{Eder2013} that serves as our toy model for which we evaluate the Green's function. 
Following the general framework outlined in Sec.~\ref{sec:VCA}, it can be seen as the smallest non-trivial cluster, 
so that the variational cluster approximation is able to deliver physically reasonable results. 
Specifically, within VCA such minimal cluster is sufficient to reproduce the Mott insulating transition
in the Hubbard model. On the other hand, in order to recover the d-wave superconducting phase, one needs at least 
four sites per cluster~\cite{Senechal2010}. 

The two-site dimer model consists of a Hubbard site that is coupled to a bath site. Its Hamiltonian at half-filling given by
\begin{equation}
\label{eq:dimer_hamiltonian}
H' = H_0 + H_U = -t \sum_\sigma (c^\dagger_\sigma b_\sigma + b^\dagger_\sigma c_\sigma) + \frac{U}{2}(n^2_c - 2n_c),
\end{equation}
where $t$ is the hopping energy and $U$ is the Coulomb repulsion. 
The field operators $c_\sigma^\dagger, c_\sigma$ respectively creates or destroys a fermion with spin $\sigma$ at the Hubbard site, while $b_\sigma^\dagger, b_\sigma$ respectively creates or destroys a fermion with spin $\sigma$ at the bath site and $n_c = \sum_\sigma c_\sigma^\dagger c_\sigma$. 
Lastly, the linear $n_c$ term in $H_U$ stems from a chemical potential $\mu=U$ at half-filling.

\begin{figure}[b]
	\centering
	\includegraphics[scale=0.15]{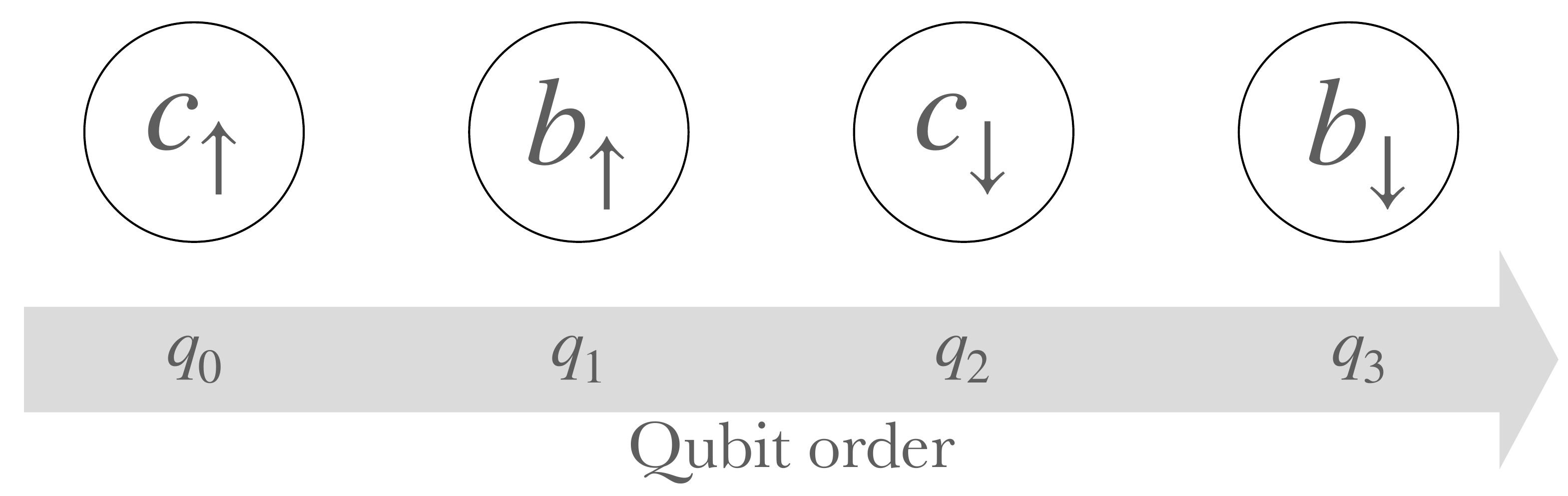}
	\bigskip
	\caption{Indexing qubits over one Hubbard site and and bath site.}
	\label{fig:qubit_ordering}
\end{figure}

Qubit ordering is shown in Fig.\ \ref{fig:qubit_ordering} and is chosen in a way that allows for decreasing circuit depth, see section \ref{ch: section_VHA}.

Investigating the dimer at half-filling allows us to concentrate on just six out of 16 possible state configurations. With $\lvert 0 \rangle$ being the vacuum state, following states $\lvert 1 \rangle, \lvert 2 \rangle, \dots, \lvert 6 \rangle$ are possible as two electrons reside at the dimer:
\begin{equation}
\begin{aligned}
|1\rangle &= c^\dagger_\downarrow c^\dagger_\uparrow |0\rangle, \quad |2\rangle = c^\dagger_\uparrow b^\dagger_\uparrow |0\rangle, 
\quad |3\rangle = b^\dagger_\downarrow b^\dagger_\uparrow |0\rangle\\
|4\rangle &= b^\dagger_\downarrow c^\dagger_\uparrow |0\rangle, \quad
|5 \rangle = b_\uparrow^\dagger c_\uparrow^\dagger \lvert 0 \rangle, \quad
|6 \rangle = b_\downarrow^\dagger c_\downarrow^\dagger \lvert 0 \rangle.
\end{aligned}
\label{eq:eigenstates}
\end{equation}
In matrix form, the Hamiltonian (\ref{eq:dimer_hamiltonian}) thus reads
\begin{equation}
H'=  - \begin{pmatrix}
0 & -t & 0 & t & 0 & 0\\
-t & \frac{U}{2} & -t & 0 & 0 & 0 \\
0 & -t  & 0 & t & 0 & 0  \\
t & 0 & t & \frac{U}{2} & 0 & 0 \\
0 & 0 & 0 & 0 & \frac{U}{2} & 0 \\
0 & 0 & 0 & 0 & 0 & \frac{U}{2}
\end{pmatrix},
\end{equation}
with the ground state energy 
\begin{equation}
    E_0=-\frac{1}{4}(U+\sqrt{U^2+64t^2}),
\label{eq:eigenenergies}
\end{equation}
corresponding to the eigenstate
\begin{equation}
\label{eq:Psi_0}
  |\Psi\rangle \propto C(|1\rangle + |3\rangle) + |2\rangle + |4\rangle) ,\quad
  C=\frac{\sqrt{U^2 + 64 t^2} - U}{8 t},
\end{equation}
up to a normalization factor.
Lastly, we find the expectation values of $H_0$ and $H_U$ to be
\begin{equation}
\langle H_0 \rangle = -\frac{16t^2}{\sqrt{U^2 + 64t^2}},
\end{equation}
and
\begin{equation}
\langle H_U \rangle = -\frac{U}{4} \Big( 1+\frac{U}{\sqrt{U^2 + 64t^2}} \Big).
\end{equation}

As required by linear response theory, we need to start from an equilibrium state, i. e. the ground state. Consequently, we present a route to find the ground state via the variational Hamiltonian ansatz (VHA).

\subsection{Ground state preparation}
\label{ch: section_VHA}
For the sake of finding the ground state of the dimer system, we employ a technique known as the variational Hamiltonian ansatz \cite{Reiner2019}. Starting from the ground state of
the cluster $|\Psi_0\rangle$ in the non-interacting limit $U=0$,
we aim to find the interacting system's ground state $\lvert \Psi \rangle$ as
\begin{equation}
\lvert \Psi(\alpha, \beta) \rangle = \prod_{j=1}^p 
e^{i\beta_k H_0^{t \to 1}} e^{-i\alpha_k H_U^{U \to 1}} |\Psi_0\rangle
\label{eq:VHA}
\end{equation}
where $\alpha_j, \beta_j \in \mathbb{R}$ are the variational parameters. The ground state $|\Psi_0\rangle$ 
is a Slater determinant. For instance, Eq.~(\ref{eq:Psi_0}) at $U=0$ can be written in terms of two $f$-fermions,
\begin{equation}
\label{eq:Psi_0_fs}
|\Psi_0\rangle = f^\dagger_\downarrow f^\dagger_\uparrow |0\rangle, \qquad 
f^\dagger_\sigma = \frac{1}{\sqrt{2}} (c^\dagger_\sigma + b^\dagger_\sigma),
\end{equation} 
which are linear superpositions of $b$ and $c$. 

\begin{figure}[t]
	\centering
	\includegraphics[scale=0.11]{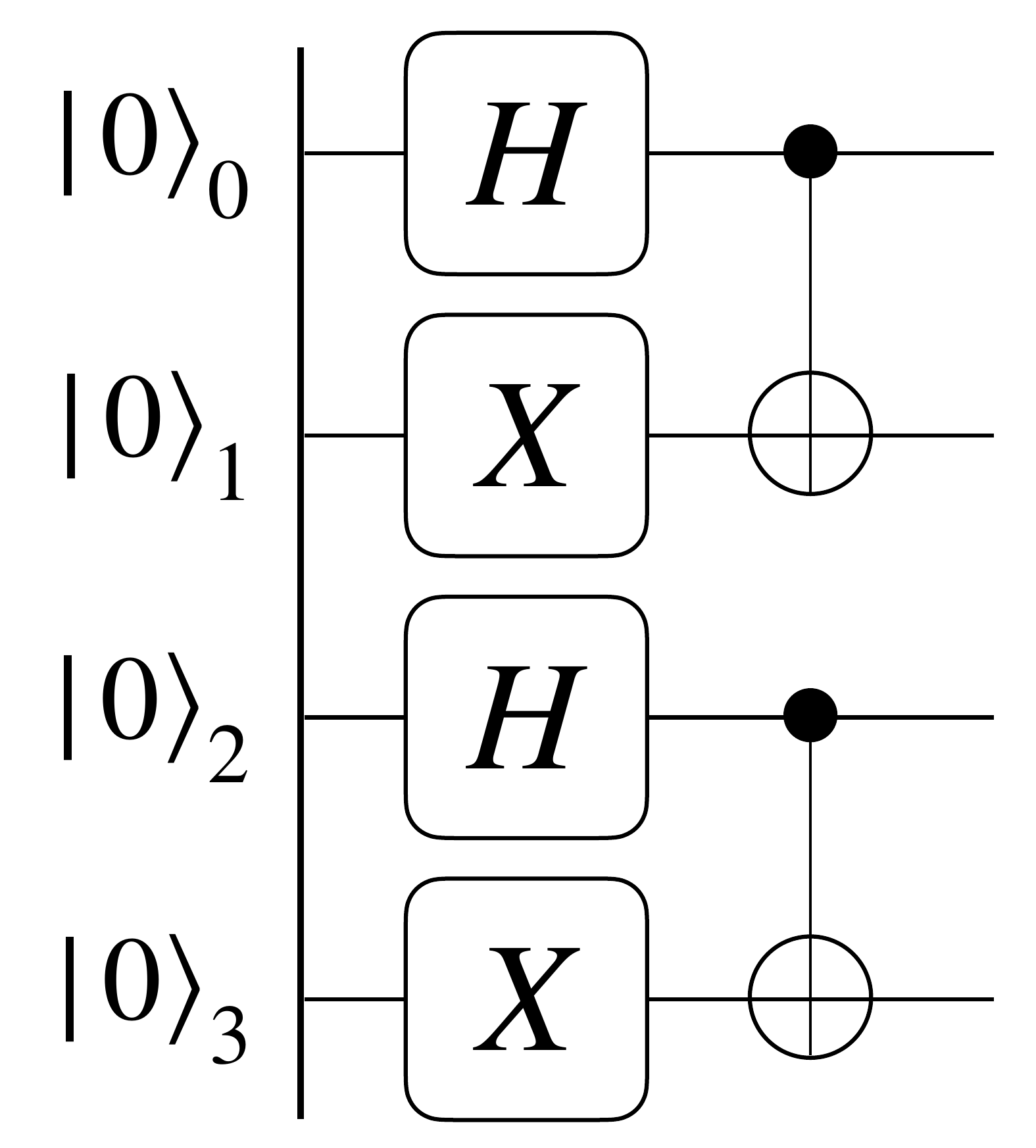}
	\caption{Quantum circuit to prepare a Slater determinant as a trial Ansatz for the VHA.}
	\label{fig:slater_prep}
\end{figure}

\begin{figure}[b]
	\centering
	\includegraphics[scale=0.1]{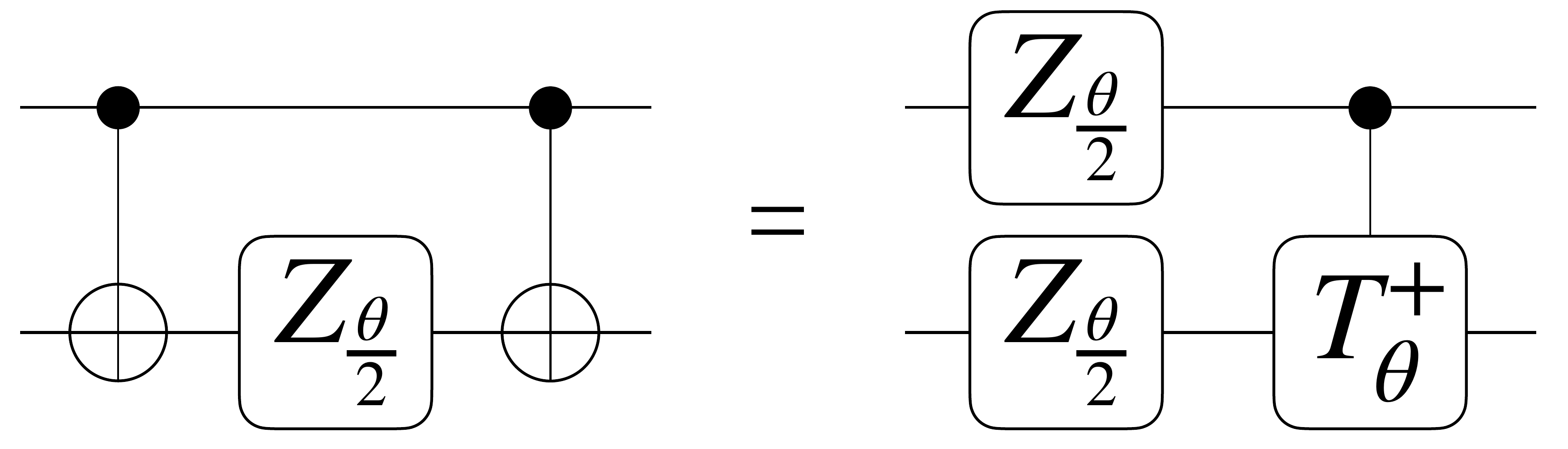}
	\caption{Circuit identity as used in the two-site dimer.}
	\label{fig:2nd_simplification}
\end{figure}

\begin{figure*}[t]
	\centering
	\includegraphics[scale=0.18]{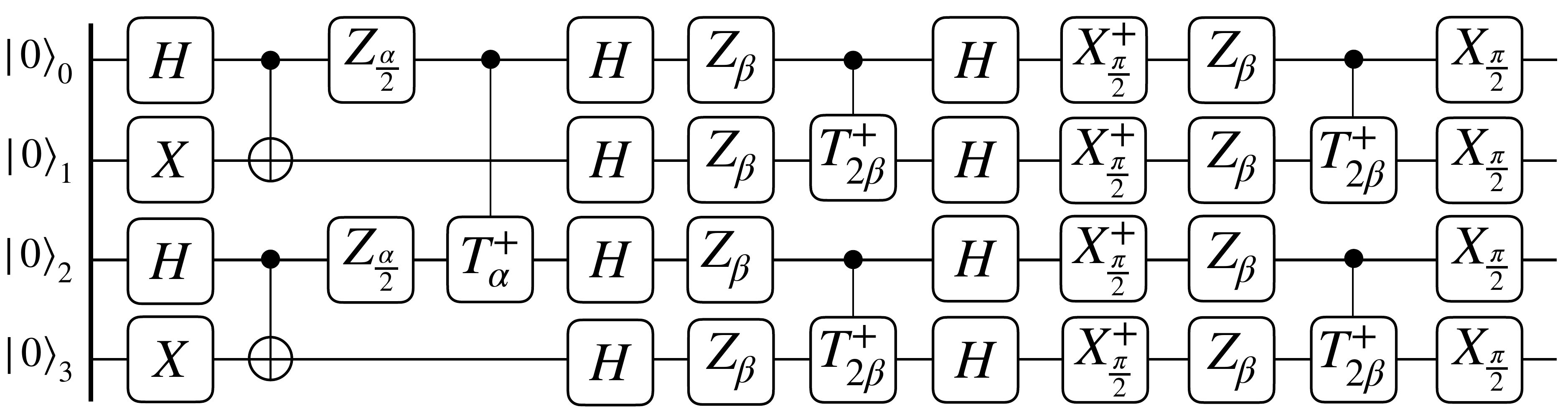}
	\caption{Reduced quantum circuit for finding the ground state of a correlated system. 
		At the end of the circuit we are left with the interacting ground state of the two-site dimer.}
	\label{fig:interacting_GS_prep}
\end{figure*}

\begin{figure}
	\includegraphics[scale=0.45, center]{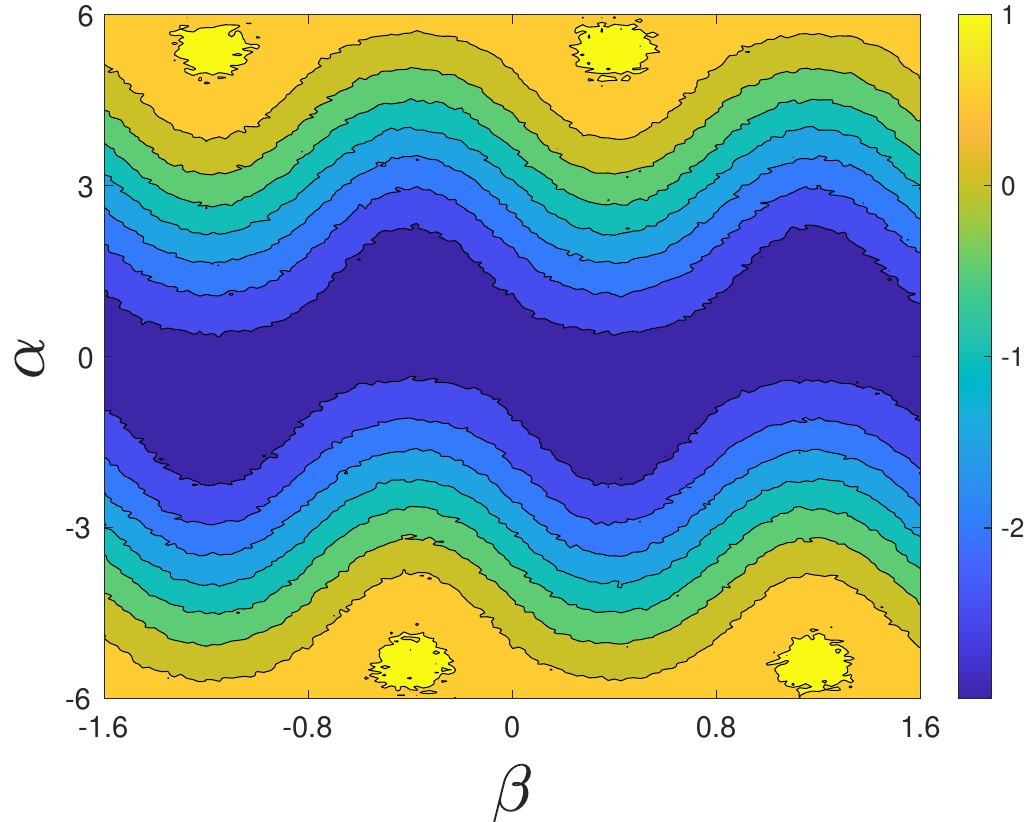}
	\caption{(Color online) Energy landscape for the two-site dimer for finding its ground state via the variational Hamiltonian ansatz with angles $\alpha$ and $\beta$ shown for $t=1$ and $U=4$. The energies were evaluated on Qiskit's noisy \textit{Aer} simulator of the \textit{FakeKolkataV2()} backend, an open-access simulator of the corresponding superconducting device \textit{ibmq\_kolkata} provided by IBM. The optimum is lying very close to theoretical values $\alpha_* = -0.92$ and $\beta_* = 0.39$ found from Eq.~(\ref{eq:GS_Energy_Dimer}).}
	\label{fig:e_landscape}
\end{figure} 

An initialization of the quantum chip in a Slater determinant state most generally can be constructed from an initial one, $|0, 0, ..., 0\rangle$, by the so-called Given's rotations, \cite{Wecker2015}. However, for the two-fermion state~(\ref{eq:Psi_0_fs}) a much simpler circuit is sufficient.
One can verify by direct inspection that the circuit shown in Fig.~\ref{fig:slater_prep} transforms the vacuum state into $|\Psi_0\rangle$. 
Moreover, it is possible to find the exact matching between the trial and actual ground state functions already for the minimal depth VHA, i.e with $p=1$. Therefore, we find the full circuit to prepare the ground state is a single sequence of the circuit shown in Fig. \ref{fig:slater_prep}, followed by a variation of the hopping circuit, cf. Fig. \ref{fig:circuit_hopping} and finalized by a simpler variation of the repulsion circuit, Fig. \ref{fig:circuit_repulsion}. 

At this point we note that the evolution operator over a time step $\Delta \tau$ under the interaction Hamiltonian $H_U$, see Eq.~(\ref{eq:dimer_hamiltonian}), reads 
\begin{equation}
\label{eq:U_int}
U(\theta) = \text{CNOT}^{(13)} \cdot Z^{(3)}_{\theta/2} \cdot \text{CNOT}^{(13)}, 
\end{equation}
with angle $\theta = U\cdot \Delta \tau$. A difference from the repulsion circuit on Fig.~\ref{fig:circuit_repulsion} comes from the extra term $U n_c$ in the present choice of $H_U$,
which effectively leads to the reduction of two single-qubit $Z$-rotation gates. Furthermore, 
if one of the natural two-qubit gates on the hardware is a controlled-phase gate,
$\text{CT}(\theta)$, then the unitary (\ref{eq:U_int}) can be simplified to
\begin{equation}
U(\theta) =  Z^{(1)}_{\theta/2}   \cdot Z^{(3)}_{\theta/2} 
({\theta}/{2}) \cdot \text{CT}^{(13)}_{-\theta},
\label{eq:2nd_simplification}
\end{equation}
up to a global phase. The equivalence of the two circuits in Eqs. (\ref{eq:U_int}) and (\ref{eq:2nd_simplification}) for $U(\theta)$ is presented in Fig.\ \ref{fig:2nd_simplification}. 

The full circuit to prepare the VHA ground state is presented in Fig.~\ref{fig:interacting_GS_prep}.
Comparing Fig.~\ref{fig:interacting_GS_prep} to Fig. \ref{fig:circuit_repulsion} and \ref{fig:circuit_hopping}, we reach a reduction of two two-qubit gates for the repulsion compared to the scheme in Fig.\ \ref{fig:circuit_repulsion} and another reduction of eight two-qubit gates for the new hopping scheme. Finally, the variational energy reads
\begin{equation}
\label{eq:GS_Energy_Dimer}
E(\alpha,\beta)=-2t\cos{\frac{\alpha}{2}}-\frac{U}{4}\Big(1-\sin{\frac{\alpha}{2}}\sin{4\beta}\Big),
\end{equation}
which depends on just two parameters, $\alpha$ and $\beta$.  
Fig.~\ref{fig:e_landscape} shows the corresponding energy landscape. For direct comparison to analytical values, we refer the reader to Appendix \ref{sec:e_landscape_analytical}, Fig. \ref{fig:e_landscape_analytical}.

\subsection{Analytical formulae for the Green's function}
Next we discuss the analytical results for the Green's functions of the two-site model in order to benchmark them with our circuit simulations which we review in the next subsection.
Following the ordering of fermion states under Jordan-Wigner mapping shown in Fig.~\ref{fig:qubit_ordering}, we introduce a set of eight Majorana operators $\{x_n, y_n\}$ so that
\begin{eqnarray}
    c_\uparrow &=& \tfrac 12 (x_0 - i y_0), \qquad b_\uparrow =  \tfrac 12 (x_1 - i y_1), \nonumber \\
    c_\downarrow &=& \tfrac 12 (x_2 - i y_2), \qquad b_\downarrow = \tfrac 12 (x_3 - i y_3),
    \end{eqnarray}
with $n$ being a composite index accounting for both site and spin.
Correlation functions of interest take the form given by Eqs.~(\ref{eq:superposition_pauli}) and  (\ref{eq:g_Majorana}). Because of spin symmetry we find them to be block-diagonal
w.r.t. spin indices, 
\begin{equation}
\begin{aligned}
i g^{\uparrow\uparrow}(\tau) &=
\begin{pmatrix}
\langle x_0(\tau) x_0(0)\rangle & \langle x_0(\tau) y_1(0) \rangle \\
\langle y_1(\tau) x_0(0) \rangle  & \langle y_1(\tau) y_1(0)\rangle
\end{pmatrix} ,
\end{aligned}
\end{equation}
and
\begin{equation}
\begin{aligned}
i g^{\downarrow\downarrow}(\tau) &=
\begin{pmatrix}
\langle x_2(\tau) x_2(0)\rangle & \langle x_2(\tau) y_3(0) \rangle \\
\langle y_3(\tau) x_2(0) \rangle  & \langle y_3(\tau) y_3(0)\rangle
\end{pmatrix} ,
\end{aligned}
\end{equation}
with two blocks being mutually equal, $ g^{\uparrow\uparrow}(\tau)  = g^{\downarrow\downarrow}(\tau)$.
Other non-zero correlators follow from the symmetries 
\begin{equation}
\label{eq:corr_symmetries}
\begin{aligned}
   \langle x_i(\tau) x_i(0) \rangle &= \langle y_i(\tau) y_i(0) \rangle, \qquad \forall \, i \\
    \langle x_i(\tau) y_{i+1}(0) \rangle &= \langle x_{i+1}(\tau) y_{i}(0) \rangle, \quad i=0,2.
\end{aligned}
\end{equation}
Additionally, the self-adjoint property of Majorana operators implies that
\begin{equation}
 \langle x_i(\tau) y_{j}(0) \rangle^* =  \langle y_{j} (0)  x_{i}(\tau) \rangle,   
\end{equation}
and the same for $x\!-\!x$ and $y\!-\!y$ correlators. Therefore, the retarded
correlator reads
\begin{equation}
    \langle \{x_i(\tau), y_{j}(0)\} \rangle = 2 {\rm Re}\, \langle x_i(\tau) y_{j}(0) \rangle, \quad \tau>0.
\end{equation}

\begin{figure}[b]
	\centering
	\includegraphics[scale=0.14]{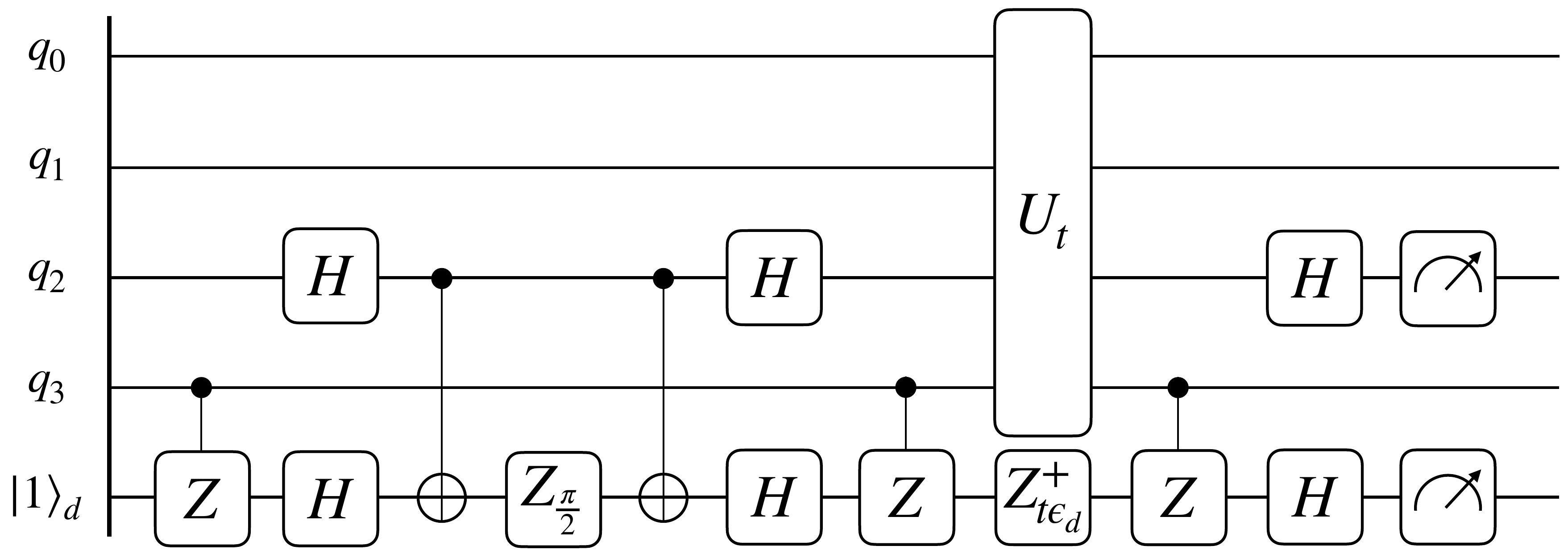}
	\caption{Circuit for evaluating the $\langle \{y_2(\tau), y_2(0) \} \rangle $ correlator.}
	\label{fig:y2y2}
\end{figure}
\begin{figure}
	\centering
	\includegraphics[scale=0.12]{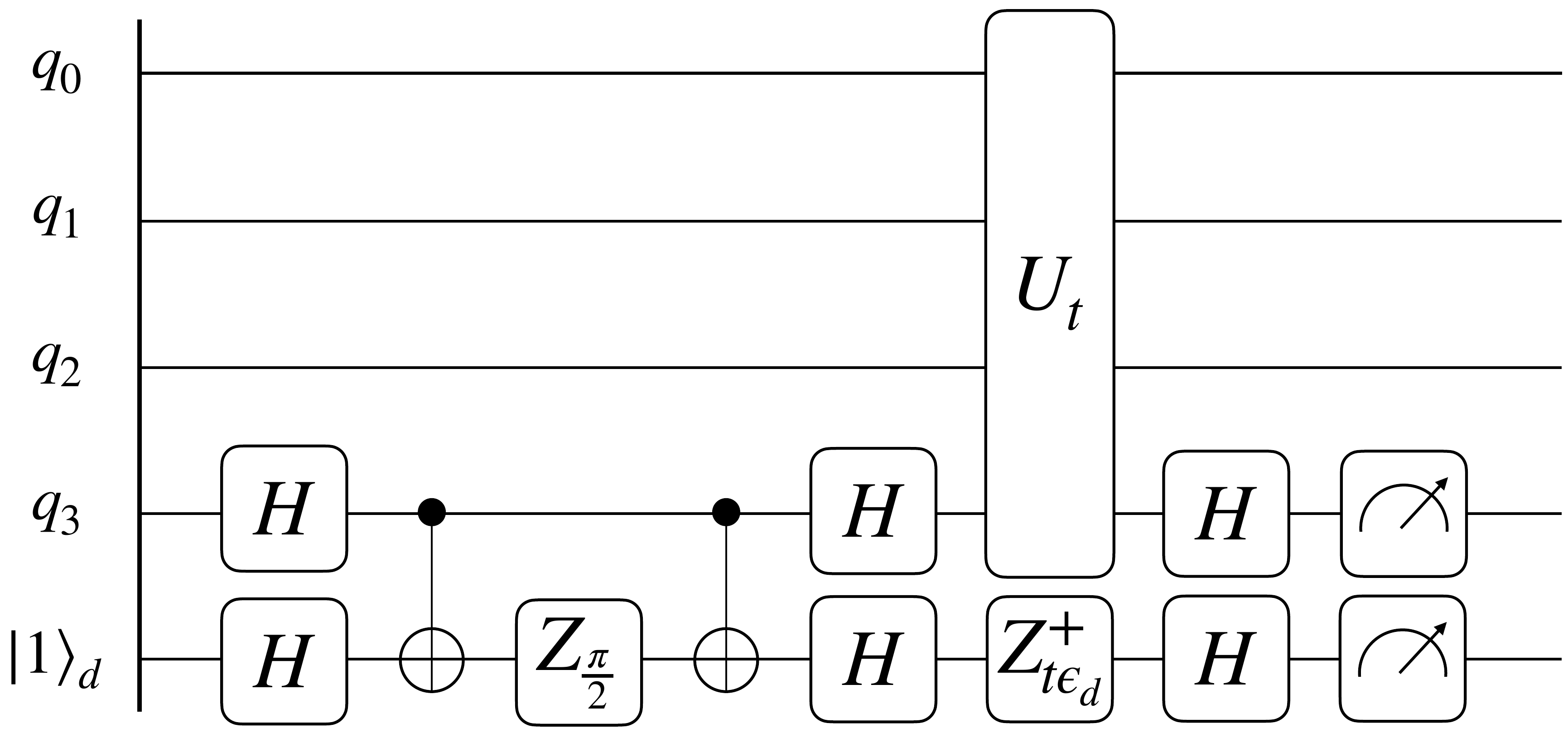}
	\caption{Circuit for evaluating the $\langle \{y_3(\tau), y_3(0) \} \rangle $ correlator.}
	\label{fig:y3y3}
\end{figure}
\begin{figure}
	\centering
	\includegraphics[scale=0.127]{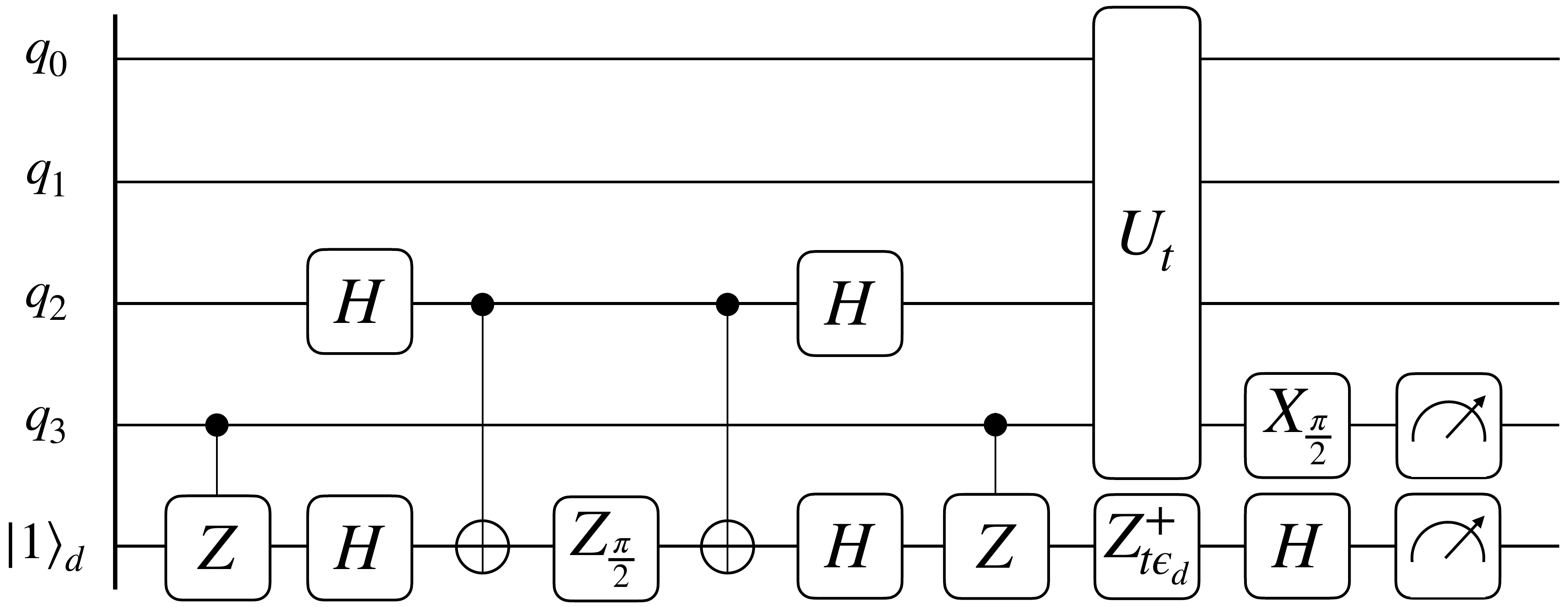}
	\caption{Circuit for evaluating the $\langle \{ x_3(\tau), y_2(0) \}\rangle $ correlator.}
	\label{fig:x3y2}
\end{figure}

\begin{figure*}[t!]
	\centering
	\includegraphics[scale=0.32]{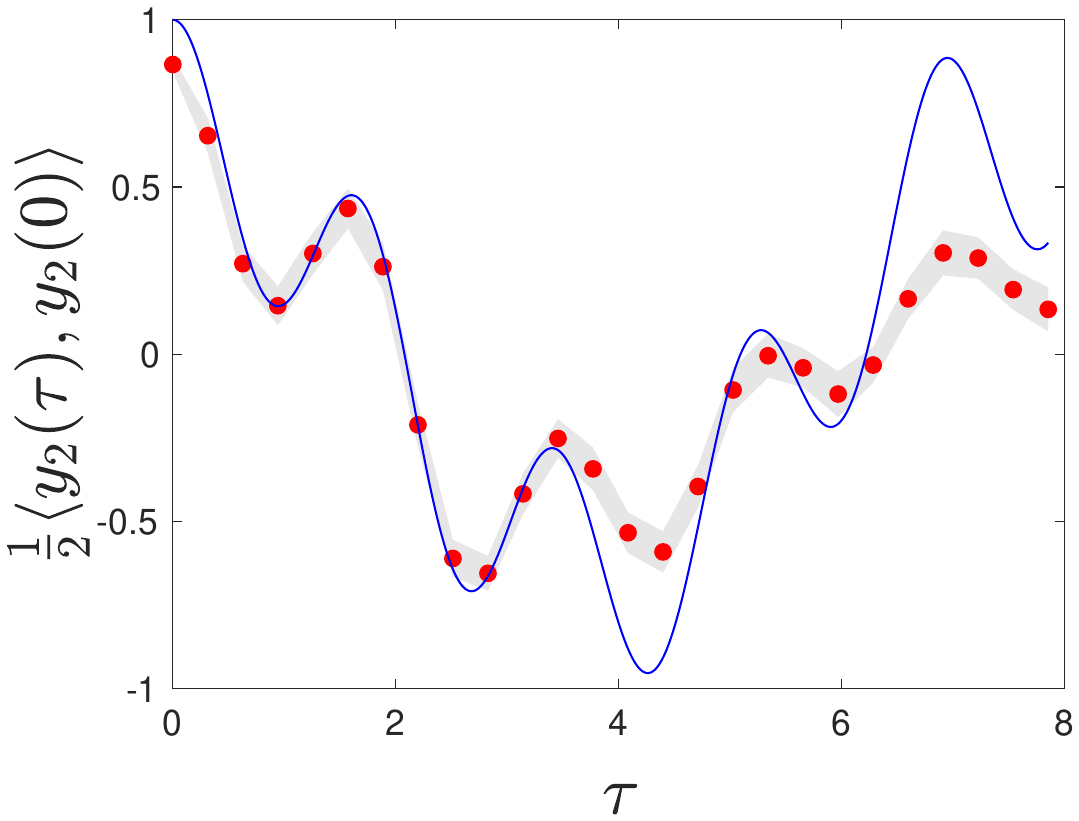}
	\includegraphics[scale=0.32]{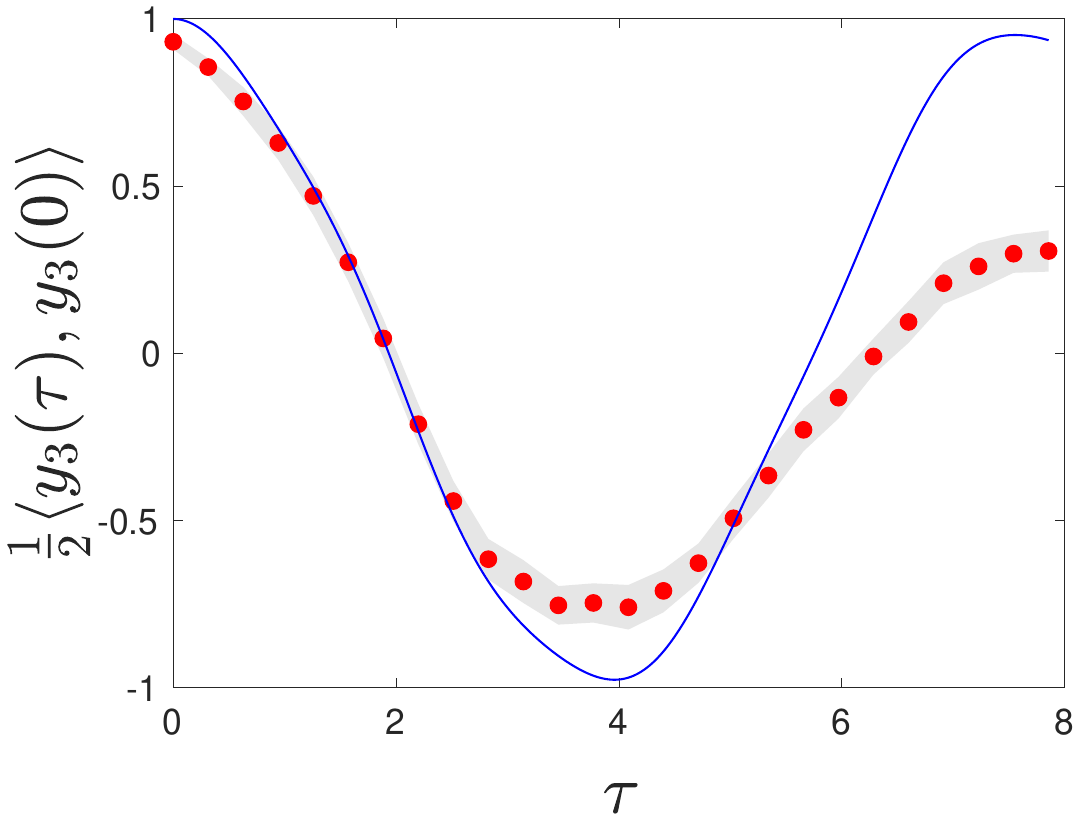}
	\includegraphics[scale=0.32]{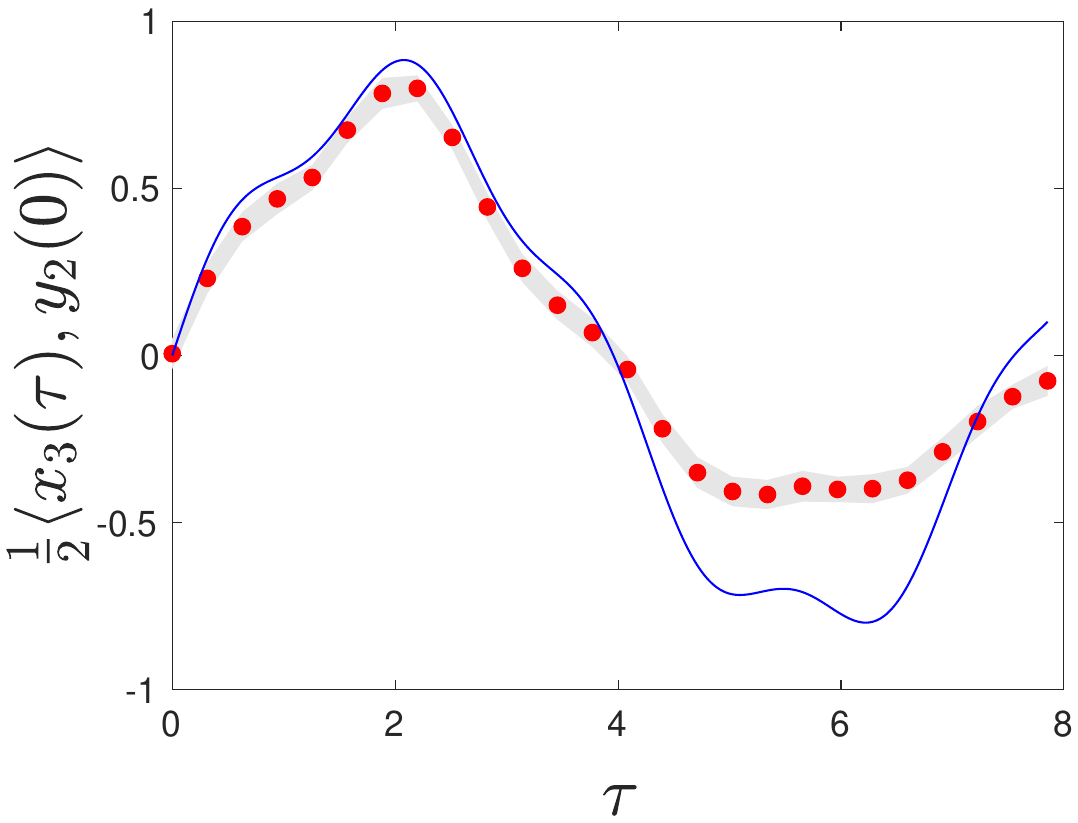}
	\caption{(Color online) Correlators $y_2\!-\!y_2$, $y_3\!-\!y_3$ and $x_3\!-\!y_2$ evaluated in 25 Trotter steps (red dots) of the duration $\Delta \tau =0.314$. A shaded area (grey) indicates the standard deviation over 100 repetitions of executing quantum circuits. The solid blue curves correspond to the analytical results, cf.\ Eqs.~(\ref{eq:analytical_X0X0}--\ref{eq:analytical_X0Y1}), where we account for the symmetry relations~(\ref{eq:corr_symmetries}). System parameters are chosen to be $V=1$ and $U=4V$. The correlators were evaluated on Qiskit's noisy \textit{Aer} simulator of the \textit{FakeKolkataV2()} backend, an open-access simulator of the corresponding superconducting device \textit{ibmq\_kolkata}  provided by IBM.}
	\label{fig:Y2Y2_result}
\end{figure*}

We can evaluate the above correlation functions exactly using the Lehmann representation. Starting from $i G^>(t) =\langle \Psi_* \lvert U^\dagger(t) x_i U(t) y_j \lvert \Psi_* \rangle$, where $\lvert \Psi_* \rangle$ is the ground state, we use the eigenstate decomposition of the evolution operator, $U(t) = \sum_m^{D} \lvert m \rangle e^{-iE_mt} \langle m \lvert$, with $D=4^{N_c}$ being the Hilbert space dimension of a cluster. 
It then follows that $ i G^>(t) =\sum_m^{D} e^{-iE_0t} \langle \Psi_* \lvert x_i \lvert m \rangle e^{-iE_mt} \langle m \lvert y_j \lvert \Psi_* \rangle$. 
The eigenstates $\lvert m \rangle$ can be obtained by exact diagonalization of $H'$, cf. Eq.~(\ref{eq:dimer_hamiltonian}), which renders the two-site dimer Green's function amenable to analytic treatment. 
The correlation functions evaluated in this manner read
\begin{equation}
\begin{aligned}
\langle x_0(\tau)&x_0(0)\rangle = \\
& e^{-\frac{i }{4} \tau U_2} \left(\cos \frac{\tau  U_1 }{4}  -\frac{i \left(U^2-32 t^2\right)}{U_1 U_2 } \sin \frac{\tau  U_1}{4}\right), 
\label{eq:analytical_X0X0}
\end{aligned}
\end{equation}
\begin{equation}
\begin{aligned}
\langle x_1(\tau)&x_1(0)\rangle = \\
& e^{-\frac{i }{4} \tau U_2} \left(\cos \frac{\tau  U_1 }{4}  +\frac{i \left(U^2+32 t^2\right)}{U_1 U_2 } \sin \frac{\tau  U_1}{4}\right) 
\label{eq:analytical_X1X1}
\end{aligned}
\end{equation}
and
\begin{equation}
\begin{aligned}
\langle x_0(\tau)&y_1(0)\rangle = \\
& 4 e^{-\frac{i }{4} \tau U_2} \left( 
\frac{2it}{ U_2} \cos \frac{\tau  U_1 }{4} 
- \frac{t}{U_1} \sin \frac{\tau  U_1 }{4}
\right), 
\label{eq:analytical_X0Y1}
\end{aligned}
\end{equation}
where two additional energy scales, 
\begin{equation}
U_1=\sqrt{U^2+16 t^2}, \quad U_2=\sqrt{U^2+64 t^2},
\end{equation}
were introduced to shorten the results.

\subsection{Quantum circuits for direct measurement of the two-site dimer Green's function}
In accordance to section \ref{direct_measurement}, we provide quantum circuits for the direct measurement of the Green's function. 
The $y_2\!-\!y_2$ correlator is directly measured when employing the circuit shown in Fig.\ \ref{fig:y2y2}. 
In turn, circuits for measuring $y_3\!\!-\!\!y_3$ and $y_3\!\!-\!\!x_2$ correlators are shown in Fig.~\ref{fig:y3y3} and Fig.~\ref{fig:x3y2}, respectively. 
To implement the unitary evolution operator, $U_t$, one applies the Trotterization scheme with a duration of a single step being $\Delta \tau$. The one-step
evolution operator can be constructed following the same sequence of gates used in VHA circuit, see Fig.~\ref{fig:e_landscape}, with angles $\alpha = U \Delta \tau$ 
and $\beta = - t \Delta \tau$.

\section{Results and discussion}
\label{sec:results}
We have tested the proposed algorithm for the direct measurement of fermionic correlation functions using an open access Qiskit's noisy simulator of the superconducting IBM chip (\textit{ibmq\_kolkata}). 
The numerical results of these simulations are presented in Fig.~\ref{fig:Y2Y2_result}, where
data points correspond to the Green's function of the two-site dimer model in the range of 25 Trotter steps plotted against the analytical Green's function, 
see Eqs.~(\ref{eq:analytical_X0X0}--\ref{eq:analytical_X0Y1}). The chosen parameters are $V=1$, $U=4V$, $\Phi = \pi/2$ and $\epsilon_d t=\pi/2$. 
We have relegated the description of implementation details (including an overview of used error mitigation techniques) of the simulated quantum processor to Appendix~\ref{sec:Implementation_details}.

It can be seen that 25 Trotter steps are sufficient to have a decent overlap of measurement points and the analytic correlator up to time $\tau \sim 8$. 
Single-qubit gates take typical operation times of~$20$ns, two- and three-qubit gates may be designed with operation times of typically~$100$ns. 
It is thus crucial to reduce two- and multi-qubit gates as they appear when a standard Hadamard test to measure the Green's function is implemented as much as possible to keep operation time short and quantum state fidelity large. One needs to stress here that a discrepancy between analytical results for
the Green's function and the outcome of numerical experiment (see Fig.~\ref{fig:shotnumber}), which becomes pronounced at long times, originates mainly from the 
infidelities of two-qubit entangling gates incorporated into the model of Qiskit's noisy simulator.
By itself, the algorithm delivers the exact Green's function under a sufficient degree of Trotterization,
as proved in Appendix~\ref{appendix_nonlin_kubo}.

It is worth noting the important universality feature of the proposed algorithm. Its formal construction,  which is provided in full details in Appendix~\ref{appendix_nonlin_kubo}, 
is based entirely on the algebra of Majorana operators. Therefore the algorithm can be utilized within any compact fermion mapping scheme. While the local encoding due to Li and Po~\cite{bosonization}
was used in this paper for the sake of illustration, the Derby-Klassen approach~\cite{DerbyKlassen} reviewed recently in Ref.~\cite{jafarizadeh2024} can be applied on equal footing~\footnote{
In the 'compact' encoding by Derby and Klassen the odd fermionic subspace can be either accessible or not for the open boundary conditions depending on particular implementation details within this scheme.}.
For any of these two mappings an estimation of the Green's function via the direct measurements involves
manipulations with the Jordan-Wigner string operators (see Fig.\ref{fig:Measurement_XY}).
Their length, however, grows only as the distance between two lattice sites, ${\bf r}$ and ${\bf r}'$,
which has to be contrasted with the conventional Jordan-Wigner mapping. 
In the latter case the advanced Hadamard test and the direct measurement scheme are of the same complexity. This can be seen from a simple comparison of two quantum circuits shown, respectively, in Fig.~\ref{fig:advanced_hadamard_test} and Fig.~\ref{fig:measurement_full_correlator}. Both algorithms require a single uncontrolled unitary evolution operator together with either four or three Jordan-Wigner strings to implement two measured Majorana operators.

It would be also very interesting to extend our scheme of direct measurements to multi-point fermionic correlation functions following recent ideas of Ref.~\cite{Lorenzo:2024}, including so-called
OTOC correlators~\cite{Swingle:2016, Mitarai:2019}, which are extensively used to characterize many-body quantum chaos in Majorana SYK models~\cite{kitaev2020} and beyond.  

To conclude, we have presented the quantum algorithm that is motivated by the linear response theory to evaluate the Green's function of the Fermi-Hubbard model on a quantum computer. The number of required measurements scales polynomially in the number of sites $N_c$ within the cluster and the number of qubits scales linearly in $N_c$. The new algorithm is superior in terms of two-qubit gates as well as operation time when compared to the standard Hadamard test,
and has the potential to outperform the latter in terms of gate count, operation time and thereby fidelity of measured quantities.
 
Directly measuring the Green's function is also a powerful alternative to the advanced Hadamard test, provided any compact fermion-to-qubit encoding scheme is used,
since it obviates the need of constructing ancilla-controlled single fermion operators. 

\vspace{0.5cm}
\section*{Acknowledgements} The authors gratefully acknowledge funding, support and computational resources from Mercedes-Benz AG. We further acknowledge support from OpenSuperQPlus100 (101113946) and German Federal Ministry of Education and Research in the funding program “Quantum technologies – from basic research to market”, contract number 13N15584 (DAQC). Furthermore, we acknowledge useful discussions with Clemens Possel. 

\bibliography{ref.bib} 

\appendix

\section{Variational cluster approach}
\label{app:VCA_approach}
\begin{figure}[t]
	\centering
	\includegraphics[scale=0.13]{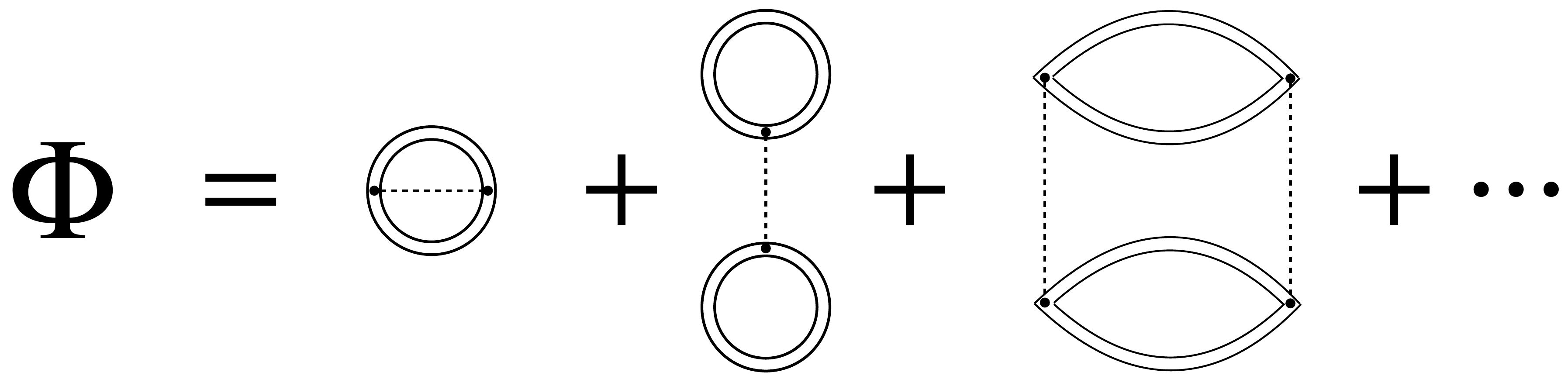}
	\caption{The Luttinger-Ward functional $\Phi[\textbf{G}]$ is a sum over closed two-particle skeleton diagrams. The first summand is a particle-hole pair interacting with itself, the second summand are two particle-hole pairs interacting with each other once, the third summand are two particle-hole pairs interacting with each other twice.}
	\label{fig:feynman_diagrams}
\end{figure}

In this Appendix we outline foundations of the VCA.
Following Luttinger and Ward \cite{Luttinger1960a}, one can consider the grand canonical potential of interacting fermions to
be a functional of the Green's function $\textbf{G}$ and the self-energy~$\boldsymbol{\Sigma}$,
\begin{equation}
\Omega_\textbf{t}[\textbf{G}, \boldsymbol{\Sigma}] =  
- \text{Tr}\ln(\textbf{G}_{0}^{-1} - \boldsymbol{\Sigma}) - \Tr ( \textbf{G} \boldsymbol{\Sigma}) + \Phi[\textbf{G}],
\label{eq:LW_functional}
\end{equation}
where $\textbf{G}_{0}^{-1}$ is the non-interacting Green's function and $\Phi[\textbf{G}]$ is the Luttinger-Ward functional. Diagrammatically, the latter can be defined as a sum over all irreducible two-particle diagrams, referred to as skeleton diagrams, Fig.~\ref{fig:feynman_diagrams}.
In the expression above both the Green's function, $G^{\alpha\beta}_{\tau_1\tau_2}$, and the self-energy,
$\Sigma^{\alpha\beta}_{\tau_1\tau_2}$, have to be understood as matricies in position, spin and
(Matsubara) time domains, with Greek letters, $\alpha=(i,\sigma)$, being used as combined indices in lattice and spin spaces. 
The functional $\Omega_\textbf{t}[\textbf{G}, \boldsymbol{\Sigma}]$ achieves its stationary value at the physical $\textbf{G}$
and $\boldsymbol{\Sigma}$.
In particular, a functional derivative of the Luttinger-Ward functional gives the diagrammatic expansion for the self-energy, 
\begin{equation}
\label{eq:self_energy_expansion}
    {\delta \Phi[\textbf{G}]}/{\delta \textbf{G}}  =  \boldsymbol{\Sigma}[\textbf{G}].
\end{equation}
This relation guarantees that $\delta\Omega_\textbf{t}[\textbf{G}, \boldsymbol{\Sigma}]/\delta\textbf{G} = 0$.
On the other hand, optimization over the self-energy yields the exact Dyson equation,
\begin{equation}
\label{eq:Dyson}
\frac{\delta \Omega_\textbf{t}[\textbf{G}, \boldsymbol{\Sigma}]}{\delta \boldsymbol{\Sigma}} = 0\, \Rightarrow\,
(\textbf{G}_{0}^{-1} - \boldsymbol{\Sigma}) \textbf{G}  = \mathds{1}.
\end{equation}
The variational principle outlined above can be simplified if one assumes that one can resolve~(\ref{eq:self_energy_expansion}) by
defining the Green's function $\textbf{G}=\textbf{G}[\boldsymbol{\Sigma}]$ in terms of the self-energy.
The functional~(\ref{eq:LW_functional}) then reduces to
\begin{equation}
 \Omega_\textbf{t}[\boldsymbol{\Sigma}] =  
- \text{Tr}\ln(\textbf{G}_{0}^{-1} - \boldsymbol{\Sigma}) + F[\boldsymbol{\Sigma}],
\end{equation}
where we have introduced the Legendre transform of the Luttinger-Ward functional, 
\begin{equation}
    F[\boldsymbol{\Sigma}] = \Phi[\textbf{G}[\boldsymbol{\Sigma}]] - \Tr(\boldsymbol{\Sigma} \textbf{G}[\boldsymbol{\Sigma}]),
\end{equation}
which satisfies $\delta F[\boldsymbol{\Sigma}] / \delta \boldsymbol{\Sigma} = \textbf{G}[\boldsymbol{\Sigma}]$. 
It follows that $\Omega_\textbf{t}[\boldsymbol{\Sigma}]$ is stationary at the physical self-energy,
since the condition $\delta \Omega_\textbf{t}[\boldsymbol{\Sigma}]/\delta \boldsymbol{\Sigma} = 0$
constitutes the Dyson equation~(\ref{eq:Dyson}).

The Luttinger-Ward functional, and hence $F[\boldsymbol{\Sigma}]$, is not known in general~\footnote{
One exception is the so-called SYK model with random all-to-all two-body interaction, where the Luttinger functional can be found on average,  $\Phi[\boldsymbol{G}] = J \int d^2 \tau\,  G^4_{\tau_1 \tau_2}$, with $J$ being an interaction strength.}, cf.~\cite{kitaev2020}. 
However it is universal in the sense that it is defined only by the interaction part of the Hamiltonian, $H_1({\bf U})$, and is independent of $H_0({\bf t})$. This observation has motivated Potthoff to restrict the class of variational self-energies to those which optimize
the functional $\Omega_\textbf{t'}[\boldsymbol{\Sigma}]$ for the reference system of disjoint clusters described by the Hamiltonian $H=H_0(\textbf{t'})+H_1(\textbf{U})$. Denoting the (exact) solution of this optimization problem by $\boldsymbol{\Sigma}_{\textbf{t'}}$, 
one can relate the original functional $\Omega_\textbf{t}[\boldsymbol{\Sigma}]$ of the physical system to the reference one,
$\Omega_{\textbf{t'}} \equiv \Omega_\textbf{t'}[\boldsymbol{\Sigma}_{\textbf{t'}}]$,
by a simple relation 
\begin{equation}
\Omega_\textbf{t}[\boldsymbol{\Sigma}_{\textbf{t'}}] = \Omega_{\textbf{t'}}\! -\!
 \text{Tr}\ln(\textbf{G}_{0}^{-1} - \boldsymbol{\Sigma}_{\textbf{t'}}) +  
 \text{Tr}\ln(\textbf{{G}'}_{0}^{-1} - \boldsymbol{\Sigma}_{\textbf{t'}}), \nonumber
\end{equation}
where $\textbf{G}'_{0}$ represents the free fermion propagator of a cluster. 
The above approximate functional can be rewritten in the Fourier space
with the help of Matsubara sums and treating a single cluster as a unit cell of the infinite size physical system, 
\begin{equation}
\label{eq:Omega_t_prime}
    \Omega_\textbf{t}[\boldsymbol{\Sigma}_{\textbf{t'}}]  = 
    \Omega_{\textbf{t'}}\! -\!
    T \sum_{\omega_n, {\bf k}} {\rm Tr} \ln \left( 1- V_{\bf k} \textbf{{G}'}(\omega_n) \right),
\end{equation}
where $\textbf{G'}(\omega_n) = \bigl(i\omega_n - {\bf t}' -  \boldsymbol{\Sigma}_{\textbf{t'}}(\omega_n) \bigr)^{-1}$ 
is the cluster Green's function expressed through the corresponding self-energy with $V_{\bf k} = \textbf{t'} - {\bf t}_{\bf k}$
being a matrix of inter-cluster hopping terms.

The self-consistency scheme of the VCA then substitutes the variational principle in Eq.~(\ref{eq:Dyson})
by optimizing (\ref{eq:Omega_t_prime}) over inter-cluster parameters ${\bf t'}$. The latter may include 
different mean-field order parameters related to expected patterns of symmetry breaking, 
which are not a part of the microscopic Hamiltonian~(\ref{eq:FH_Hamiltonian}).
Such optimization procedure over ${\bf t}'$ requires an efficient evaluation of the Green's function $\mathbf{G}$, 
which turns out to be the computationally most demanding task and is thereby delegated to quantum hardware. 
Indeed, let $N_c$ be the number of sites per cluster. On taking into account spin, the dimension of its Hilbert space grows exponentially as
$D = 4^{N_c}$, thus brute-force classical computations need to operate with $D\times D$ linear problems. 
On the other hand, an alternative computation of a cluster Green's function $\textbf{G'}$ using a quantum circuit requires
measurements of only ${\mathcal{O}}(N_c^2)$ site-to-site correlation functions, while the number of qubits scales as $N_c$. 
Based on this idea, we show in the remainder of this paper a new efficient route how to evaluate the Green's function on a quantum computer. 

\section{Set of rules for hopping operators}
\label{app:bosonization_rules}
The hopping operators differ based on the hopping direction. We distinguish between hopping $x$-direction and in $y$-direction.
\vspace{1em}
\paragraph{Hopping in $x$-direction:}
Within the bosonization framework~\cite{bosonization}, the elementary bilinear hopping operators along the $x$-axis are represented in terms of qubit Pauli operators 
in the following way: 
\begin{align}
	&\hspace{0.4em} i \, \left\{ 
	x_{\bf r}^\uparrow \bar{x}_{\bf r}^\uparrow, \, 
	x_{\bf r}^\uparrow \bar{y}_{\bf r}^\uparrow, \, 
	y_{\bf r}^\uparrow \bar{x}_{\bf r}^\uparrow, \, 
	y_{\bf r}^\uparrow \bar{y}_{\bf r}^\uparrow 
	\right\} \notag \\
	\mapsto &\left\{ 
	- X_{{\bf r} \uparrow }^{(1)} \bar{Y}_{{\bf r} \uparrow }^{(1)}, \, 
	- X_{{\bf r} \uparrow }^{(1)} \bar{X}_{{\bf r} \uparrow }^{(1)}, \,
	Y_{{\bf r} \uparrow }^{(1)} \bar{Y}_{{\bf r} \uparrow }^{(1)}, \, 
	Y_{{\bf r} \uparrow }^{(1)} \bar{X}_{{\bf r} \uparrow }^{(1)}
	\right\} \notag \\
	&\otimes \bar{Z}_{{\bf r} \uparrow }^{(2)}
	\label{eq:x_axis_maps}
\end{align}

With minimal changes similar mappings also work for spin down hopping operators along the $x$-axis. For instance, the bilinear $\bar x_{\bf r}^\downarrow x_{\bf r}^\downarrow$ is mapped onto  $- \bar X_{{\bf r} \downarrow }^{(1)}  Y_{{\bf r} \downarrow }^{(1)}  Z_{{\bf r} \downarrow}^{(2)}$. The difference to Eq.~(\ref{eq:x_axis_maps}) stems from the opposite way of ordering physical versus auxiliary qubits (, representing spin down fermions, as shown in Fig.~\ref{fig:2DCluster}).

\vspace{1em}
\paragraph{Hopping in $y$-direction:}
Elementary bilinear hopping operators along the $y$-axis involve physical and auxiliary fermions with opposite spins. Again, bosonization rules provide the following correspondence.

\begin{align}
	&\hspace{0.4em} i \, \left\{ 
	x_{\bf r}^\uparrow \bar{x}_{\bf r}^\downarrow, \, 
	x_{\bf r}^\uparrow \bar{y}_{\bf r}^\downarrow, \, 
	y_{\bf r}^\uparrow \bar{x}_{\bf r}^\downarrow, \, 
	y_{\bf r}^\uparrow \bar{y}_{\bf r}^\downarrow 
	\right\} \notag \\
	\mapsto &- \left\{ 
	Y_{{\bf r} \uparrow }^{(1)} \bar{Y}_{{\bf r} \downarrow }^{(1)}, \, 
	Y_{{\bf r} \uparrow }^{(1)} \bar{X}_{{\bf r} \downarrow }^{(1)}, \,
	X_{{\bf r} \uparrow }^{(1)} \bar{Y}_{{\bf r} \downarrow }^{(1)}, \,
	X_{{\bf r} \uparrow }^{(1)} \bar{X}_{{\bf r} \downarrow }^{(1)}
	\right\} \notag \\
	&\otimes Y_{{\bf r} \uparrow }^{(2)} \bar{X}_{{\bf r} \downarrow }^{(2)}
	\label{eq:y_axis_maps1}
\end{align}

and similar mappings for the other choice of vertical links,

\begin{align}
	&\hspace{0.4em} i \, \left\{ 
	\bar{x}_{\bf r}^\uparrow x_{\bf r}^\downarrow, \, 
	\bar{x}_{\bf r}^\uparrow y_{\bf r}^\downarrow, \, 
	\bar{y}_{\bf r}^\uparrow x_{\bf r}^\downarrow, \, 
	\bar{y}_{\bf r}^\uparrow y_{\bf r}^\downarrow 
	\right\} \notag \\
	\mapsto &- \left\{ 
	\bar{Y}_{{\bf r} \uparrow }^{(1)} Y_{{\bf r} \downarrow }^{(1)}, \, 
	\bar{Y}_{{\bf r} \uparrow }^{(1)} X_{{\bf r} \downarrow }^{(1)}, \,
	\bar{X}_{{\bf r} \uparrow }^{(1)} Y_{{\bf r} \downarrow }^{(1)}, \,
	\bar{X}_{{\bf r} \uparrow }^{(1)} X_{{\bf r} \downarrow }^{(1)}
	\right\} \notag \\
	&\otimes \bar{Y}_{{\bf r} \uparrow }^{(2)} X_{{\bf r} \downarrow }^{(2)}
	\label{eq:y_axis_maps2}
\end{align}

\section{Derivation of the Kubo formula}

\label{appendix_kubo}

In this Appendix we summarize the basics of linear response theory and derive the generalized susceptibility~(\ref{eq:chi_ij_t}). Consider
the Hamiltonian of a system, $H'(t) = H + V(t)$, with a perturbation $V(t) = \sum_j \Phi_j(t) A_j $ acting at times $t>0$ and given by the sum of hopping operators $A_j$ defined in Eq.~(\ref{eq:hopping_direct_meas}). Let also $\rho_0 = \rho(t=0)$ be an initial density matrix. Presently, we have
$\rho_0 =  |\Psi(\alpha_*, \beta_*)\rangle\langle \Psi(\alpha_*, \beta_*)|$, with $(\alpha_*, \beta_*)$ being the optimal parameters
of the VHA. However the exact form of $\rho_0$ is not important for what follows. 

By introducing the Heisenberg operators, $\widetilde A_i(t) = e^{iH't} A_i e^{-iH't}$, we are aiming to find how their averages, 
\begin{equation}
 \langle \widetilde A_i(t) \rangle_\Phi = {\rm tr} (\rho_0 A_i(t)),
 \label{eq:average_A}
\end{equation}
change in time in response to the perturbation $V(t)$. Here, a subscript in the average, $\langle \dots \rangle_\Phi$, indicates that 
the latter is a functional of generalized forces $\Phi_j(t)$.
To this end, we switch to the interaction picture by defining $A_i(t) = e^{iHt} A_i e^{-iHt}$ such that
the average in Eq.~\ref{eq:average_A} becomes
\begin{equation}
 \langle \widetilde A_i(t) \rangle_\Phi = {\rm tr}\left[\rho_0 U^+(t) A_i(t) U(t) \right],  
\end{equation}
where
\begin{equation}
U(t) = e^{i H t } e^{- i H' t } \equiv T_t \exp\left\{-i\int_0^t V_I(t')dt' \right\}
\end{equation}
is an evolution operator in the interaction picture expressed via $V_I(t) = \sum_j \Phi_j(t) A_j(t)$.
At this point we may expand $U(t)$ up to first order in perturbation $V_I(t)$ and obtain
\begin{equation}
\langle \widetilde A_i(t) \rangle_{\Phi}   = 
\langle {A}_i(t)\rangle + i \int\displaylimits_{0}^t \text{d}t' \langle [V_I(t'), {A}_i(t)] \rangle + \dots,
\label{eq:expansion}
\end{equation}
where $\langle \dots \rangle$ denotes an average with the initial density matrix $\rho_0$. For our choice of operators $A_i$ the 0th order term in Eq.~\ref{eq:expansion} vanishes. On introducing the response function
\begin{equation}
   \chi_{ij}(t,t') = -i \Theta(t-t') \langle[A_i(t), A_j(t')]\rangle, 
\end{equation}
we finally find the Kubo formula. It states that in linear order the response of a system to the perturbation $\Phi_j(t)$ is given by
\begin{equation}
 \label{eq:kubo}
 \delta \langle  \widetilde {A}_i(t)\rangle = \int\limits_0^t  \chi_{ij}(t-t') \Phi_j(t') dt'.
\end{equation}
In particular, if the perturbation is localized in time at $t=0$, i.e. $\Phi_j(t) = \Phi_j\, \delta(t)$, then Eq.~(\ref{eq:kubo}) yields 
\begin{equation}
\delta \langle  \widetilde {A}_i(t)\rangle = \sum_j\chi_{ij}(t) \Phi_j.
\end{equation}
In this form it can be used to construct the corresponding quantum circuits as described in the main text.

\section{Non-linear response}
\label{appendix_nonlin_kubo}
In this Appendix we reconsider the measurement circuit on Fig.~\ref{fig:measurement_full_correlator} at arbitrary strength of perturbation
characterized by the angle $\theta = \Phi_j^{\sigma'}$ and derive Eqs.~(\ref{eq:measurement}) and (\ref{eq:measurement_K}). 

To this end we introduce a wave function 
$|\Psi_*\rangle$ to denote the initial state of a simulated Hubbard cluster at $t=0$. It may or may not be equal
to the VHA ground state $|\Psi(\alpha_*, \beta_*)\rangle$ --- the presented quantum algorithm for the Green's function measurement is independent of this.   
With the ancilla qubit taken into account, an initial state of the quantum circuit on Fig.~\ref{fig:measurement_full_correlator} then reads 
$|\tilde{\Psi} \rangle = |\Psi_* \rangle \otimes |1\rangle_d$. When the perturbation Eq.~(\ref{eq:perturbation_direct_measurement}) followed by
the unitary evolution operator is applied to this state, it evolves into
\begin{equation}
|\Psi_t \rangle = ( U_t \otimes e^{- \tfrac i2 \lambda ( i x_d y_d) } )\, e^{\tfrac 12 \Phi_j^{\sigma'} y_{j\sigma'} x_d} |\tilde{\Psi} \rangle.
\end{equation} 
Here, the angle $\lambda = \epsilon_d t$ and we have expressed the Hamiltonian of the $d$-fermion via Majorana operators,
\begin{equation}
\label{eq:H_d_M}
H_d = \epsilon_d \left(d^+ d -\tfrac 12\right) \equiv  \tfrac i2 \epsilon_d x_d y_d.
\end{equation}
With the state $|\Psi_t \rangle $ at hand our aim is to evaluate the response to the generalized force $\Phi_j^{\sigma'}$ at time $t$, 
\begin{equation}
\label{eq:A_i_average}
\langle A_i^\sigma(t) \rangle = \frac i2 \langle \Psi_t | x_{i\sigma} x_d |\Psi_t \rangle, 
\end{equation}
beyond the linear regime discussed in the main text. Hence, we introduce Majorana operators of a cluster in the Heisenberg representation,
\begin{equation}
\label{eq:x_i_t}
x_{ i\sigma}(t):= U_t^\dagger x_{i\sigma} U_t
\end{equation}
and use the anti-commutation relation of the ancilla Majorana $x_d$ with its Hamiltonian, namely  $\{x_d, H_d\}~=~0$. 
The latter, in turn, yields a simple time evolution for $x_d$, 
\begin{equation}
\label{eq:x_d_t}
x_d(t) : = e^{ \tfrac i2 \lambda ( i x_d y_d) }  x_d e^{- \tfrac i2 \lambda ( i x_d y_d) }  = x_d e^{\lambda ( x_d y_d) },
\end{equation}
where we reiterate that $\lambda = \epsilon_d t$. 
Definitions~(\ref{eq:x_i_t}) and (\ref{eq:x_d_t}) enable one to rewrite the expectation value~(\ref{eq:A_i_average}) in a more
transparent form,
\begin{eqnarray}
\label{eq:A_i_simplified}
\langle A_i^\sigma(t) \rangle = \langle\Psi_0 | e^{ - \tfrac 12 \Phi_j^{\sigma'} y_{j\sigma'} x_d} x_{ i\sigma}(t) x_d(t) 
e^{ \tfrac 12 \Phi_j^{\sigma'} y_{j\sigma'} x_d}  |\Psi_0 \rangle.  \nonumber \\
\end{eqnarray}
Further progress is feasible by observing that any bilinears of (unequal) Majoranas square to identity, e.g. 
$( i y_{j\sigma'} x_d)^2 = ( i y_d x_d)^2 = \mathds{1}$.
As such, one may simplify the exponentials in (\ref{eq:A_i_simplified}) as
\begin{equation}
e^{ \tfrac 12 \Phi_j^{\sigma'} y_{j\sigma'} x_d} = \cos \tfrac 12 \Phi_j^{\sigma'} + y_{j\sigma'} x_d \sin \tfrac 12 \Phi_j^{\sigma'},
\end{equation}
and use the analogous transformation to rearrange Eq.~(\ref{eq:x_d_t}). With the help of these relations the expectation value~(\ref{eq:A_i_simplified})
can be decomposed into eight terms. For each of these terms one performs an average in the ancilla's Hilbert space using the following relations
\begin{eqnarray}
{}_d\langle 1| x_d |1\rangle_d &=& {}_d\langle 1| y_d |1\rangle_d  = 0, \nonumber \\
 {}_d\langle 1| i x_d y_d |1\rangle_d &=& {}_d\langle 1| 2 d^+ d - 1 |1\rangle_d = 1.  
\end{eqnarray}
As the result only four non-zero terms remain, which after straightforward algebra are simplified to
\begin{eqnarray}
\label{eq:A_i_av}
\!\!\!\langle A_i^\sigma(t) \rangle &=&  \frac 12 \sin \Phi_j^{\sigma'} \sin\lambda \, \langle \Psi_*| \{x_{i\sigma} (t), y_{j\sigma'}\}|\Psi_*\rangle
\nonumber \\
&-& \frac i 2 \sin \Phi_j^{\sigma'} \cos\lambda \, \langle \Psi_*| [x_{i\sigma} (t), y_{j\sigma'}]|\Psi_*\rangle.
\end{eqnarray}
To arrive at this expression one makes use of anti-commuting properties of Majoranas, i.e. $\{x_{i\sigma}(t), x_d\} = \{x_{i\sigma}(t), y_d\} = 0$. As one can see from Eq.~(\ref{eq:A_i_av}), by choosing the angle $\lambda$ in the quantum circuit on Fig.~\ref{fig:measurement_full_correlator} to be $\pi/2$ or $0$, the measurement outcome reproduces, respectively, either the retarded or Keldysh Green's function. Thereby we 
confirm one of our main results given by Eqs.~(\ref{eq:measurement}) and (\ref{eq:measurement_K}).  

\section{Analytical energy landscape of the two-site dimer}
\label{sec:e_landscape_analytical}
\begin{figure}
	\includegraphics[scale=0.45, center]{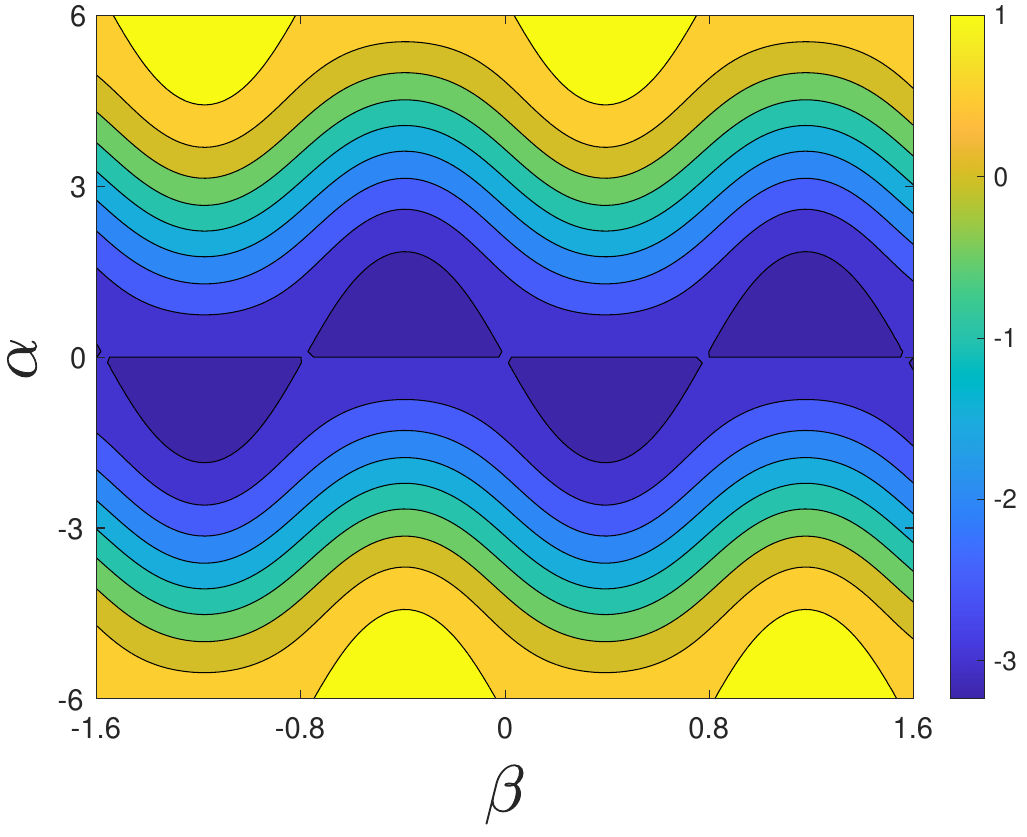}
	\caption{(Color online) Analytical energy landscape for the two-site dimer for finding its ground state via the variational Hamiltonian ansatz with angles $\alpha$ and $\beta$ shown for $t=1$ and $U=4$.}
	\label{fig:e_landscape_analytical}
\end{figure}

Fig. \ref{fig:e_landscape_analytical} is the analytical version of Fig. \ref{fig:e_landscape}. The objective is to find the angles that yield minimum energies. The analytical angles qualitatively match the angles from the simulation.

\begin{figure*}[t]
	\centering
	\includegraphics[scale=0.26]{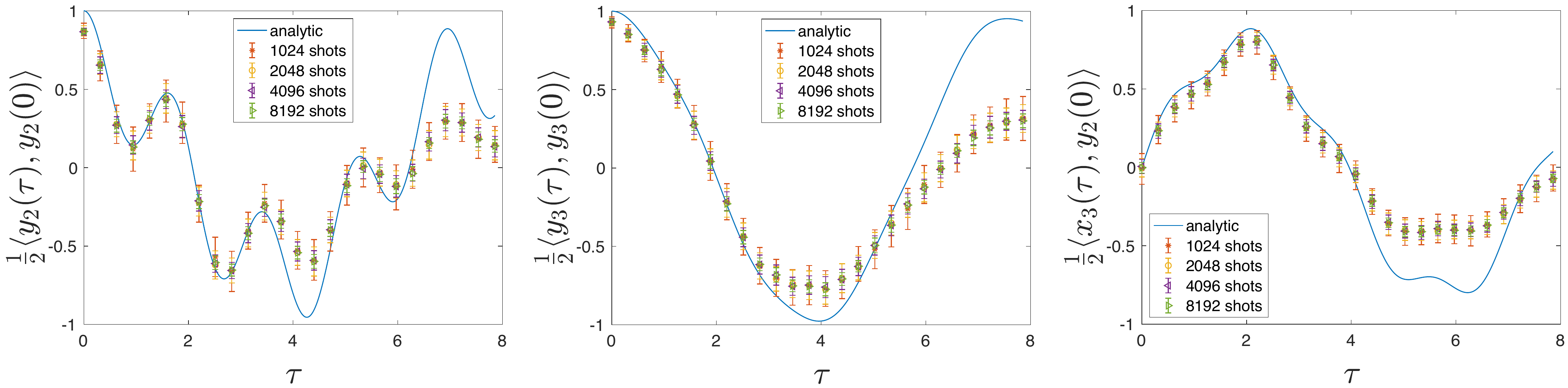}
	\caption{(Color online) Correlators $y_2\!-\!y_2$, $y_3\!-\!y_3$ and $x_3\!-\!y_2$ evaluated in 25 Trotter steps of the duration $\Delta \tau =0.314$. The solid blue curves correspond to the analytical results, cf.\ Eqs.~(\ref{eq:analytical_X0X0}--\ref{eq:analytical_X0Y1}). System parameters are chosen to be $V=1$ and $U=4V$. Errors bars indicating the standard deviation of outcomes decrease with greater number of shots. The correlators were evaluated on Qiskit's noisy \textit{Aer} simulator of the \textit{FakeKolkataV2()} backend, an open-access simulator of the corresponding superconducting device \textit{ibmq\_kolkata}  provided by IBM.}
	\label{fig:shotnumber}
\end{figure*}

\section{Implementation details}
\label{sec:Implementation_details}
In this section we give details on the simulated quantum processor used for showcasing our proposed algorithm. Moreover, we give details on the execution and error mitigation schemes for the presented quantum circuits. Throughout this work, we have worked solely with Qiskit \cite{Qiskit}, an open-source software kit provided by IBM. Qiskit allows users to design, refine and execute quantum software on either simulated quantum hardware, or on quantum hardware made available through IBM.
\subsection{Simulated quantum processor}
\label{sec:simulated-quantum-processor}
\begin{figure}
	\centering
	\includegraphics[scale=0.13]{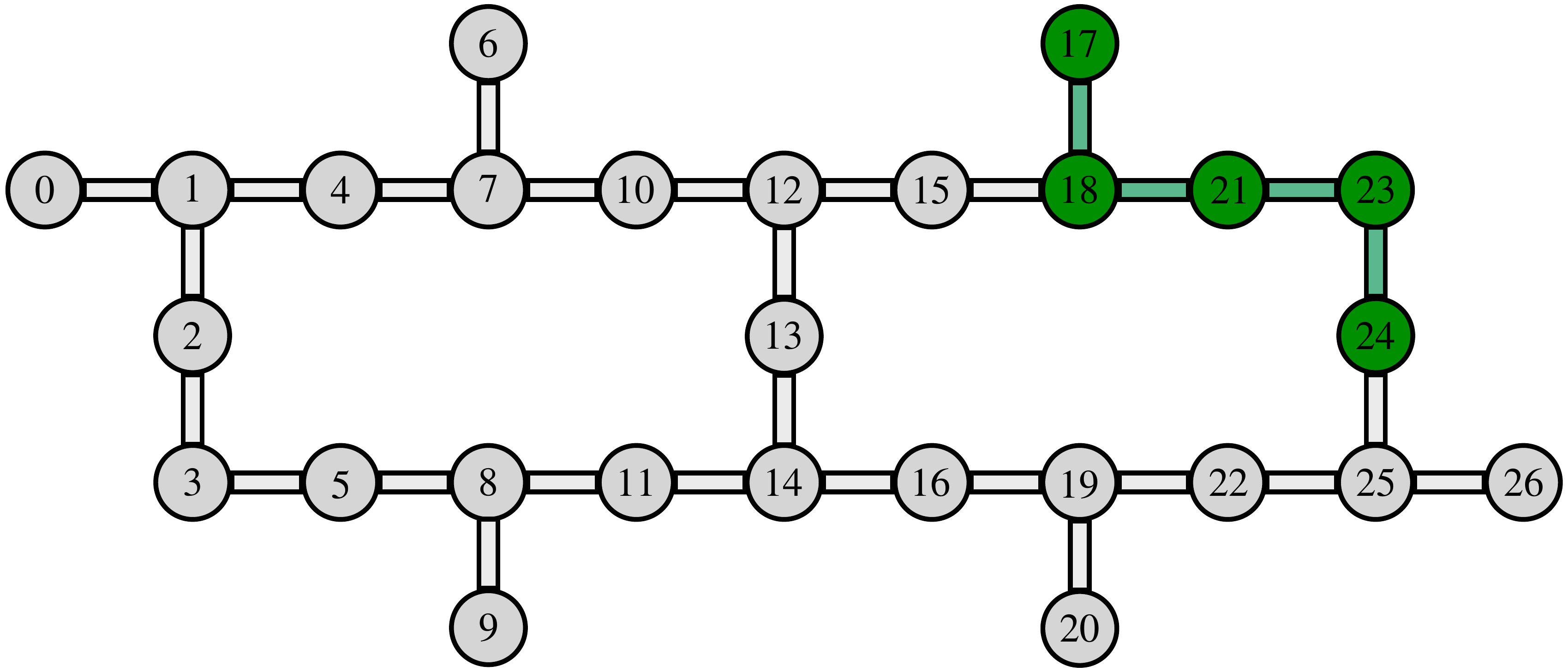}
	\caption{(Color online) Layout of the \textit{ibmq\_kolkata} quantum device, consisting of 27 superconducting qubits and their couplings. The set of physical qubits $\{18, 17, 21, 23, 24\}$ (green) was used for demonstrating our algorithm.}
	\label{fig:kolkatalayout}
\end{figure}

The demonstrations take place on the \textit{FakeKolkataV2()} backend, resembling the characteristics of the \textit{ibmq\_kolkata} quantum device, consisting of 27 superconducting qubits. The layout of the chip as well as couplings between the qubits are depicted in Fig. \ref{fig:kolkatalayout}, where the five qubits required for executing proposed algorithm are highlighted in green, alongside their couplings. For the following analysis, we will denote the simulated qubits, which have been used for our demonstrations as $q^{(l)}_{p}$, for which the exponent is the logical qubit number, and $p$ is the number of the physical qubit, which arises from the mapping of logical to physical qubits. Let $\mathcal{L}=\{0,1,2,3,4\}$ be the set of logical qubits and $\mathcal{P}=\{18, 17, 21, 23, 24\}$ be the set of physical qubits. Then, a mapping from logical to physical qubits is performed via $f:\mathcal{L}_i \mapsto \mathcal{P}_i$, where $i$ points to the $i$th element of the corresponding set.
At the time of performing the simulations, the device has the following set of basis gates: $ID, X, SX, RZ \text{ and } CX$. In Tab. \ref{tab:qubit-properties} one finds qubit calibration data at the time of the simulation. Furthermore, Tab. \ref{tab:gate-errors} shows single-qubit gate errors and measurement errors. Note, that $RZ$ is a virtual $Z$ rotation and thereby has neither an error, nor a duration. Finally, Tab. \ref{tab:cnot-error} gives the $CX$ errors for qubit pairs $(q^{(l)}_{p}, q^{(l+1)}_{p})$.

\begin{table}
    \begin{tabular}{| c | c | c | c |}
    \hline
     & Frequency $[\text{GHz}]$ & $T_1 [10^{-5}\text{s}]$ & $T_2 [10^{-5}\text{s}]$ \\ \hline
    $q^{(0)}_{18}$ & $5.09$ & $10.93$ & $6.99$ \\ \hline
    $q^{(1)}_{17}$ & $5.24$ & $9.08$ &  $3.53$ \\ \hline
    $q^{(2)}_{21}$ & $5.27$ & $10.13$ &  $10.98$\\ \hline
    $q^{(3)}_{23}$ & $5.14$ & $8.45$ &  $10.78$\\ \hline
    $q^{(4)}_{24}$ & $5.0$ & $11.41$ & $2.61$\\ \hline
    \end{tabular}
   \caption{Calibration data of qubits $q^{(l)}_{p}$, at the time of the simulation.}
    \label{tab:qubit-properties}
\end{table}

\begin{table}[ht]
\centering
\begin{tabular}{|c|c|c|}
\hline
\multirow{2}{*}{} & \multicolumn{2}{c|}{Single-qubit operation error (and duration)} \\
\cline{2-3}
& $\{ID, X, SX\}$ & Measurement \\
\hline
$q^{(0)}_{18}$ & $1.98 \cdot 10^{-4} \hspace{1mm} (3.56 \cdot 10^{-8}\text{s})$ & $7.4 \cdot 10^{-3} \hspace{1mm} (6.76 \cdot 10^{-7}\text{s})$ \\
$q^{(1)}_{17}$ & $4.22 \cdot 10^{-4} \hspace{1mm} (3.56 \cdot 10^{-8}\text{s})$ & $6.1 \cdot 10^{-3} \hspace{1mm} (6.76 \cdot 10^{-7}\text{s})$ \\
$q^{(2)}_{21}$ & $2.59 \cdot 10^{-4} \hspace{1mm} (3.56 \cdot 10^{-8}\text{s})$ & $6.8 \cdot 10^{-3} \hspace{1mm} (6.76 \cdot 10^{-7}\text{s})$  \\
$q^{(3)}_{23}$ & $1.73 \cdot 10^{-4} \hspace{1mm} (3.56 \cdot 10^{-8}\text{s})$ & $7.9 \cdot 10^{-3} \hspace{1mm} (6.76 \cdot 10^{-7}\text{s})$  \\
$q^{(4)}_{24}$ & $1.65 \cdot 10^{-4} \hspace{1mm} (3.56 \cdot 10^{-8}\text{s})$ & $5.3 \cdot 10^{-3} \hspace{1mm} (6.76 \cdot 10^{-7}\text{s})$  \\
\hline
\end{tabular}
\caption{Example table with merged cells in the first column.}
\label{tab:gate-errors}
\end{table}

\begin{table}
    \begin{tabular}{| c | c |}
    \hline
    qubit pair & $CX$ error (and duration)  \\ \hline
    $(q^{(0)}_{18}, q^{(1)}_{17})$ & $1.62 \cdot 10^{-2}  \hspace{1mm} (5.05 \cdot 10^{-7}\text{s})$ \\ \hline
    $(q^{(1)}_{17}, q^{(2)}_{21})$ & $8.77 \cdot 10^{-3}  \hspace{1mm} (4.91 \cdot 10^{-7}\text{s})$\\ \hline
    $(q^{(2)}_{21}, q^{(3)}_{23})$ & $5.39 \cdot 10^{-3}  \hspace{1mm} (3.63 \cdot 10^{-7}\text{s})$ \\ \hline
    $(q^{(3)}_{23}, q^{(4)}_{24})$ & $5.34 \cdot 10^{-3} \hspace{1mm} (2.84 \cdot 10^{-7}\text{s})$ \\ \hline
    \end{tabular}
   \caption{$CX$ gate errors on qubit pairs $(q^{(l)}_{p}, q^{(l+1)}_{p})$, at the time of the simulation.}
    \label{tab:cnot-error}
\end{table}

\subsection{Quantum simulation details}
\label{app:Simulation_Details}
For performing noisy simulations, we employ Qiskit's \textit{Aer}, a high performance simulator for executing quantum circuits within Qiskit. Next, we chose the \textit{FakeKolkataV2()} backend, which is supposed to mimick the behavior of the quantum device \textit{ibmq\_kolkata}, as described in \ref{sec:simulated-quantum-processor}. These two ingredients allow for noisy simulations based on the \textit{ibmq\_kolkata} device with given gate fidelities, operations and durations. This is particularly useful to gauge the accuracy of results, if demonstrations took place on the real quantum device.
\subsubsection{Notation for quantum gates}
We summarize the notation for quantum gates as they appear in our algorithms in Table \ref{tab:applied_gates}.

\begin{table}
	\centering
	\begin{tblr}{
			cells={valign=m,halign=c},
			row{1}={bg=lightgray,font=\bfseries,rowsep=8pt},
			colspec={QQQ},
			hlines,
			vlines
		}
		Gate & Symbol & Matrix\\
		Hadamard & \includegraphics[scale=0.04,valign=c]{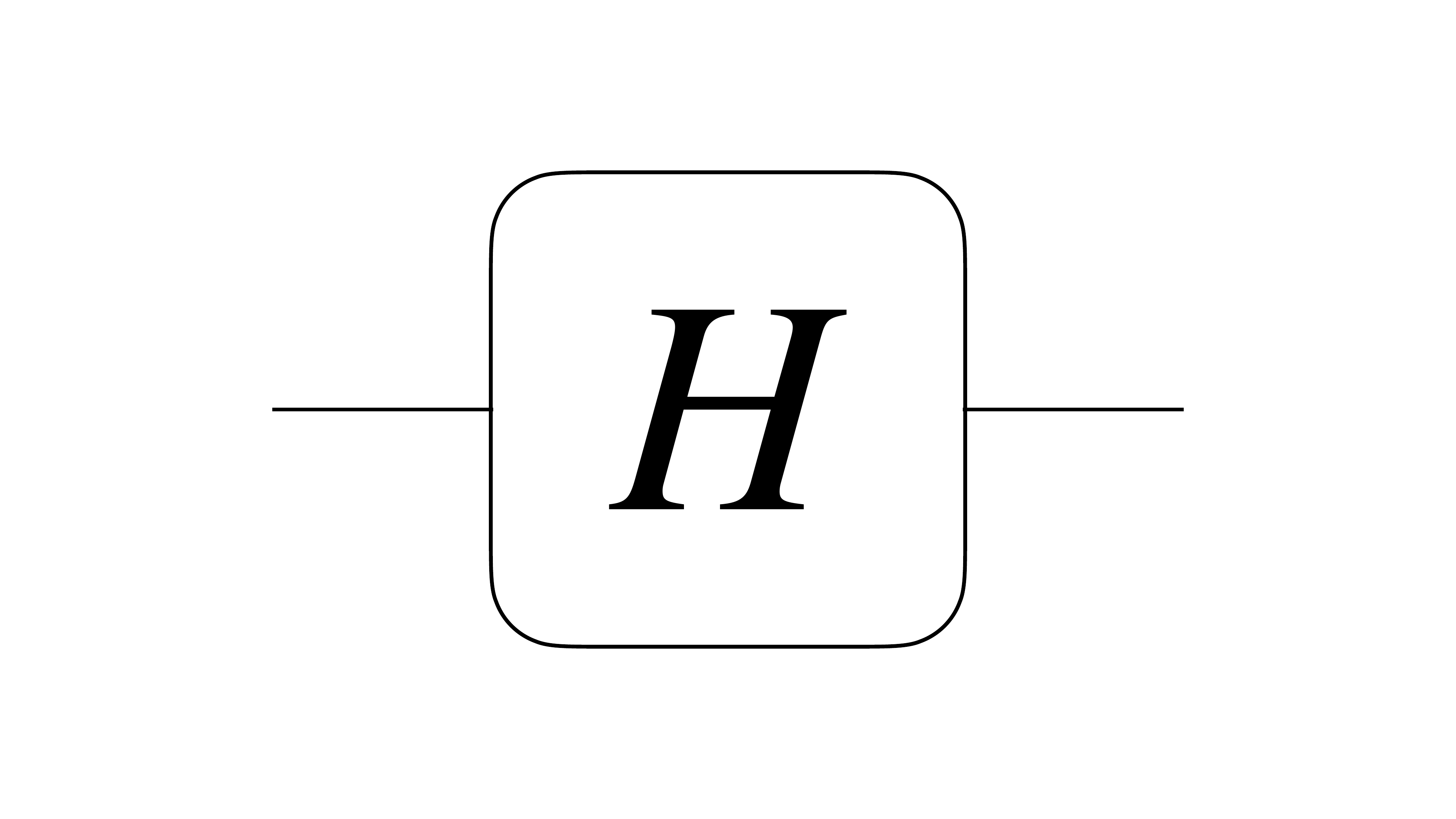}  
		& $\frac{1}{\sqrt{2}}\begin{pmatrix}
			1 & 1 \\
			1 & -1
		\end{pmatrix}$ \\ 
		X & \includegraphics[scale=0.04,valign=c]{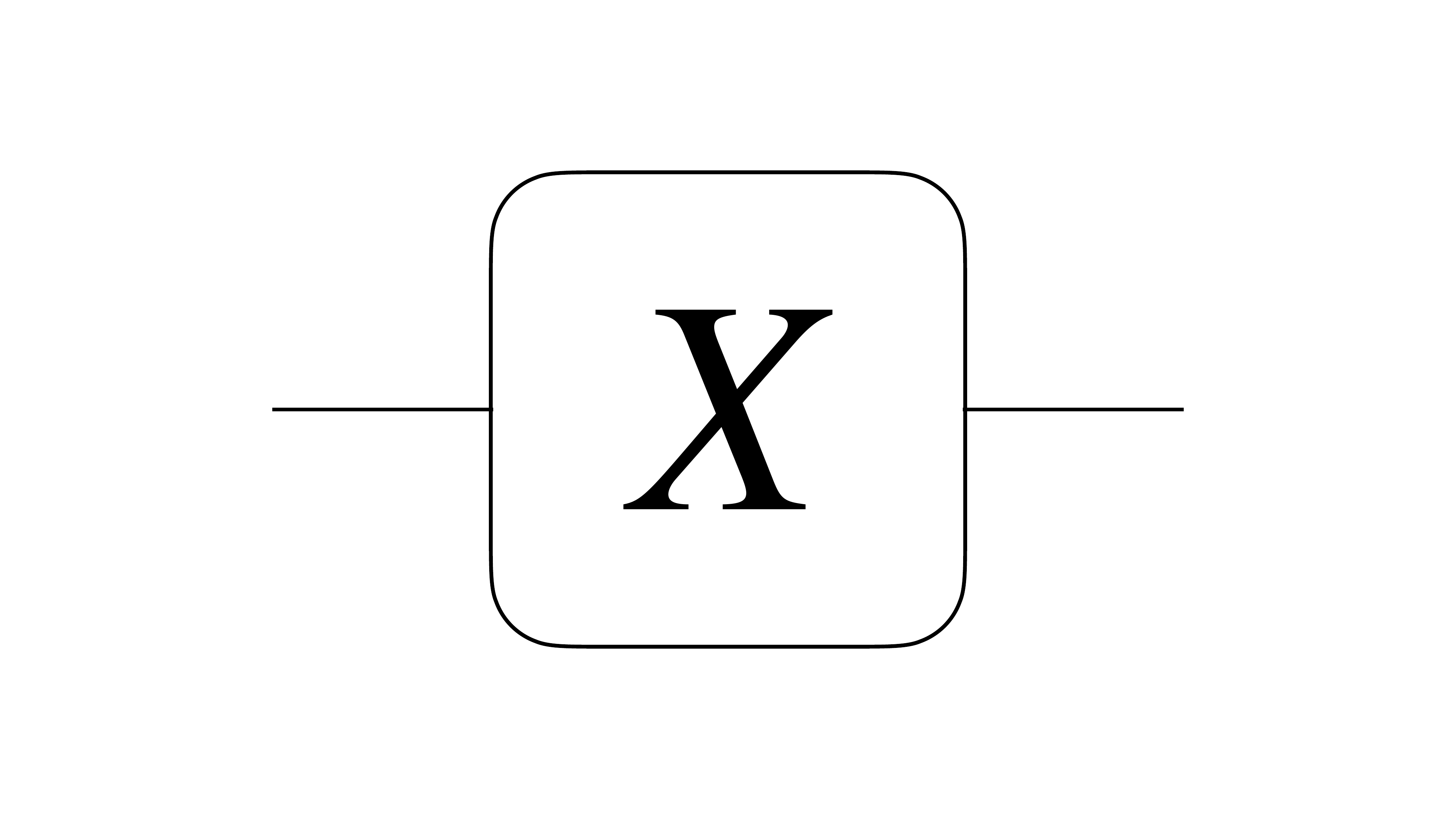}
		& $\begin{pmatrix}
			0 & 1 \\
			1 & 0
		\end{pmatrix}$
		\\ 
		Z & 
		\includegraphics[scale=0.04,valign=c]{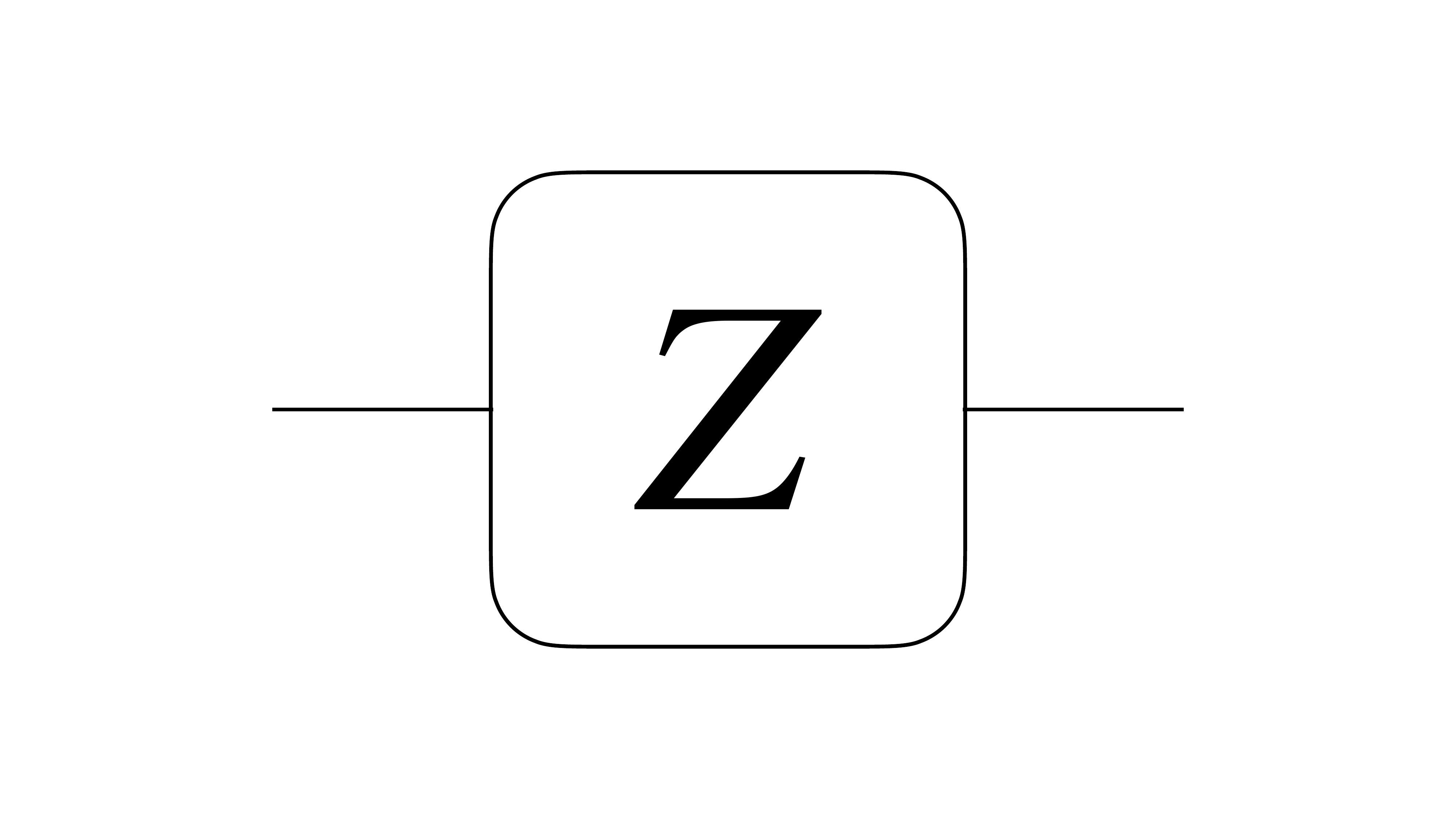}
		& $\begin{pmatrix}
			1 & 0 \\
			0 & -1
		\end{pmatrix}$
		\\ 
		Z-rotation & 
		\includegraphics[scale=0.04,valign=c]{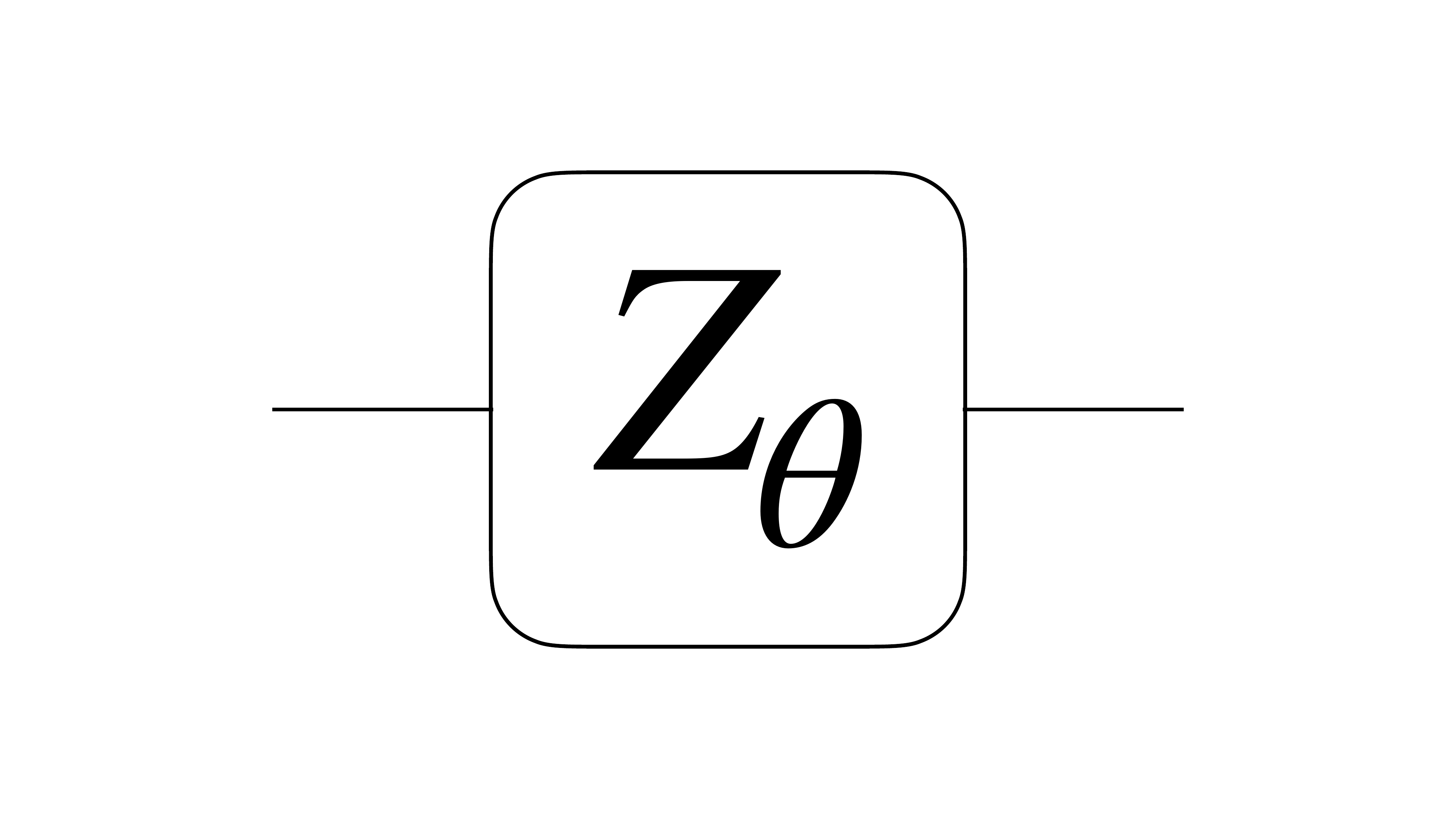}
		&   $\begin{pmatrix}
			e^{-i\theta/2} & 0 \\
			0 & e^{i\theta/2}
		\end{pmatrix}$
		\\ 
		Y-basis change & 
		\includegraphics[scale=0.04,valign=c]{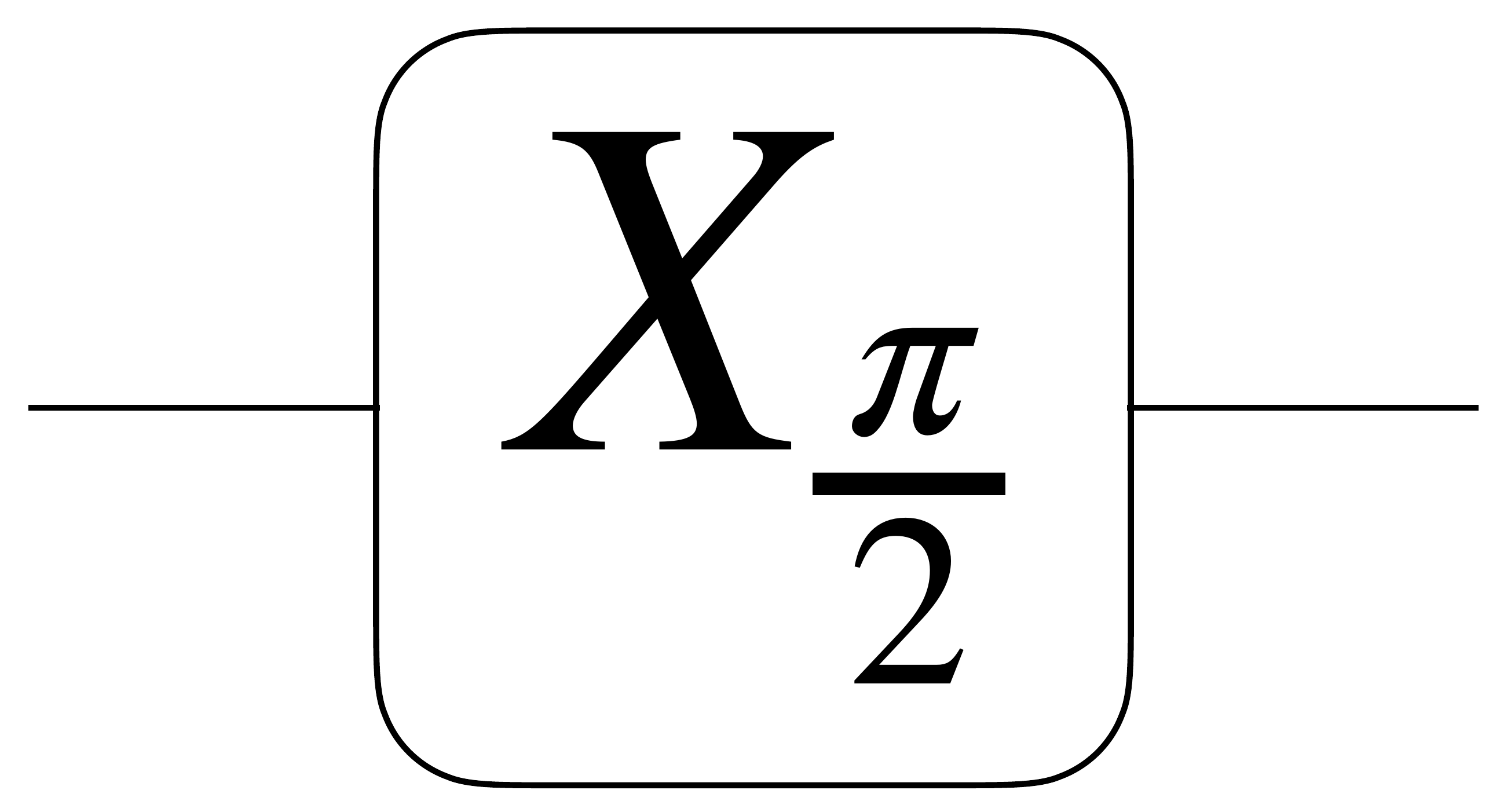}
		&  $\frac{1}{\sqrt{2}}\begin{pmatrix}
			1 & -i \\
			-i & 1 
		\end{pmatrix}$
		\\ 
		Phase & 
		\includegraphics[scale=0.04,valign=c]{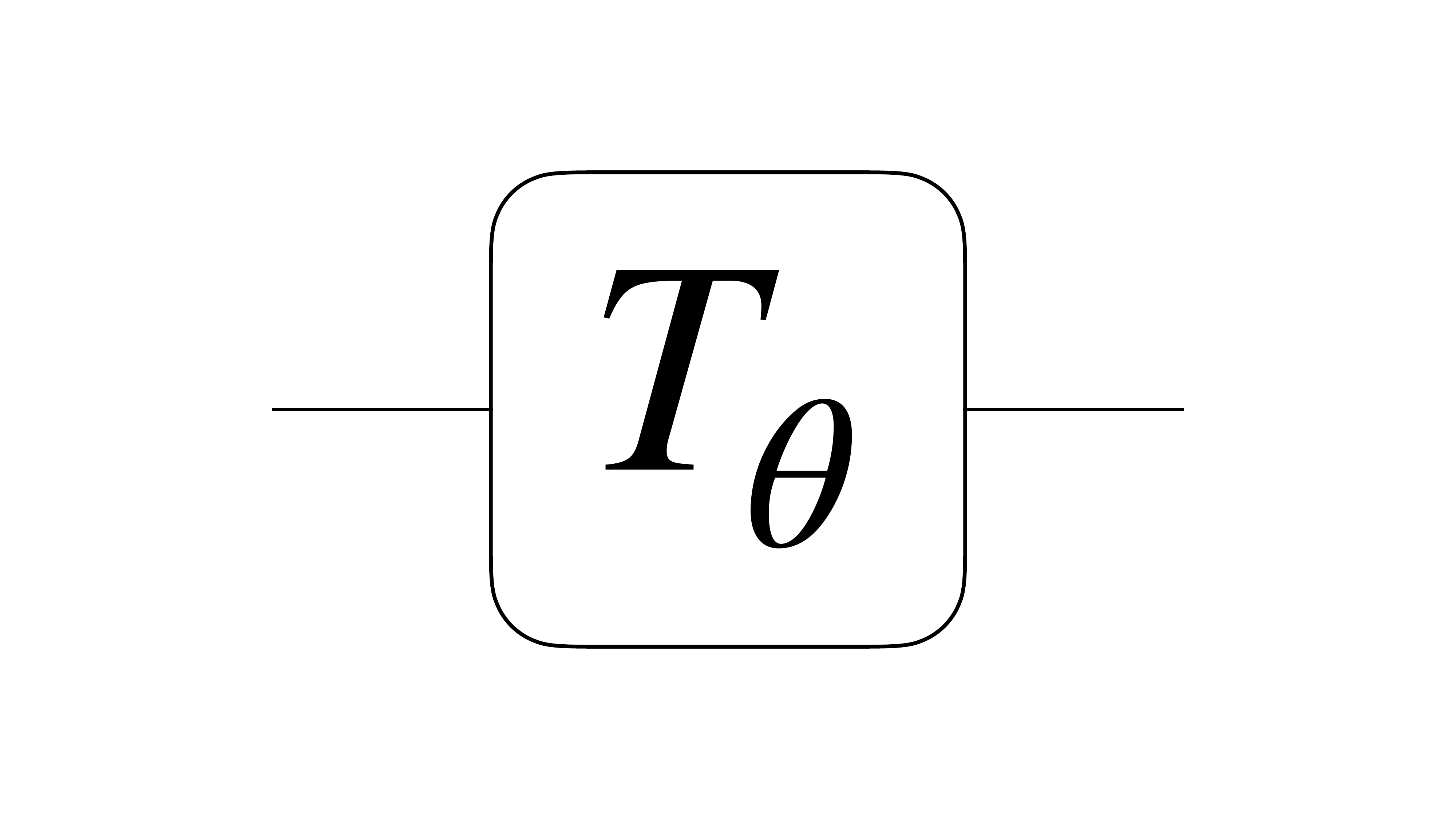}
		& $\begin{pmatrix}
			1 & 0 \\
			0 & e^{i\theta} 
		\end{pmatrix}$
		\\ 
		Controlled-Not & 
		\includegraphics[scale=0.06,valign=c]{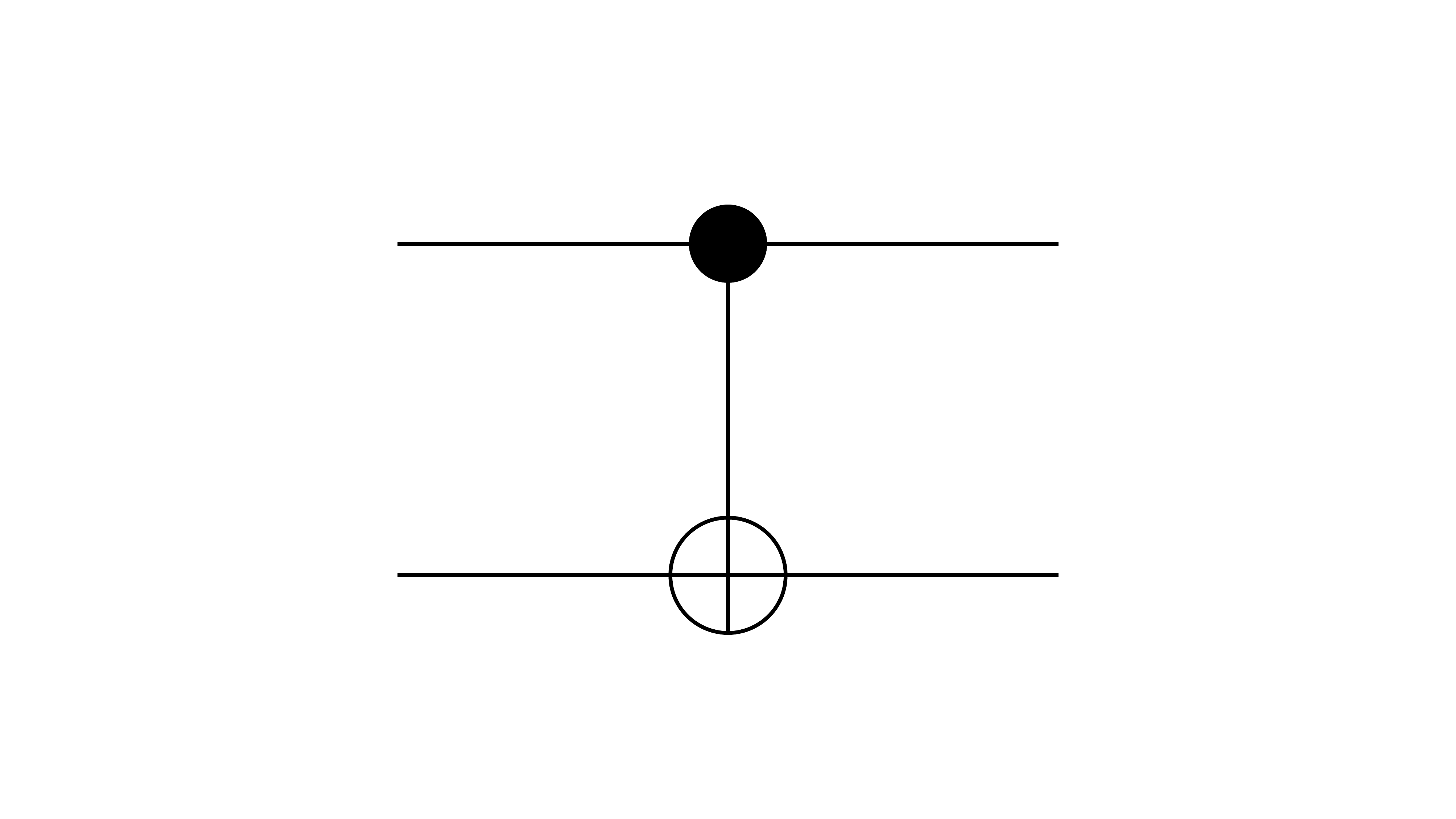}
		& $\begin{pmatrix}
			1 & 0 & 0 & 0\\
			0 & 1 & 0 & 0 \\
			0 & 0 & 0 & 1 \\
			0 & 0 & 1 & 0
		\end{pmatrix}$
		\\ 
		Controlled-Z & \includegraphics[scale=0.06,valign=c]{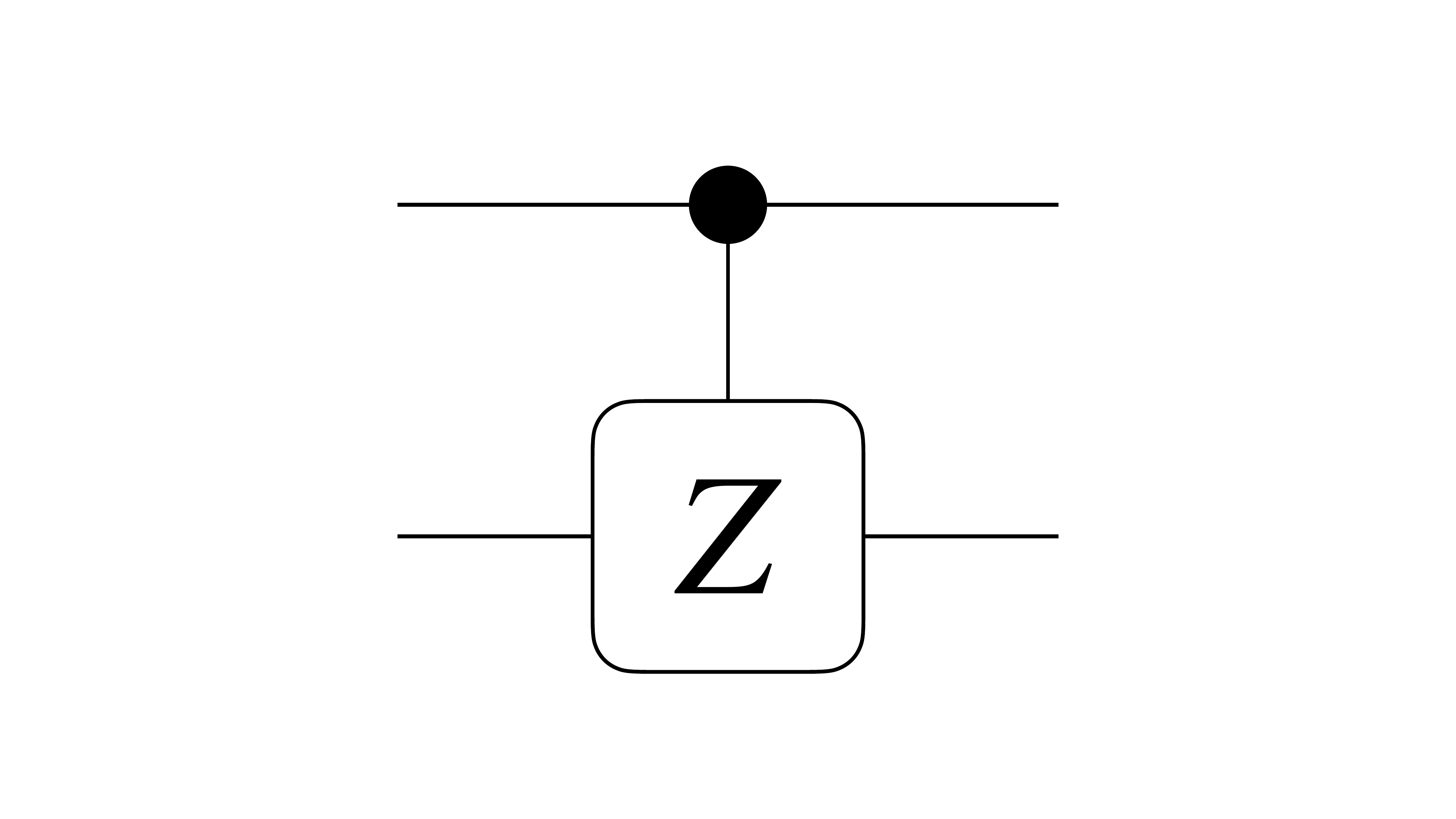}
		& $\begin{pmatrix}
			1 & 0 & 0 & 0\\
			0 & 1 & 0 & 0 \\
			0 & 0 & 1 & 0 \\
			0 & 0 & 0 & -1
		\end{pmatrix}$
	\end{tblr}
	\caption{This table summarizes our conventions for elementary quantum gates used throughout the paper.}
	\label{tab:applied_gates}
\end{table}  

\subsubsection{Error mitigation techniques}
We made use of several error mitigation techniques \cite{QEM}, implemented natively in Qiskit. In the following, we briefly outline the techniques, which have been proven useful in our calculations.

\begin{itemize}
\item \textit{Readout error mitigation} aims to reduce errors in measurements of bit strings. In its standard form, one prepares the computational states of a given system size and measures the frequency of associated bit strings. Discrepancies in the statistics can then be inversed by simple algebra in successive quantum computations.
\item \textit{Pauli twirling} is used to transform coherent errors into incoherent errors in form of a Pauli channel. Due to the relative simplicity of a Pauli channel, these errors can be mitigated more effectively. Technically, two-qubits gates $CX$ and $CZ$ gates are sandwiched between two pairs of single qubit gates, chosen randomly from a set of Pauli gates, which satisfy the equivalence of a two-qubit gate and its sandwiched counterpart. This procedure is repeated for a number of circuits, each governed by equivalent physics, however different in their gate symphony. Lastly, the results for the list of circuits are averaged out.
\item \textit{Dynamical decoupling} is applied on idle qubits during the computation to protect against decoherence. Usually, one applies a sequence of repetitive Pauli gates, e. g. $XX$ or $ZZ$, on idle qubits, which are timed such that possible phase flip and bit flip errors are reduced.
\item \textit{Zero-noise extrapolation.}
Quantum computers in the NISQ era are known for being error prone. Zero-noise extrapolation (ZNE) \cite{ZNE} respects the noisy nature of such computers by artificially increasing the noise tractably and systematically. One then extrapolates the results of increasingly noisy computations back to the zero-noise limit.
\end{itemize}

\subsection{Employment of error mitigation techniques}
\enlargethispage{\baselineskip}
The methods used for generating the results shown in Fig. \ref{fig:e_landscape} are readout error mitigation and dynamical decoupling with $XX$ sequences on idle qubits. On the other hand, the evaluation of correlators, Fig. \ref{fig:Y2Y2_result}, requires a more extensive use of error mitigation techniques. In addition to readout error mitigation and dynamical decoupling, Pauli twirling with 100 repetitions was applied. On top of that, we have employed zero-noise extrapolation at the noise scaling factors of $[1.0, 1.5, 2.0, 2.5, 3.0]$ with a polynomial extrapolator of second order. Finally, the shot number for all demonstrations was set to 4096, as Fig. \ref{fig:shotnumber} suggests this number balances out computational effort and accuracy of a result.

\end{document}